\documentclass[12pt,draftcls, onecolumn]{IEEEtran}
\usepackage[utf8]{inputenc}
\usepackage{amsmath,amsfonts,amssymb}
\usepackage{multicol}
\usepackage{comment}
\usepackage{subfig}
\usepackage{booktabs}
\usepackage{float}
\usepackage{tikz}
\tikzset{>=stealth}
\usepackage{braket}
\usepackage{pgfplots}
\pgfplotsset{compat=1.18}
\usepgfplotslibrary{fillbetween}
\usetikzlibrary{spy}
\usepackage{physics}
\usepackage{cases}
\usepackage{bbm}
\usetikzlibrary{patterns}
\tikzstyle{vertex}=[auto=left,circle,fill=black!25,minimum size=20pt,inner sep=0pt]
\usetikzlibrary{decorations.pathreplacing,calligraphy}
\usepackage[linesnumbered,ruled,vlined]{algorithm2e}
\usepackage[short]{optidef}
\SetKwRepeat{Do}{do}{while}
\renewcommand{\vec}[1]{\ensuremath{\boldsymbol{#1}}}
\usepackage[thinc]{esdiff}
\allowdisplaybreaks
\title{Automatic Link
Selection in Multi-Channel Multiple Access with Link Failures}
\author{Mevan Wijewardena, Michael J. Neely, Haipeng Luo
\thanks{Mevan Wijewardena, and Michael J. Neely are with the Electrical Engineering department at the University of Southern California.}
\thanks{Haipeng Luo is with the Computer Science department at the University of Southern California.}
}
\date{March 2026}
\usepackage{pgfplots}
\newtheorem{theorem}{Theorem}
\newtheorem{lemma}{Lemma}
\newtheorem{corollary}{Corollary}[theorem]
\begin{document}
\maketitle
\title{Automatic Link
Selection in Multi-Channel Multiple Access with Link Failures}

\author{Mevan Wijewardena, Michael J. Neely, Haipeng Luo
\thanks{Mevan Wijewardena and Michael J. Neely are with the Electrical Engineering department at the University of Southern California.}
\thanks{Haipeng Luo is with the Computer Science department at the University of Southern California.}
\thanks{This work was presented in part at the WiOpt conference in Linköping, Sweden, May 2025~\cite{Wijewardena_Wiopt2025}.}
}

\markboth{Journal of \LaTeX\ Class Files,~Vol.~14, No.~8, 
}%
{Shell \MakeLowercase{\textit{et al.}}: A Sample Article Using IEEEtran.cls for IEEE Journals}


\maketitle

\begin{abstract}
This paper focuses on the problem of automatic link selection in multi-channel multiple access control using bandit feedback. In particular, a controller assigns multiple users to multiple channels in a time-slotted system, where in each time slot, at most one user can be assigned to a given channel, and at most one channel can be assigned to a given user. Given that user $i$ is assigned to channel $j$, the transmission fails with a fixed unknown probability \(1-q_{i,j}\). The assignments are made dynamically using success/failure feedback. The goal is to maximize the time-average utility, where we consider an arbitrary (possibly nonsmooth) concave, entrywise nondecreasing utility function. The first proposed algorithm has fast \(\mathcal{O}(\sqrt{\log(T)/T})\) convergence. However, this algorithm requires solving a convex optimization problem within each iteration, which can be computationally expensive. The second algorithm has slower \(\mathcal{O}(\sqrt[3]{\log(T)/T})\) convergence, while avoiding the costly inner optimization. Both of these algorithms are adaptive. In particular, the convergence guarantee holds for any interval of \(T\) consecutive slots during which the success probabilities do not change. We further study several special cases. In the single-channel setting, we obtain both fast $\mathcal{O}(\sqrt{\log(T)/T})$ convergence and efficient implementation via a simpler adaptive mechanism. We also consider a UCB-based non-adaptive algorithm with max-weight-type decisions. Simulations highlight intriguing performance trade-offs and demonstrate rapid adaptation of the proposed adaptive schemes.
\end{abstract}

\allowdisplaybreaks
\section{Introduction}
We consider the Multiple Access Control (MAC)  problem with $n$ users and $m$ channels in slotted time $t \in \mathbb{N}$. In each time slot, a controller assigns users to channels such that at most one user is assigned to a given channel and at most one channel is assigned to a given user. The transmissions over each channel may fail. In particular, there exist $q_{i,j} \in [0,1]$ for each $(i,j) \in \{1,2,\dots,n\} \times \{1,2,\dots,m\}$, where in time slot $t$, given that the controller decided to assign user $i$ to channel $j$, the transmission fails independently with probability $1-q_{i,j}$. The controller does not know the probabilities $q_{i,j}$. Instead, at the end of every slot, it receives feedback on the successes.

Define the matrices $\vec{Y}(t),\vec{S}(t) \in \{0,1\}^{n \times m}$ and vector $\vec{X}(t) \in \{0,1\}^n$, where
\begin{align}
    S_{i,j}(t) = \begin{cases}
        1 & \text{ link } i,j \text{ is successful in time slot } t\\
        0 & \text{ otherwise},
    \end{cases}\nonumber
\end{align}
\begin{align}
    Y_{i,j}(t) = \begin{cases}
        1 & \text{ user } i \text{ is  assigned to channel } j \text{ in time slot } t\\
        0 & \text{ otherwise} .
    \end{cases} \nonumber
\end{align}
and $X_{i}(t) = \sum_{j=1}^mY_{i,j}(t)S_{i,j}(t)$ for all $i \in \{1,2,\dots,n\}$. The goal is to maximize $\lim_{T \to \infty}\phi(\mathbb{E}\{\vec{\overline{X}}(T)\})$ using feedback on the transmission failures, where $\phi: \mathbb{R}^n \to \mathbb{R}$ is a concave entrywise nondecreasing utility function known to the controller and  $\vec{\overline{X}}(T) = \frac{1}{T}\sum_{t=1}^T \vec{X}(t)$. \footnote{The limit is assumed to exist for simplicity of this introduction; the precise goal is to maximize a $\lim\inf_{T \to\infty}\phi(\vec{\overline{X}}(T))$}

We also focus on establishing finite time bounds. In particular, for given a finite time horizon $T \in \mathbb{N}$, we require the algorithm to satisfy
\begin{align}
    \phi^{\text{opt}}  - \phi\left(\frac{1}{T}\mathbb{E}\left\{\sum_{t= 1}^{T}\vec{X}(t)\right\}\right)  \leq g(T), \nonumber
\end{align}
where $\phi^{\text{opt}}$ is the optimal utility of the original problem and $g$ is a nonnegative function such that
\begin{align}
    \lim_{T \to \infty} g(T) = 0 \nonumber
\end{align}

In addition, we are looking for algorithms that are adaptive. Formally, consider a system in which the channel success probabilities $q_{i,j}$ may change. In such a system, given a $T \in \mathbb{N}$, we require
\begin{align}\label{eqn:first_goal}
    \phi^{\text{opt}}  - \phi\left(\frac{1}{T}\mathbb{E}\left\{\sum_{t= T_0}^{T+T_0-1}\vec{X}(t)\right\}\right) \leq g(T)
\end{align}
for any $T_0 \in \mathbb{N}$, irrespective of the success probabilities outside of the time frame $[T_0:T_0+T-1]$, given that success probabilities remained constant in the frame. Here, $\phi^{\text{opt}}$ is the optimal utility of the original problem that uses the constant success probabilities in $[T_0:T_0+T-1]$ of the above scenario. Note that $g$ is the same function regardless of $T_0$. In fact, it can be shown that given $T^{'} \in \mathbb{N}$, our algorithms can achieve
\begin{align}
    \phi^{\text{opt}}  - \phi\left(\frac{1}{T}\mathbb{E}\left\{\sum_{t= T_0}^{T+T_0-1}\vec{X}(t)\right\}\right) \leq g(T^{'})\nonumber
\end{align}
for any $T_0 \in \mathbb{N}$ and $T \geq T^{'}$, where the algorithm parameters only depend on $T^{'}$ (not on $T$ or $T_0$). Hence, given an error tolerance $\epsilon >0$, we can find $T^{'}$ such that $g(T) \leq \epsilon$ for all $T \geq T^{'}$. This will ensure adaptation in the case where channel probabilities remain fixed for windows longer than $T^{'}$.
\subsection{Utility functions}
One utility function that can be used in our work is $\phi(\vec{x}) = \min\{x_1,\dots, x_n\}$. This is a nonsmooth utility function that seeks to maximize the minimum time-average success rate across all users. However, if there is one user with very low success probability, this utility function can cause almost all the resources to be devoted to that user, resulting in poor performance for all users. 

Another choice is $\phi(\vec{x}) =\sum_{i=1}^n \log(1+\beta x_i)$, where $\beta$ are given nonnegative weights. The logarithmic term introduces a form of proportional fairness~\cite{kelly-charging,kelly-shadowprice}. See also discussion of different utility functions in~\cite{kelly-charging,kelly-shadowprice,mo-walrand-fair,altman-alpha-fair,chiang-axiomatic-fairness} 
\subsection{Related work}
Network scheduling in stochastic environments is a widely considered problem in the literature on communication networks. Some examples include scheduling in wireless networks~\cite{Fattah2002,Jung2007}, computer networks~\cite{Maguluri2012StochasticMO}, vehicular networks~\cite{Cai2024} and unmanned aerial vehicle networks~\cite{Kong2021}. These works consider different goals such as optimal scheduling for queue-stability~\cite{Huang2024}, power minimization~\cite{Cruz2003}, utility maximization~\cite{Palomar2006}.

Multiple access, where many users access a limited number of communication channels, is an important problem in network scheduling. Here, it is desirable for users to be scheduled to avoid collisions. Transmission failures occur when the receiver is unable to decode the packet transmissions. This can occur, for example, when a fixed transmission power is used, but channel conditions have random and unknown fluctuations so that the received signal strength is insufficient for decoding. Different links can have different properties (such as different geographic distances to the receiver), so they can also have different success probabilities. The naive approach of assigning the users to links with the least failure probabilities leads to unfairness since, in such a scenario, users with high transmission failure probabilities will never be assigned. A common approach to solve this problem is to maximize a utility function of the time-averaged success. This problem has been well studied in the full information scenario when either the fluctuating channel conditions are known before transmission (called opportunistic scheduling) or when channel success probabilities are known in advance. Opportunistic scheduling has been considered using utility functions~\cite{shroff-opportunistic}, Lyapunov drift~\cite{tass-server-allocation}, Frank-Wolfe~\cite{prop-fair-down,vijay-allerton02,neely-frank-wolfe-ton}, primal-dual~\cite{atilla-primal-dual-jsac,stolyar-greedy}, and drift-plus-penalty~\cite{Neely2010coml}. The case when success probabilities are known in advance can be solved offline as a convex optimization problem using the mirror descent technique~\cite{duchi2018introductory}. 

The problem becomes challenging when the success probabilities are unknown and only bandit feedback on transmission successes is available. As a result, the problem must be addressed by combining ideas from utility optimization over time averages with multi-armed bandit learning. Existing work on bandits with vector rewards and concave utility functions can be adapted to the single-channel case (\(m = 1\)) of our problem (see~\cite{Agrawal2014,do2023contextual,agrawal2019bandits}). However, these approaches have two main drawbacks. First, they do not consider the matching constraints present in our setting. Second, they rely on upper confidence bound (UCB) techniques and are not adaptive. In particular, these algorithms cannot adapt to changes in transmission failure probabilities without being explicitly informed of such changes. Hence, the goal of this paper is to develop adaptive algorithms for utility optimization with matching constraints under bandit feedback. This is achieved by combining the EXP3.S algorithm~\cite{Auer2002}, which enables adaptive bandit learning, with Lyapunov optimization~\cite{neely2022stochastic} to address the utility maximization objective. Although EXP3.S was originally developed for adversarial bandits, it has also been successfully applied in nonstationary stochastic environments~\cite{Huang2024}.

Existing work on using learning for scheduling problems in communications includes spectrum sharing~\cite{Bande2019,Dutta2022}, network utility maximization with unknown utility functions under full feedback~\cite{fu2020learningnumnetworkutilitymaximization}, and queue stability~\cite{Huang2024}. The literature on bandit learning-based scheduling primarily focuses on maximizing throughput~\cite{Taheri2020}. The work of~\cite{fu2020learningnumnetworkutilitymaximization} considers network utility maximization where the utility functions are learned online, and~\cite{Salem2022} studies online resource allocation with the objective of maximizing $\alpha$-fairness. Neither of these works considers a bandit reward structure or matching-type constraints, and both differ from our problem formulation. The work of~\cite{Ballu2023} studies an online optimization problem on transport polytopes. While their framework can be adapted to the Birkhoff polytope to incorporate matching constraints, it does not extend to bandit feedback settings. Moreover, none of the above works explicitly addresses the adaptiveness of the proposed algorithms.

It should be noted that the concept of adaptiveness considered in this paper is different from the adversarial setting~\cite{auer1995gambling,pmlr-v75-wei18a, Auer2002}, although they have similarities. In adversarial settings, the guarantee of convergence is defined with respect to the rewards received throughout the time horizon. This is useful when no stochastic assumptions can be placed on the rewards. On the other hand, the adaptiveness considered in this paper is useful when we can place stochastic assumptions on the rewards, but the distributions may change from time to time. In this case, we can define an optimal strategy for each time frame in which the reward distribution remained constant. Hence, the goal in this frame is to learn the aforementioned strategy irrespective of the reward distributions of the past. An approach commonly used in problems of this flavor is minimizing dynamic regret~\cite{pmlr-v99-auer19a,pmlr-v99-chen19b,pmlr-v89-cheung19b}. Here, the regret is modified to account for the changing environments and the regret bounds are in terms of some measure that captures the degree of change. Various algorithms are developed for the setting with linear utility functions using optimizing in phases/episodes~\cite{pmlr-v99-auer19a,pmlr-v99-chen19b}, and sliding window-based algorithms~\cite{garivier2008upperconfidenceboundpoliciesnonstationary,pmlr-v89-cheung19b}. We utilize a simpler notion of adaptiveness and develop an algorithm for the case with general utility functions.

Multi-armed bandit (MAB) learning~\cite{Lai1985,Auer2002FinitetimeAO} provides a principled framework for sequential decision-making under uncertainty. Classical stochastic bandit models focus on learning unknown mean rewards, often using upper confidence bound (UCB)–type algorithms~\cite{Bubeck2012,lattimore_szepesvári_2020}. In contrast, adversarial bandit models make no stochastic assumptions on rewards and are commonly addressed using randomized algorithms such as EXP3~\cite{Auer2002}. Beyond these settings, numerous extensions have been studied, including linear, contextual, and combinatorial bandits~\cite{Bubeck2012}. 

Table~\ref{tab:algs} positions our work relative to the most closely related literature. To the best of our knowledge, no prior work simultaneously addresses utility optimization under bandit feedback with matching constraints while providing interval-based adaptiveness guarantees. Unlike classical bandit algorithms, which maximize scalar time-averaged reward over independent arms, our setting involves combinatorial matching constraints and optimization of a concave utility function of the time-average throughput vector. These structural features introduce nonlinear coupling across users and channels and prevent a direct application of standard bandit techniques. Furthermore, achieving adaptiveness under changing channel statistics requires incorporating EXP3.S-style exploration mechanisms within a Lyapunov optimization framework.

\begin{table*}[t]
\centering
\caption{Comparison of work on Network Utility Optimization and Adaptive Bandits.}
\label{tab:algs}
\small
\begin{tabular}{lcccccc}
\toprule
\textbf{Work} &\textbf{Utility} & \textbf{Feedback} & \textbf{Environment}& \textbf{Matching}&\textbf{Adaptive}& \textbf{Bound}\\
& && & \textbf{Constraints}&&\\
\midrule
\textbf{Adaptive MAC} &General Lipschitz& Bandit  & Stochastic&Yes & Yes &$\mathcal{O}\left(\sqrt{\frac{\log(T)}{T}}\right)$ \\[3pt]
\textbf{(This Paper)}& & & & & &\\
\hline

\textbf{Adaptive} & & & & & &\\
\textbf{MAC.CF}& General Lipschitz &Bandit   &Stochastic & Yes & Yes & $\mathcal{O}\left(\sqrt[3]{\frac{\log(T)}{T}}\right)$ \\[3pt]
 \textbf{(This Paper)}& & & & & &\\
\hline
\textbf{UCB MAC}&General Lipschitz & Bandit   & Stochastic & Yes&No &$\mathcal{O}\left(\sqrt{\frac{\log(T)}{T}}\right)$ \\[3pt] 
\textbf{(This Paper)}& & & & & &\\
\hline
Neely~\cite{Neely2010coml}&General Lipschitz & Full & Stochastic &  No & Yes &$\mathcal{O}\left(\sqrt{\frac{1}{T}}\right)$ \\[3pt]\hline
Agarwal et al~\cite{Agrawal2014}&General Lipschitz& Bandit  &Stochastic& No&No &  $\mathcal{O}\left(\sqrt{\frac{\log(T)}{T}}\right)$ \\[3pt]
\hline
Neely~\cite{neely-frank-wolfe-ton} & General Smooth& Full & Stochastic &  No & No& $\mathcal{O}\left(\frac{\log(T)}{T}\right)$\\[3pt]
\hline
Garivier et al~\cite{garivier2008upperconfidenceboundpoliciesnonstationary} & Linear& Bandit & Stochastic &  No & Yes& $\mathcal{O}\left(\sqrt{\frac{\log(T)}{T}}\right)$\\[3pt]
\hline
Ballu et al~\cite{Ballu2023} & General Lipschitz& Full & Stochastic & Yes & No & $\mathcal{O}\left(\sqrt{\frac{1}{T}}\right)$\\[3pt]
\bottomrule
\end{tabular}
\end{table*}
\subsection{Our Contributions}
\begin{enumerate}
\item{\textbf{Novel Problem Formulation:}}
We develop and analyze algorithms to solve the problem of automatic link selection in multi-channel multiple access by combining ideas from multi-armed bandit learning and Lyapunov optimization. Although the classical multi-armed bandit (MAB) problem has been widely studied for maximizing linear utilities, such approaches can suffer from a lack of fairness in assignments when applied to the problem considered here. In contrast, our framework allows for smooth or nonsmooth concave utility functions.
\item{\textbf{Adaptive Algorithm with Performance Guarantees:}}
We develop two adaptive algorithms that jointly address bandit learning, matching constraints, and utility maximization. The algorithms combine the EXP3.S algorithm for adversarial bandits~\cite{Auer2002} with Lyapunov optimization and techniques for online matching problems developed in~\cite{Chen2022SimultaneouslyLS}. We show that the first algorithm (Adaptive MAC) algorithm achieves an \(\mathcal{O}(\sqrt{\log(T)/T})\) performance gap over any interval of \(T\) consecutive time slots during which the (unknown) success probabilities remain fixed. This regret bound matches, up to logarithmic factors, the minimax lower bound for non-stationary multi-armed bandits established in~\cite{garivier2008upperconfidenceboundpoliciesnonstationary}. Moreover, if the success probabilities change at some time \(T_0\), the performance guarantees over the interval \(\{T_0, T_0+1, \dots, T_0+T-1\}\) are independent of the system behavior prior to \(T_0\), even though the algorithm does not know the change point \(T_0\). Hence, the algorithm is adaptive. Adaptive MAC achieves fast convergence but requires solving a convex optimization problem in each iteration, which is computationally expensive. To reduce computational complexity, we propose a second adaptive algorithm (Adaptive MAC.CF) whose iterations admit closed-form solutions. This approach trades off a slower convergence rate for significantly lower per-slot computational cost.

\item{\textbf{Special Cases:}}
In the single-channel special case, we recover both fast convergence and closed form solutions for iterations via a simpler adaptive algorithm. When adaptation is not required, we also consider a UCB-based algorithm that achieves the same asymptotic convergence rate while relying only on simple max-weight-type decisions. The UCB algorithm highlights the advantages of adaptive algorithms in non-stationary environments. We further study the single-channel case with utility function $\phi(\vec{x}) =min\{x_1, \dots, x_n\}$,
where a simple estimation-free mechanism achieves both convergence and adaptation.

\item \textbf{Empirical evaluation:}
Through simulations, we demonstrate the fast convergence of the proposed adaptive algorithms and their ability to quickly adjust to changes in link success probabilities. In the conducted simulations, the Adaptive MAC.CF algorithm reduces the per-iteration computational complexity by approximately 36\% compared to the Adaptive MAC algorithm. We also empirically compare our methods with prior work in the single-channel setting.
\end{enumerate}

\subsection{Notation}
For integers $a,b$, we use $[a:b]$ to denote the set of integers between $a,b$ inclusive. We use $[a] = [1:a]$. 
For $\vec{a} \in \mathbb{R}^k$, $\lVert \vec{a}\rVert = \sqrt{\sum_{i=1}^k a_{i}^2}$, $\lVert \vec{a}\rVert_1 = \sum_{i=1}^k |a_{i}|$, and $[\vec{a}]_+ \in \mathbb{R}^k$ is the vector with $i$-th entry $\max\{a_i,0\}$. We use $\vec{1}_k$ to denote the $k$-dimensional vector of ones. When the dimension is clear from context, we use $\vec{1}$ instead of $\vec{1}_k$. For vectors $\vec{a,b} \in \mathbb{R}^k$, $\vec{c} = \vec{a} \odot \vec{b} \in \mathbb{R}^k$ is defined such that $c_i = a_ib_i$ for all $i \in [k]$. For matrices $\vec{A,B} \in \mathbb{R}^{k \times l}$ we use $\lVert \vec{A}\rVert = \sqrt{\sum_{i=1}^k \sum_{j=1}^l A_{i,j}^2}$, $\lVert \vec{A}\rVert_1 = \sum_{i=1}^k\sum_{j=1}^l |A_{i,j}|$, and $\vec{C} = \vec{A} \odot \vec{B} \in \mathbb{R}^{k \times l}$ is defined such that $C_{i,j} = A_{i,j}B_{i,j}$ for all $i \in [k]$ and $j \in [l]$.
\subsection{Definitions and Preliminaries}\label{sec:def_prel}
In this subsection, we define some quantities that will be useful throughout. Define $s = \max\{n,m\}$. Also,
\begin{align}
    &\Delta^l_{\varepsilon} = \left\{\vec{p} \in \mathbb{R}^{l}:  \sum_{i=1}^l p_{i} = 1, p_i \geq \varepsilon \ \forall i\in [l]\right\}, \text{ where } l \in \mathbb{N},\nonumber\\
&\mathcal{S}^{\text{row}}_{\varepsilon} = \Bigg{\{}\vec{P} \in \mathbb{R}^{s \times s}_+: \sum_{k=1}^s P_{i,k} = 1, P_{i,j} \geq \varepsilon \ \forall i,j \in [s]\Bigg{\}}, \nonumber\\
&\mathcal{S}^{\text{col}}_{\varepsilon} = \Bigg{\{}\vec{P} \in \mathbb{R}^{s \times s}_+: \sum_{k=1}^s P_{k,j} = 1, P_{i,j} \geq \varepsilon \ \forall i,j \in [s]\Bigg{\}}, \nonumber\\
&\mathcal{S}^{\text{doub}}_{\varepsilon} = \mathcal{S}^{\text{col}}_{\varepsilon} \cap \mathcal{S}^{\text{row}}_{\varepsilon}
    , \ \mathcal{S}_{\varepsilon} = \mathcal{S}^{\text{col}}_{\varepsilon} \cup \mathcal{S}^{\text{row}}_{\varepsilon}. \nonumber
\end{align}
We also denote $\Delta^l = \Delta^l_0$,    $\mathcal{S}^{\text{row}} = \mathcal{S}^{\text{row}}_{0}$, $\mathcal{S}^{\text{col}} = \mathcal{S}^{\text{col}}_{0}$, $\mathcal{S}^{\text{doub}} = \mathcal{S}^{\text{doub}}_{0}$, and $\mathcal{S} = \mathcal{S}_{0}$. Also, let $\Delta^l_+ = \Delta^l \cap (0,1]^{l}$, $\mathcal{S}^{\text{doub}}_{+} = \mathcal{S}^{\text{doub}} \cap (0,1]^{s\times s}$ and  $\mathcal{S}_{+} = \mathcal{S} \cap (0,1]^{s\times s}$. We hide the dependence on $s$ in the notation for sets for clarity. 

\paragraph{$\beta$-approximation}Given $\beta>0$, and matrix $\vec{P} \in \mathcal{S}^{\text{doub}}_+$, we say matrix $\hat{P}  \in \mathcal{S}^{\text{doub}}_+$ is a $\beta$-approximation of $\vec{P}$ if 
\begin{align}\label{eqn:beta_approx}
    e^{-\beta} \leq \frac{\hat{P}_{j,j}}{P_{i,j}}
\end{align}
hold for all $i,j \in [s]$.

\paragraph{KL-divergence}For $\vec{p} \in \Delta^s$ and $\vec{q} \in \Delta^s_+$, we define the Kullback--Leibler (KL) divergence as
\begin{align}
    D_{\mathrm{KL}}(\vec{p}\lVert\vec{q})
    = \sum_{i=1}^s p_i \ln\left(\frac{p_i}{q_i}\right),
\end{align}
with the convention that $0 \ln(0/q) = 0$ for any $q > 0$. For matrices $\vec{P} \in \mathcal{S}$ and $\vec{Q} \in \mathcal{S}_+$, we define the (entrywise) matrix KL divergence by
\begin{align}\label{eqn:matrix_KL}
    D(\vec{P}\lVert \vec{Q})
    = \sum_{i=1}^s \sum_{j=1}^s
    P_{i,j}\ln\left(\frac{P_{i,j}}{Q_{i,j}}\right),
\end{align}
where again $0 \ln(0/q) = 0$ for $q>0$.

\paragraph{Sampling from $\vec{P} \in \mathcal{S}^{\text{doub}}$} We define the process of sampling a permutation matrix $\vec{Y}$ from a doubly stochastic matrix $\vec{P} \in \mathcal{S}^{\text{doub}}$. By the Birkhoff--von Neumann theorem~\cite{birkhoff1946three}, there exists an integer $r \in \mathbb{N}$ and permutation matrices $\vec{M}^1, \dots, \vec{M}^r$ such that $\vec{P} = \sum_{l=1}^{r} s_l \vec{M}^l$, 
where $\vec{s} \in \Delta^r$ is a probability vector over $[1:r]$. Consider the following sampling procedure: draw an index $a \in [1:r]$ according to the distribution $\vec{s}$ and set $\vec{Y} = \vec{M}^a$. Then, $\mathbb{E}\!\left\{\vec{Y} \mid \vec{P}\right\} = \vec{P}$.
\subsection{Assumptions}
We state our main assumptions.
\begin{itemize}
\item[\textbf{A1}] The function $\phi$ is concave, entrywise nondecreasing, and has bounded subgradients in $[0,1]^n$, i.e, $|\phi^{'}_i(\vec{x})| \leq B \ \forall i\in [n]$, and $\vec{x} \in [0,1]^n$. Hence, $\phi$ is $\sqrt{n}B$-Lipschitz continuous. Also, let $\phi^{\text{max}} = \max_{\vec{x} \in [0,1]^n} |\phi(\vec{x})|$.
 \item[\textbf{A2}] We have access to the solution of the problem $\max_{\vec{x} \in [0,1]^n}[\phi(\vec{x}) + \sum_{i=1}^n c_i x_i]$, for all $\vec{c} \in \mathbb{R}^n_+$.

\noindent
\textbf{Note: }This is valid for most separable functions $\phi$. For instance, when $\phi$ is a proportionally fair utility function type of the form $\phi(\vec{x}) = \sum_{i=1}^n \log(1+ \beta x_i)$, where $\beta \in \mathbb{R}_+$, the problem in \textbf{A2} has an explicit solution.
\end{itemize}
\section{Problem Setup}
We formalize the problem of interest below. 
\begin{align}
    \text{(P1:) } &\underset{\vec{Y}(1),\vec{Y}(2),\dots}{\text{max}} \lim\inf_{T \to \infty}\phi\left(  \frac{1}{T}\sum_{t=1}^T\mathbb{E}\{\vec{X}(t)\}\right) \\& \text{s.t.  } \vec{Y}(t), \text{ and }\vec{S}(\tau) \text{ are independent for all }  \nonumber\\&\ \ \ \ \ \ \ t ,\tau \in \mathbb{N} \text{ and } \tau \geq t \label{cons:ind_fut}\\& \ \ \ \ \ \vec{Y}(t), \text{ and }S_{i,j}(\tau) \text{ are independent for all } t,\tau \in \mathbb{N}, \nonumber\\& \ \ \ \ \ \ \  (i,j) \in [s] \times [s], \tau < t,
    Y_{i,j}(\tau) \neq 1 \label{cons:ind_past}\\& \ \ \ \ \  \vec{Y}(t)\in \{0,1\}^{s \times s}  \  \forall t \in \mathbb{N}\label{cons:zer_one_Y}\\& \ \ \ \ \ \sum_{i=1}^s Y_{i,j}(t) \leq 1 \  \forall t \in \mathbb{N}, j \in [s] \label{cons:sum_Y_j} \\& \ \ \ \ \ \sum_{j=1}^s Y_{i,j}(t) \leq 1 \  \forall t \in \mathbb{N}, i \in [s] \label{cons:sum_Y_i}\\& \ \ \ \ \ X_i(t) = \sum_{j=1}^m Y_{i,j}(t)S_{i,j}(t)  \forall t \in \mathbb{N}, \ i \in [n], \label{cons:def_X}
\end{align}
where constraint \eqref{cons:ind_fut} ensures transmission decisions do not know success/failures before they happen; \eqref{cons:ind_past} ensures we cannot use information that is never observed.
Define $\phi^{\text{opt}}$ as the optimal objective value of (P1). 

\subsection{Deterministic Problem}

Consider the deterministic problem given by
\begin{align}
  \text{(P2:) }  &\underset{\vec{P},\vec{\gamma}}{\text{max }}\phi\left(\vec{\gamma}\right)\\& \text{s.t. }
   \vec{P}\in \mathcal{S}^{\text{doub}}\label{cons:simp}\\& \ \ \ \ \
   \vec{\gamma}\in [0,1]^n\label{cons:R}\\& \ \ \ \ \
   \sum_{j=1}^m q_{i,j}P_{i,j}  \geq \gamma_{i} \ \forall i \in \{1,\dots,n\} \label{cons:pq_geq_gam},
\end{align}
Let $\phi^*$ denote the optimal objective value of the above problem. In Lemma~\ref{lemma:equiv_prob_lemma}, we show that $\phi^* \geq \phi^{\text{opt}}$. This lemma will be useful in establishing the performance guarantees of our algorithms. Let us define $(\vec{P}^*,\vec{\gamma}^*)$ as the optimal solution Problem (P2).

\noindent
\textbf{Note: } More generally, it can be shown that $\phi^* = \phi^{\text{opt}}$.
If $q_{i,j}$ were known, Problem~(P2) could be solved to obtain 
$(\vec{P}^*, \vec{\gamma}^*)$. The stationary policy that samples a 
permutation matrix from $\vec{P}^*$ in each time slot 
(see Section~\ref{sec:def_prel}-c) and uses it for assignment 
achieves an objective value of $\phi^*$ for Problem~(P1). However, even when $q_{i,j}$ is unknown, 
Section~\ref{sec:fast_adaptive} establishes that our algorithm 
attains an objective value arbitrarily close to $\phi^*$, 
thereby implying that $\phi^{\text{opt}} \geq \phi^*$.

\begin{lemma}\label{lemma:equiv_prob_lemma}
We have $\phi^* \geq \phi^{\text{opt}}$, where $\phi^{\text{opt}}$ and $\phi^*$ are the optimal objective values of Problems~(P1) and ~(P2), respectively.
\begin{proof}
Fix $\varepsilon>0$. Then there is a positive integer $T$ and decisions $\vec{Y}^*(1),\vec{Y}^*(2),\dots,\vec{Y}^*(T)$ that respect constraints \eqref{cons:ind_fut}-\eqref{cons:sum_Y_i}, such that
\begin{align}\label{eqn:geq_eps}
    \phi\left(\frac{1}{T}\sum_{t=1}^T \mathbb{E}\{\vec{X}^*(t)\}\right) \geq \phi^{\text{opt}} - \varepsilon,
\end{align}
where $X^*_i(t) = \sum_{j=1}^m Y^*_{i,j}(t)S_{i,j}(t)$ for all $t \in \mathbb{N}$ and $i \in [n]$. Define the matrix $\vec{\tilde{P}}^* \in \mathbb{R}^{s \times s}$ such that
\begin{align}
    \tilde{P}^*_{i,j} = \begin{cases}
        \frac{1}{T}\sum_{t=1}^T\mathbb{E}\{Y^*_{i,j}(t)\} & \text{ if $(i,j) \in [n] \times [m]$}\\
        0 & \text{ otherwise}
    \end{cases}
\end{align}
Constraints~\eqref{cons:zer_one_Y}--\eqref{cons:sum_Y_i} ensure the $s \times s$ matrix $\vec{\tilde{P}}^*$ has nonnegative entries with all row and column sums less than or equal to 1. A well known result of Von-Neumann ensures that there is a matrix $\vec{P}^* \in \mathcal{S}^{\text{doub}}$ such that $\tilde{P}^*_{i,j} \leq P^*_{i,j}$ for all $i,j \in [s]$. Let $\vec{\gamma}^* \in [0,1]^n$ such that $\gamma^*_i = \sum_{j=1}^m q_{i,j}P^*_{i,j}$ for $i \in [n]$. It is easy see that $(\vec{P}^*,\vec{\gamma}^*)$ satisfy constraints \eqref{cons:simp}-\eqref{cons:pq_geq_gam}. Also, we have 
\begin{align}\label{eqn:ent_non}
\gamma^*_i = \sum_{j=1}^m q_{i,j}P^*_{i,j}  \geq \sum_{j=1}^m q_{i,j}\tilde{P}^*_{i,j} = \frac{1}{T}\sum_{t=1}^T\mathbb{E}\{X^*_{i}(t)\}
\end{align}
since \eqref{cons:ind_fut} ensures $\mathbb{E}\{Y^*_{i,j}(t)S_{i,j}(t)\} = \mathbb{E}\{Y^*_{i,j}(t)\}q_{i,j}$. Hence,
\begin{align}
    \phi^* &\geq_{(a)} \phi(\vec{\gamma}^*) 
    \geq_{(b)}    \phi\left(\frac{1}{T}\sum_{t=1}^T \mathbb{E}\{\vec{X}^*(t)\}\right) \geq_{(c)} \phi^{\text{opt}} - \varepsilon, \nonumber
\end{align}
where (a) follows since $\phi^*$ is the optimal objective value of (P2) and $(\vec{P}^*,\vec{\gamma}^*)$ is feasible for (P2), (b) follows from \eqref{eqn:ent_non} and the entrywise nondecreasing property of $\phi$ (see assumption~\textbf{A1}) and (c) follows from \eqref{eqn:geq_eps}. The above is true for all $\varepsilon>0$. Hence, we have $\phi^* \geq \phi^{\text{opt}}$ as desired.
\end{proof}
\end{lemma}
\section{Intuition Behind Algorithms}\label{sec:intution}
Before moving onto the algorithms, we describe the general intuition unifying all the algorithms. Consider the partial Lagrangian of problem~(P2) defined as
\begin{align}\label{eqn:f_defffff}
f(\vec{Q},\vec{P},\vec{\gamma})
= -V\phi(\vec{\gamma})
+ \sum_{i=1}^n Q_i\!\left(\gamma_i - \sum_{j=1}^m q_{i,j}P_{i,j}\right),
\end{align}
obtained by dualizing constraint~\eqref{cons:pq_geq_gam},
where $Q_i \ge 0$ are the Lagrange multipliers associated with~\eqref{cons:pq_geq_gam}, and $V>0$ is a tunable parameter that scales the utility term without affecting the optimal solution of~(P2). If the success probabilities $q_{i,j}$ were known, then for a suitably 
chosen parameter $V$, the iterative primal update
\begin{align}\label{eqn:primal}
(\vec{\gamma}(t+1), \vec{P}(t))
&= \arg\min_{\substack{\vec{P} \in \mathcal{S}^{\text{doub}}\\
\vec{\gamma} \in [0,1]^n}}
f(\vec{Q}(t), \vec{P}, \vec{\gamma}) \\
&= \arg\min_{\vec{P}, \vec{\gamma}}
\left\{
- V \phi(\vec{\gamma})
+ \sum_{i=1}^n Q_i(t)
\left[
\gamma_i - \sum_{j=1}^m q_{i,j} P_{i,j}
\right]
\right\}, \nonumber
\end{align}
together with the dual update
\begin{align}\label{eqn:dual}
Q_i(t+1)
= \left[
Q_i(t)
+ \gamma_i(t+1)
- \sum_{j=1}^m q_{i,j} P_{i,j}(t)
\right]_+,
\end{align}
for all $i \in [n]$,
guarantees
\begin{align}\label{eqn:gamma_gurr}
\phi(\bar{\vec{\gamma}}(T))
&\ge
\phi^{\text{opt}}
- \frac{o(T)}{T},
\end{align}
and
\begin{align}\label{eqn:p_gurr}
\bar{\gamma}_i(T)
&\le
\sum_{j=1}^m q_{i,j} \bar{P}_{i,j}(T)
+ \frac{o(T)}{T},
\quad \forall i \in [n],
\end{align}
where the time averages are defined as $(\bar{\vec{\gamma}}(T), \bar{\vec{P}}(T))
=
\frac{1}{T}
\sum_{t=1}^{T}
(\vec{\gamma}(t), \vec{P}(t))$. 
The result follows from standard primal--dual convergence arguments~\cite{NedicOzdaglar2009}. Hence, we can get arbitrarily close to the optimal solution of (P2). 

We can either find $\vec{\bar{P}}(T)$ offline using the above approach, and implement the fixed policy of sampling an assignment from $\vec{\bar{P}}(T)$ (see Section~\ref{sec:def_prel}-c) in each time slot, or we can implement an online procedure where  we sample an assignment from $\vec{P}(t)$ in time slot $t$. Both of these approaches will guarantee~\eqref{eqn:first_goal} given that $q_{i,j}$ values are known. 

All of our algorithms follow the structure described above. However, in our scenario, we do not know the values of $q_{i,j}$. Hence, they have to be estimated using the bandit feedback. For this task, in the adaptive algorithms, we use the EXP3.S importance sampling based estimation, whereas in the non-adaptive algorithm, we use UCB-based estimation. For each $i \in [n]$, the sequence $Q_i(1), Q_i(2), \dots$ is referred to as a 
\emph{virtual queue}, since the dual update~\eqref{eqn:dual} resembles a 
standard queueing recursion. Observe that the primal update~\eqref{eqn:primal} decomposes into separate 
decisions over $\vec{\gamma}$ and $\vec{P}$. This decomposition is the motivation behind 
the introduction of the auxiliary variable $\vec{\gamma}$. In particular, 
it enables us to decouple the matching decisions from the utility 
optimization.
\section{Adaptive MAC Algorithm}\label{sec:fast_adaptive}
Now, we develop our first algorithm. Similar to the description of the known $q_{i,j}$ setting in Section~\ref{sec:intution}, the idea is to implement an online algorithm to find a doubly stochastic matrix $\vec{P}(t)$ in each time using the success/failure feedback. However, this is more challenging since we do not know $q_{i,j}$, and we only receive bandit feedback on success/failures. In addition, we require the algorithm to adapt when the success failure probabilities change. To achieve this, we combine EXP3.S based importance sampling idea with Lyapunov optimization.

In each time slot, we solve a convex optimization problem to obtain an intermediate matrix $\vec{\tilde{P}}(t)$. The problem cannot be solved exactly. Hence, we find an approximation $\vec{\hat{P}}(t)$ of $\vec{\tilde{P}}(t)$. We then find the doubly stochastic matrix $\vec{P}(t)$ by mixing $\vec{\hat{P}}(t)$ with the all one matrix. This step is vital in ensuring adaptiveness. We then sample $\vec{Y}(t)$ from $\vec{P}(t)$ using Birkhoff-von Neumann Decomposition~\cite{birkhoff1946three}. We assume that if $i>n$ or $j>m$, $Y_{i,j}(t)$ corresponds to a \emph{fictional link}. Formally, we extend the notation by setting $S_{i,j}(t) = 0$ and $q_{i,j} = 0$ for $i>n$ and $j>m$. In this algorithm, we use the loss-based importance sampling estimator 
\begin{align}
\hat{S}_{i,j}(t) = 1-\frac{(1-S_{i,j}(t))Y_{i,j}(t)}{P_{i,j}(t)}
\end{align}
of $q_{i,j}$. This enables fast convergence guarantees.

In Algorithm~\ref{algo:61}, we provide the algorithm for the task. The algorithm uses parameters $V>0$, $\theta >0$, $\varepsilon \in (0,1/2]$, and $\eta>0$. 
\begin{algorithm}
Initialize $\vec{P}(1)   = \frac{1}{s}\vec{1}\vec{1}^{\top} \in \mathcal{S}^{\text{doub}}$ and the virtual queues $\vec{Q}(1) = \vec{0} \in \mathbb{R}^n$.\\
\For{each time slot $t \in \mathbb{N}$}{
Sample $\vec{Y}(t)$ from $\vec{P}(t)$ (see Section~\ref{sec:def_prel}-c), and receive $\vec{S}(t) \odot \vec{Y}(t)$ as feedback.\\
Compute the estimator $\vec{\hat{S}}(t)$ for $\vec{S}(t)$ using $\hat{S}_{i,j}(t) = 1 - \frac{(1-S_{i,j}(t))Y_{i,j}(t)}{P_{i,j}(t)}$ for all $i,j\in [s]$ where we define $S_{i,j}(t) = 0$ whenever $i>n$ or $j > m$.\\ 
Define
\begin{align}\label{eqn:decision_1_gamma}
    &\vec{\gamma}(t+1)= \arg\min_{\vec{\gamma} \in [0,1]^n}\Big{[} -V\phi(\vec{\gamma}) + \sum_{i=1}^n Q_{i}(t)\gamma_{i}\Big{]}.
\end{align}\\
Find $\vec{\hat{P}}(t+1)$ as a $(\theta/s)$-approximation of $\vec{\tilde{P}}(t+1)$ where
\begin{equation}\label{eqn:decision_1}
\begin{aligned}
    \vec{\tilde{P}}(t+1) &= \arg\min_{\vec{P} \in \mathcal{S}^{\text{doub}}}\Big{[} -\sum_{i=1}^n\sum_{j=1}^s Q_{i}(t) \hat{S}_{i,j}(t)P_{i,j}  + \frac{1}{\eta}D(\vec{P} \lVert \vec{P}(t)) \Big{]}, 
\end{aligned}
\end{equation}
where $\beta$-approximation is defined in \eqref{eqn:beta_approx} and the divergence  $D(\cdot\lVert \cdot)$ is defined in \eqref{eqn:matrix_KL}.\\
Mix
\begin{align}
    \vec{P}(t+1) = (1-\varepsilon)\vec{\hat{P}}(t+1) + \frac{\varepsilon}{s} \vec{1}\vec{1}^{\top}
\end{align}\\

Update the virtual queues
\begin{align}\label{eqn:queue_eqn}
    \vec{Q}(t+1) = \left[\vec{Q}(t) + \vec{\gamma}(t+1) - \vec{X}(t)\right]_+,
\end{align}
where $X_i(t) = \sum_{j=1}^m Y_{i,j}(t)S_{i,j}(t)$ for $i \in [n]$.\\
}
\caption{Adaptive MAC (Parameters $V>0$, $\theta >0$, $\varepsilon \in (0,1/2]$, and $\eta>0$)}\label{algo:61}
\end{algorithm}
The following Theorem establishes the performance bound of Algorithm~\ref{algo:61}.
\begin{theorem}\label{theorem:final_bound_1_}
Consider $T, T_0 \in \mathbb{N}$ and suppose that the success probabilities 
$q_{i,j}$ remain fixed on the interval $[T_0,\, T_0 + T - 1]$. Then, independently of their behavior outside this interval, for any algorithm parameters  
$\varepsilon \in (0,1/2]$, $\eta>0$, $\theta>0$, and $V>0$, Algorithm~\ref{algo:61} yields
\begin{align}
    &\phi^{\text{opt}} -\phi\left(\mathbb{E}\left\{\frac{1}{T}\sum_{t=T_0}^{T+T_0-1}\vec{X}(t)\right\}\right) \nonumber\\&\leq \frac{n}{V}+\frac{2\varepsilon s}{\eta V}  + \frac{\theta }{ V\eta}+\frac{\eta ns(BV+1)^2 }{V}  + \frac{s}{\eta VT}\log\left(\frac{s}{\varepsilon}\right)+ \frac{n(BV+1)^2}{2 VT} +\frac{nB(BV+1)}{T}  \nonumber
\end{align}
where $B$ is the bound on the subgradients of $\phi$ defined in assumption~\textbf{A1}.
In particular, choosing $\eta = \Theta(\sqrt{\log(T)}/T)$, $V = \Theta(\sqrt{T})$, $\varepsilon= \Theta(1/T)$, and $\theta = \Theta(1/T)$, we have 
\begin{align}
    \phi^{\text{opt}} -\phi\left(\mathbb{E}\left\{\frac{1}{T}\sum_{t=T_0}^{T+T_0-1}\vec{X}(t)\right\}\right) =  O\left(\sqrt{\frac{\log(T)}{T}}\right). \nonumber
\end{align}
\end{theorem}
\subsection{Discussion of Algorithm~\ref{algo:61}}
In this section, we connect Algorithm~\ref{algo:61} with the Intuition described in Section~\ref{sec:intution}.
Define the history $\mathcal{H}(t)$ as the sigma algebra generated by
\begin{align}\label{eqn:history_adap}
    \mathcal{H}(t)=\{\vec{Y}(\tau),\;\vec{Y}(\tau)\odot\vec{S}(\tau):\ \tau\in[t-1]\}.
\end{align}
Using the importance-sampling estimator $\hat S_{i,j}(t)$ (see line~5 of Algorithm~\ref{algo:61}), we can form unbiased estimates of the quantities $q_{i,j}$ appearing in the primal objective~\eqref{eqn:primal}. In particular, for any $\mathcal{H}(t)$-measurable $\vec P\in\mathcal S^{\mathrm{doub}}$, $\vec\gamma\in[0,1]^n$, and $\vec Q\ge 0$, the random variable
\begin{align}\label{eqn:unb_est}
    -V\phi(\vec\gamma)
    +\sum_{i=1}^n Q_i\!\left[\gamma_i-\sum_{j=1}^m \hat S_{i,j}(t)P_{i,j}\right]
\end{align}
is an unbiased estimator of $f(\vec Q,\vec P,\vec\gamma)$ defined in~\eqref{eqn:f_defffff}.

However, we cannot directly replace the primal update~\eqref{eqn:primal} at time $t$ by the estimator~\eqref{eqn:unb_est}, since $\hat S_{i,j}(t)$ depends on the decision $\vec P(t)$ over which we are optimizing (see line~5 of Algorithm~\ref{algo:61}). To enable an update of the choice probabilities, we instead introduce a new
decision variable $\vec{\tilde{P}}(t+1)$ and compute $(\vec\gamma(t+1), \vec{\tilde{P}}(t+1))$
by minimizing the surrogate objective
\begin{align}\label{eqn:surrogate}
    &-V\phi(\vec\gamma(t+1))  + \sum_{i=1}^n Q_i(t)\!\left[\gamma_i(t+1)
    -\sum_{j=1}^m \hat S_{i,j}(t)\tilde{P}_{i,j}(t+1)\right]  + \frac{D\big(\vec{\tilde{P}(t+1)}\lVert \vec P(t)\big)}{\eta}.
\end{align}
Since the optimization problem in \eqref{eqn:surrogate} does not admit a closed-form solution, we compute $\vec{\hat{P}}(t+1)$ as an approximation to $\vec{\tilde{P}}(t+1)$. 
Although the first two terms of \eqref{eqn:surrogate} do not form an unbiased estimator of
$f(\vec Q(t), \vec{\tilde{P}}(t+1), \vec\gamma(t+1))$ (since $\vec{\tilde{P}}(t+1)$ is not
$\mathcal{H}(t)$-measurable), the divergence penalty forces $\vec{\tilde{P}}(t+1)$
to remain close to the $\mathcal{H}(t)$-measurable point $\vec P(t)$. Intuitively, this ensures that the surrogate objective remains close to
the true primal objective evaluated at $\vec P(t)$, while still allowing
the policy to evolve over time. We mix $\vec{\hat{P}}(t+1)$, the approximation of $\vec{\tilde{P}}(t+1)$, with the uniform doubly stochastic matrix $\frac{1}{s}\vec{1}\vec{1}^{\top}$ to obtain $\vec{P}(t+1)$ in line~8 of Algorithm~\ref{algo:61}.
This guarantees persistent exploration and ensures that the algorithm
remains adaptive.

Finally, note that in the virtual queue update~\eqref{eqn:dual}, 
we replace the last term with the unbiased estimator 
$\sum_{j=1}^m Y_{i,j}(t)S_{i,j}(t)$. In particular,
\begin{align}
    \mathbb{E}\left\{ 
    \sum_{j=1}^m Y_{i,j}(t)S_{i,j}(t) 
    \,\Big|\, \mathcal{H}(t)
    \right\}
    =
    \sum_{j=1}^m P_{i,j}(t)q_{i,j}.
\end{align}
This estimator can be computed under bandit feedback, 
since it only requires observing $S_{i,j}(t)$ for indices with 
$Y_{i,j}(t)=1$.

\subsection{Problems~\eqref{eqn:decision_1_gamma}-\eqref{eqn:decision_1}}
Finding $\vec{\gamma}(t+1)$ (Problem~\eqref{eqn:decision_1_gamma}) is possible due to Assumption~\textbf{A2}. 
A standard approach for computing $\vec{\hat{P}}(t+1)$ (Problem~\eqref{eqn:decision_1}) is to apply Sinkhorn procedure, as discussed in Appendix~\ref{app:intem_al_2}. 
Each Sinkhorn iteration has computational complexity $\mathcal{O}(s^2)$. 
Consequently, the overall complexity of computing $\vec{\hat{P}}(t+1)$ is $\mathcal{O}(s^2 T^{\text{inner}})$, where $T^{\text{inner}}$ denotes the number of Sinkhorn iterations performed. 
While Sinkhorn’s procedure exhibits linear convergence in our problem, the exact convergence rate depends on the matrices $\vec{P}(t)$, $\vec{\hat{S}}(t)$, and the vector $\vec{Q}(t)$. 
To mitigate this dependence on the inner convergence behavior, we propose in Section~\ref{sec:slow_adaptive} the \emph{Adaptive MAC.CF} algorithm, in which each time slot has computational complexity $\mathcal{O}(s^2)$. The rest of this section focuses on proving Theorem~\ref{theorem:final_bound_1_}. 
\subsection{Preliminary Lemmas}\label{sec:prel_lemmas}
We state the preliminary lemmas used in the proof of Theorem~\ref{theorem:final_bound_1_}. 
We begin with a standard result on minimizing strongly convex functions (see, e.g.,~\cite{Xiaohan2020}). 
\begin{lemma}\label{lemma:push_back}
Consider $\beta \in [0,1/s]$ and $\alpha >0$. Let
\begin{enumerate}
    \item $\mathcal{Q} \in \{\mathcal{S}^{\text{doub}}_{\beta}, \mathcal{S}^{\text{row}}_{\beta}, \mathcal{S}^{\text{col}}_{\beta}, \mathcal{S}_{\beta}\}$, $g: \mathbb{R}_+^{s\times s} \to \mathbb{R}$ be a convex function, $\vec{Y} \in \mathcal{S}_+$, and $\vec{X}^* \in \arg\min_{\vec{X}\in \mathcal{Q}} \left[g(\vec{X}) + \alpha D(\vec{X}\lVert \vec{Y})\right]$. We have
    for all $\vec{Z} \in \mathcal{Q}$,
\begin{align}
   g(\vec{X}^*) + \alpha D(\vec{X}^* \lVert \vec{Y}) \leq g(\vec{Z}) + \alpha D(\vec{Z}\lVert \vec{Y}) - \alpha D(\vec{Z}\lVert \vec{X}^*)\nonumber.
\end{align}
    \item $\mathcal{Q} = \Delta^s_{\varepsilon}$, $g: \mathbb{R}_+^{s} \to \mathbb{R}$ be a convex function, $\vec{Y} \in \Delta^s_+$, and $\vec{X}^* \in \arg\min_{\vec{X}\in \mathcal{Q}} [g(\vec{X}) + \alpha D_{\text{KL}}(\vec{X}\lVert \vec{Y})]$.
    We have for all $\vec{Z} \in \mathcal{Q}$,
\begin{align}
   g(\vec{X}^*) + \alpha D_{\text{KL}}(\vec{X}^* \lVert \vec{Y}) \leq g(\vec{Z}) + \alpha D_{\text{KL}}(\vec{Z}\lVert \vec{Y}) - \alpha D_{\text{KL}}(\vec{Z}\lVert \vec{X}^*)\nonumber.
\end{align}
\end{enumerate}

\end{lemma}
\begin{lemma}\label{lemma:KL_with_L1_1}
We have for $\vec{X}\in \mathcal{S}$, $\vec{Y} \in \mathcal{S}_+$,
\begin{align}\label{eqn:KL_pink}
    D(\vec{X} \lVert \vec{Y}) \geq  \frac{1}{2s}\lVert\vec{X} - \vec{Y}\rVert_1^2 \geq \frac{1}{2s}\lVert\vec{X} - \vec{Y}\rVert^2,
\end{align}
\begin{proof}
    Notice that
\begin{align}
  D(\vec{X} \lVert \vec{Y}) \geq   \sum_{i=1}^s \frac{1}{2}\left(\sum_{j=1}^s|X_{i,j} - Y_{i,j}|\right)^2 \geq \frac{1}{2s}\lVert\vec{X} - \vec{Y}\rVert_1^2 \nonumber
\end{align}
where the first inequality follows from the Pinsker's inequality and the last inequality follows from the fact that for $a_1,a_2,\dots,a_s \in \mathbb{R}$, 
    $\sum_{i=1}^s a_i^2 \geq \left(\sum_{i=1}^s a_i\right)^2/s$.
\end{proof}
\end{lemma}
\begin{lemma}\label{lemma:KL_u_bound_1}
We have $D(\vec{X} \lVert \vec{Y}) \leq s\log\left(\frac{1}{\varepsilon}\right)$, for all $\vec{X} \in \mathcal{S}$, where $\vec{Y} \in \mathcal{S}_{\varepsilon}$.
\begin{proof}
   
Notice that
\begin{align}
     D(\vec{X} \lVert \vec{Y}) &= \sum_{i=1}^s \sum_{j=1}^sX_{i,j}\log\left(\frac{X_{i,j}}{Y_{i,j}}\right) \nonumber\\&\leq_{(a)} \sum_{i=1}^s\sum_{j=1}^s X_{i,j}\log\left(\frac{1}{\varepsilon}\right) = s\log\left(\frac{1}{\varepsilon}\right), \nonumber
\end{align}
where (a) follows since $X_{i,j} \leq 1$, $Y_{i,j} \geq \varepsilon$, and log is a non-decreasing function.
\end{proof}
\end{lemma}
\begin{lemma}\label{lemma:old}
    Consider $\vec{x} \in \mathbb{R}^{s}$ and $\vec{p} \in \Delta^s$, $\vec{q} \in \Delta^s_+$. We have
    \begin{align}
        \sum_{i = 1}^s p_i x_i \leq D_{\text{KL}}(\vec{p} \lVert \vec{q})+\ln\left(\sum_{i = 1}^s q_i e^{x_i}\right)
    \end{align}
    \begin{proof}
        Let $A = \sum_{j=1}^s q_je^{x_j}$. Define the vector $\vec{r} \in \Delta_+$ by $r_i = q_ie^{x_i}/A $. Notice that
\begin{align}
    D_{\text{KL}}(\vec{p}\lVert \vec{r}) &= \sum_{i=1}^s p_i\ln\left(\frac{p_i}{r_i}\right) = \sum_{i=1}^s  p_i\ln\left(\frac{p_iA}{q_ie^{x_i}}\right)= D_{\text{KL}}(\vec{p}\lVert \vec{q})+ \ln(A) - \sum_{i=1}^s p_ix_i
\end{align}
Using $D_{\text{KL}}(\vec{p}\lVert \vec{r}) \geq 0$ and rearranging, we are done.
    \end{proof}
\end{lemma}
Now, we establish the following lemma, which will be useful in establishing a deterministic bound on $Q_i(t)$. 
\begin{lemma}\label{lemma:det_queue_bnd_1}
Consider the three vector sequences $\{\vec{A}(t) \in [0,1]^n:t \in \mathbb{N}\},\{\vec{G}(t) \in [0,1]^n:t \in \mathbb{N}\},\{\vec{F}(t) \in [0,\infty)^n:t \in \mathbb{N}\}$, where $\{\vec{A}(t):t \in \mathbb{N}\}$ is an arbitrary sequence in $[0,1]^n$, and
\begin{align}\label{eqn:def_fg}
&\vec{F}(1) = \vec{0} \in \mathbb{R}^n\nonumber\\
&\vec{G}(t+1) = \arg\min_{\vec{g} \in [0,1]^n} \left[-V\phi(\vec{g}) + \sum_{i=1}^n F_i(t)g_i\right], \ \forall t \in \mathbb{N},\nonumber\\&\vec{F}(t+1) = [\vec{F}(t) + \vec{G}(t+1)-\vec{A}(t)]_+ \ \forall t \in \mathbb{N}
\end{align}
We have for all $t \in \mathbb{N}$ and $i \in [n]$,  $F_i(t)  \leq BV + 1$,
where $B$ is the bound on the subgradients of $\phi$ defined in assumption~\textbf{A1}.
\begin{proof}
Fix $t \in \mathbb{N}$. 
We first prove the following claim.

\noindent
\textbf{Claim: } If $F_k(t) > BV$, for $k \in [n]$, then $G_k(t+1) = 0$. 
\begin{proof}

We prove the contrapositive. In particular, assume that $G_k(t+1) > 0$. We prove  $F_k(t) \leq BV$. Define the function $f:[0,1]^n \to \mathbb{R}$ by
\begin{align}
    f(\vec{g}) = -V\phi(\vec{g}) + \sum_{i=1}^n g_iF_i(t)
\end{align}

Notice that due to the definition of $\vec{G}(t+1)$ we have 
\begin{align}\label{eqn:optimality}
f(\vec{G}(t+1)) \leq f(\vec{g})
\end{align}
for any $\vec{g} \in [0,1]^n$. Using the subgradient inequality for the convex function $f$, we have for any $\vec{g} \in [0,1]^n$
\begin{align}\label{eqn:sub_g}
  f^{'}(\vec{g})[\vec{G}(t+1) - \vec{g}]  \leq f(\vec{G}(t+1))- f(\vec{g}) \leq 0
\end{align}
where the last inequality follows from \eqref{eqn:optimality}. Let us pick $\vec{g}$ as 
\begin{align}\label{eqn:g_i}
    g_i = \begin{cases}
        0 & \text{ if } i = k\\
        G_i(t+1) & \text { otherwise}
    \end{cases}
\end{align}
Using \eqref{eqn:sub_g} with the above $\vec{g}$, we have
\begin{align}
    0 &\geq f^{'}(\vec{g})[\vec{G}(t+1) - \vec{g}]  = \sum_{i=1}^n[-V\phi_i^{'}(\vec{g})+F_i(t)](G_i(t+1)-g_i)\nonumber\\&= [-V\phi_k^{'}(\vec{g})+F_i(t)]G_k(t+1)
\end{align}
where for the last equality, we have used the definition of $\vec{g}$ in \eqref{eqn:g_i}. Notice that since by assumption $G_k(t+1)>0$, the above translates to $-V\phi_k^{'}(\vec{g})+F_i(t) \leq 0$. Since we have $|\phi_k^{'}(\vec{g})| \leq B$ (Assumption \textbf{A1}), we have $F_i(t) \leq V\phi_k^{'}(\vec{g}) \leq BV$ as desired.
\end{proof}
Now, we move on to the proof of Lemma~\ref{lemma:det_queue_bnd_1}. Fix $i \in [n]$.
We use induction to prove that $F_i(t) \leq BV+1$ for all $t \in \mathbb{N}$. Notice that the result trivially follows for $t = 1$ ($\vec{F}(1) = \vec{0} \in \mathbb{R}^n$). Assume $F_i(t) \leq BV+1$. We establish $F_i(t+1) \leq BV+1$. We consider two cases.

\noindent
\textbf{Case 1: } $F_i(t) = BV+1$: Notice that from the previous claim we have $G_i(t+1) = 0$. Hence, we have from the definition of $F(t+1)$ in \eqref{eqn:def_fg}
\begin{align}
    F_i(t+1) &= [F_i(t) + G_i(t+1) - A_i(t)]_+ \leq_{(a)} [F_i(t)]_+ = BV+1 \nonumber,
\end{align}
where (a) follows since $G_i(t+1) = 0$, $A_i(t) \geq 0$, and $[\cdot]_+$ is a non-decreasing function, and the last inequality follows from $F_i(t) = BV+1$.

\noindent
\textbf{Case 2: } $F_i(t) \leq BV$: We have from \eqref{eqn:def_fg},
\begin{align}
    F_i(t+1) &= [F_i(t) + G_i(t+1) - A_i(t)]_+ \leq_{(a)} [F_i(t)+1]_+ \leq BV+1\nonumber,
\end{align}
where (a) follows since $G_i(t+1) \leq 1$, $A_i(t) \geq 0$, and $[\cdot]_+$ is a non-decreasing function, and the last inequality follows from $F_i(t) \leq BV$ and $[\cdot]_+$ is non-decreasing.
\end{proof}
\end{lemma}

\subsection{Analysis of the Algorithm}\label{sec:core_lemmas}
Now, we establish the core lemmas required to prove Theorem~\ref{theorem:final_bound_1_}.  In this section, we consider the interval $[T_0,\, T_0 + T - 1]$, where $T, T_0 \in \mathbb{N}$. We assume only that the success probabilities $q_{i,j}$ remain fixed over this interval and we impose no assumptions on their behavior outside of it.
Before proving Theorem~\ref{theorem:final_bound_1_}, we establish a deterministic bound $Q_i(t)$, which is direct consequence of Lemma~\ref{lemma:det_queue_bnd_1}.
\begin{corollary}\label{corr:queue_bound}
We have for $t \in [T_0,\, T_0 + T - 1]$ and $i \in [n]$, $Q_i(t) \leq BV + 1$.

\noindent
\textbf{Note:} This sample path result does not require fixed $q_{i,j}$ probabilities. It only requires $X_i(t) \in [0,1]$, and $\gamma_i(t) \in [0,1]$.
\end{corollary}

Now we state Lemma~\ref{lemma:t_p_1_to_t_1} that controls the error introduced by replacing the true link success probabilities with importance-sampling estimates. 
\begin{lemma}\label{lemma:t_p_1_to_t_1}
We have for all $t \in [T_0,\, T_0 + T - 1]$,
\begin{align}
    &-\sum_{i = 1}^n\sum_{j=1}^m q_{i,j}\mathbb{E}\{P_{i,j}(t) Q_i(t)\}\nonumber\\& \leq -\sum_{i=1}^n\sum_{j=1}^s \mathbb{E}\{Q_{i}(t)\hat{S}_{i,j}(t)\tilde{P}_{i,j}(t+1)\}  +\frac{1}{\eta}D_{\text{KL}}(\vec{\tilde{P}}(t+1)\lVert \vec{P}(t))  + \eta (BV+1)^2ns
\end{align}
where $\hat{S}_{i,j}(t)$ is defined in line~5 of Algorithm~\ref{algo:61}.
\begin{proof}
Consider
$\vec{x}^i, \vec{p}^i, \vec{q}^i \in \mathbb{R}^s$, where
\begin{align}
x^i_j &= -\eta Q_i(t)\bigl(1-\hat{S}_{i,j}(t)\bigr), 
p^i_j = \tilde{P}_{i,j}(t+1), q^i_j = P_{i,j}(t). \nonumber
\end{align}
Using Lemma~\ref{lemma:old} on $\vec{x}^i,\vec{p}^i,\vec{q}^i$ we get
\begin{align}\label{eqn:first_KL}
    &-\sum_{j=1}^s Q_{i}(t)(1-\hat{S}_{i,j}(t))\tilde{P}_{i,j}(t+1) \\&\leq_{(a)}\frac{1}{\eta}\Bigg{[}D_{\text{KL}}(\vec{\tilde{P}}_i(t+1)\lVert \vec{P}_i(t)) + \ln\left(\sum_{j = 1}^s {P}_{i,j}(t) e^{-\eta Q_i(t) (1-\hat{S}_{i,j}(t))}\right)\Bigg{]}
\end{align}
where $\vec{\tilde{P}}_i(t+1)$ and $\vec{P}_i(t) $ denote the $i$-th rows of $\vec{\tilde{P}}(t+1)$ and $\vec{P}_i(t)$, respectively. Notice that
\begin{align}
    D(\vec{\tilde{P}}(t+1)\lVert \vec{P}(t)) &= \sum_{i=1}^s D_{\text{KL}}(\vec{\tilde{P}}_i(t+1)\lVert \vec{P}_i(t))  \geq \sum_{i=1}^n D_{\text{KL}}(\vec{\tilde{P}}_i(t+1)\lVert \vec{P}_i(t))
\end{align}
Summing~\eqref{eqn:first_KL} over $i \in [n]$ and using the above inequality, we get
\begin{align}
    &-\sum_{i=1}^n\sum_{j=1}^s Q_{i}(t)(1-\hat{S}_{i,j}(t))\tilde{P}_{i,j}(t+1) \\&\leq_{(a)}\frac{1}{\eta}\Bigg{[}D(\vec{\tilde{P}}(t+1)\lVert \vec{P}(t)) + \sum_{i=1}^n\ln\left(\sum_{j = 1}^s {P}_{i,j}(t) e^{-\eta Q_i(t) (1-\hat{S}_{i,j}(t))}\right)\Bigg{]}
     \nonumber\\&\leq_{(b)} \frac{1}{\eta}D(\vec{\tilde{P}}(t+1)\lVert \vec{P}(t))  \nonumber\\&\ \ \ \ \ + \frac{1}{\eta}\sum_{i=1}^n\ln\Bigg{(}\sum_{j = 1}^s {P}_{i,j}(t)\Bigg{[}1 - \eta Q_i(t) (1-\hat{S}_{i,j}(t)) + \left(\eta Q_i(t) (1-\hat{S}_{i,j}(t))\right)^2\Bigg{]} \Bigg{)}
     \nonumber\\&\leq \frac{D(\vec{\tilde{P}}(t+1)\lVert \vec{P}(t))}{\eta} \nonumber\\&\ \ \ \ + \frac{1}{\eta}\sum_{i=1}^n\ln\Bigg{(}1-\sum_{j = 1}^s \eta{P}_{i,j}(t)Q_i(t)(1- \hat{S}_{i,j}(t))  +\sum_{j = 1}^s{P}_{i,j}(t)\left(\eta Q_i(t) (1-\hat{S}_{i,j}(t))\right)^2\Bigg{)}\nonumber\\&\leq_{(c)} \frac{1}{\eta}D(\vec{\tilde{P}}(t+1)\lVert \vec{P}(t)) \nonumber\\&\ \ \ \ \ - \sum_{i=1}^n\sum_{j = 1}^s  P_{i,j}(t) Q_i(t) (1-\hat{S}_{i,j}(t))  + \eta\sum_{i=1}^n\sum_{j = 1}^sP_{i,j}(t)\left( Q_i(t) (1-\hat{S}_{i,j}(t))\right)^2\nonumber\\&\leq_{(d)} \frac{1}{\eta}D(\vec{\tilde{P}}(t+1)\lVert \vec{P}(t)) \nonumber\\&\ \ \ \ \ - \sum_{i=1}^n\sum_{j = 1}^s P_{i,j}(t) Q_i(t) (1-\hat{S}_{i,j}(t)) + \eta (BV+1)^2\sum_{i=1}^n\sum_{j = 1}^sP_{i,j}(t)(1- \hat{S}_{i,j}(t))^2\nonumber\\&=_{(e)} \frac{1}{\eta}D(\vec{\tilde{P}}(t+1)\lVert \vec{P}(t)) \nonumber\\&\ \ \ \ \ - \sum_{i=1}^n\sum_{j = 1}^s P_{i,j}(t) Q_i(t) (1-\hat{S}_{i,j}(t)) + \eta (BV+1)^2\sum_{i=1}^n\sum_{j = 1}^s(1-S_{i,j}(t))Y_{i,j}(t)(1- \hat{S}_{i,j}(t))\nonumber\\&\leq \frac{D(\vec{\tilde{P}}(t+1)\lVert \vec{P}(t))}{\eta} - \sum_{i=1}^n\sum_{j = 1}^s P_{i,j}(t) Q_i(t) (1-\hat{S}_{i,j}(t)) + \eta (BV+1)^2\sum_{i=1}^n\sum_{j = 1}^s(1- \hat{S}_{i,j}(t))\nonumber
\end{align}
where (a) follows from Lemma~\ref{lemma:old}, (b) follows from $e^{-x} \leq 1- x + x^2$ for $x \geq -1$ (notice that $Q_i(t) \geq 0$, and $(1-\hat{S}_{i,j}(t)) \geq 0$ by definition), (c) follows from $\ln(1+x) \leq x$, and (d) follows from Corollary~\ref{corr:queue_bound}, (e) follows from the definition of $\hat{S}_{i,j}(t)$, and the last inequality follows from $(1- \hat{S}_{i,j}(t)) \geq 0$, and $(1-S_{i,j}(t)Y_{i,j}(t)) \in \{0,1\}$.  Taking  the expected value conditioned on $\mathcal{H}(t)$, followed by taking unconditional expectations, we have 
\begin{align}
    &-\sum_{i=1}^n\sum_{j=1}^s \mathbb{E}\{Q_{i}(t)(1-\hat{S}_{i,j}(t))\tilde{P}_{i,j}(t+1)\}\nonumber\\&\leq \frac{\mathbb{E}\{D(\vec{\tilde{P}}(t+1)\lVert \vec{P}(t))\}}{\eta} - \sum_{i=1}^n\sum_{j = 1}^s \mathbb{E}\{ (1-q_i)P_{i,j}(t) Q_i(t)\} + \eta (BV+1)^2\sum_{i=1}^n\sum_{j = 1}^s(1- q_{i,j})\nonumber\\&\leq \frac{1}{\eta}\mathbb{E}\{D(\vec{\tilde{P}}(t+1)\lVert \vec{P}(t))\}  - \sum_{i=1}^n\sum_{j = 1}^s \mathbb{E}\{ (1-q_i)P_{i,j}(t) Q_i(t)\}  + \eta (BV+1)^2ns\nonumber
\end{align}
Noticing that $\sum_{j=1}^s \tilde{P}_{i,j}(t+1) = \sum_{j=1}^s P_{i,j}(t) = 1$ and rearranging we have the result.
\end{proof}
\end{lemma}

For $t \in [T_0,\, T_0 + T - 1]$, define the drift $\Delta(t)$ as
\begin{align}
    \Delta(t)= \frac{1}{2}\mathbb{E}\{\lVert \vec{Q}(t+1) \rVert^2\} - \frac{1}{2}\mathbb{E}\{\lVert \vec{Q}(t) \rVert^2\} \nonumber.
\end{align}
Lemma~\ref{lemma:delta_lemma_1} provides an upper bound on the drift. Lemma~\ref{lemma:vital_lemma_1} shows that decisions~\eqref{eqn:decision_1_gamma} and~\eqref{eqn:decision_1} minimize an upper bound on the drift-plus-penalty term $\Delta(t) - V\phi(\vec{\gamma}(t+1))$. Intuitively, the penalty term $-V\phi(\vec{\gamma}(t+1))$ drives the algorithm toward utility maximization and enforces constraint~\eqref{eqn:gamma_gurr}, while the drift term controls the growth of the virtual queues. Keeping the drift small promotes stability of the virtual queues, which in turn enforces the long-term constraint~\eqref{eqn:p_gurr}. The parameter $V$ balances these two effects: larger $V$ places more emphasis on utility maximization at the cost of larger virtual queues.
\begin{lemma}\label{lemma:delta_lemma_1}
    We have for all $t \in [T_0,\, T_0 + T - 1]$,
    \begin{align}
        \Delta(t) &\leq  n+ \sum_{i=1}^n\mathbb{E}\{\gamma_{i}(t+1)Q_{i}(t)\}  -  \sum_{i=1}^n\sum_{j=1}^m q_{i,j}\mathbb{E}\{Q_{i}(t)P_{i,j}(t)\} \nonumber
    \end{align}
\begin{proof}
Notice that from the queuing equation \eqref{eqn:queue_eqn}, we have for all $i \in [n]$,
\begin{align}
    Q^2_{i}(t+1)&\leq \left(Q_{i}(t) + \gamma_{i}(t+1) - X_{i}(t)\right)^2 \nonumber\\& \leq Q^2_{i}(t) + \gamma_{i}^2(t+1) +X^2_{i}(t)  + 2Q_{i}(t)\left[\gamma_{i}(t+1) - X_{i}(t)\right] \nonumber\\& \leq Q^2_{i}(t) + 2+2Q_{i}(t)\left[\gamma_{i}(t+1) - X_{i}(t)\right],  \nonumber
\end{align} 
where for the last inequality, we have used $X_{i}(t)\in \{0,1\}$.
Summing the above for $i \in [n]$, we have
\begin{align}
    &\lVert \vec{Q}(t+1)\rVert^2 \leq \lVert \vec{Q}(t)\rVert^2 + 2n + 2 \sum_{i=1}^nQ_{i}(t)\left[\gamma_{i}(t+1) - X_{i}(t)\right].\nonumber
\end{align}
Taking the expectations conditioned $\mathcal{H}(t)$:
\begin{align}
    &\mathbb{E}\{\lVert \vec{Q}(t+1)\rVert^2 |\mathcal{H}(t) \}  \leq  \lVert\vec{Q}(t)\rVert^2 + 2n + 2\sum_{i=1}^nQ_{i}(t)\left[\mathbb{E}\{\gamma_{i}(t+1) |\mathcal{H}(t)\} -\sum_{j=1}^m q_{i,j}P_{i,j}(t)\right]. \nonumber
\end{align} 
 Taking expectations and performing simple algebraic manipulations, we have the result.
\end{proof}
\end{lemma}
We introduce the following lemma that combines $\vec{\tilde{P}}(t)$, and $\vec{P}(t)$, the actual and approximate solutions of \eqref{eqn:decision_1}.

\begin{lemma}\label{eqn:lemma_bound_lemma}
We have that for any $\vec{P} \in \mathcal{S}^{\text{doub}}$
\begin{align}
    D(\vec{P}\lVert \vec{P}(t)) \leq  2\varepsilon s+\theta+D(\vec{P}\lVert \vec{\tilde{P}}(t))
\end{align}
\begin{proof}
Notice that, for any $\vec{P} \in \mathcal{S}^{\text{doub}}$,
\begin{align}
   & D(\vec{P}\lVert \vec{\hat{P}}(t)) = \sum_{i=1}^s\sum_{j=1}^s P_{i,j}\ln\left(\frac{P_{i,j}}{\hat{P}_{i,j}(t)}\right) \leq_{(a)} \sum_{i=1}^s\sum_{j=1}^s P_{i,j}\ln\left(\frac{e^{\theta/s}P_{i,j}}{\tilde{P}_{i,j}(t)}\right) =  \theta +  D_{KL}(\vec{P}\lVert \vec{\tilde{P}}(t))\nonumber
\end{align}
where (a) follows since $\ln$ is non-decreasing and since $\vec{\hat{P}}(t)$ is a $(\theta/s)$-approximation of $\vec{\tilde{P}}(t)$ (See line~7 of Algorithm~\ref{algo:61}). Next,
\begin{align}
     &D(\vec{P}\lVert \vec{P}(t)) = \sum_{i=1}^s \sum_{j=1}^s P_{i,j}\ln\left(\frac{P_{i,j}}{P_{i,j}(t)}\right)\leq  \sum_{i=1}^s \sum_{j=1}^s P_{i,j}\ln\left(\frac{P_{i,j}}{(1-\varepsilon)\hat{P}_{i,j}(t)}\right) \nonumber\\&= s\ln\left(\frac{1}{1-\varepsilon}\right)+ D_{KL}(\vec{P}\lVert \vec{\hat{P}}(t))  \leq  2\varepsilon s+ D_{KL}(\vec{P}\lVert \vec{\hat{P}}(t))\nonumber
\end{align}
where for the last inequality we have used $\ln\left(\frac{1}{1-\varepsilon}\right) \leq 2\varepsilon$ whenever $\varepsilon \in [0,1/2]$. Combining the two inequalities, we are done.
\end{proof}
\end{lemma}
Now, we introduce the following lemma that will be useful in deriving the final bounds.
\begin{lemma}\label{lemma:vital_lemma_1}
We have for any $T,T_0 \in \mathbb{N}$, $\vec{\gamma} \in [0,1]^n$, and $\vec{P} \in \mathcal{S}^{\text{doub}}$,
\begin{align}
    &VT\phi(\vec{\gamma}) -V\sum_{t=T_0}^{T+T_0-1}\mathbb{E}\{\phi(\vec{\gamma}(t+1))\}\nonumber\\& \leq  \sum_{t=T_0}^{T+T_0-1}\sum_{i=1}^n\left[\gamma_{i}-\sum_{j=1}^mq_{i,j}P_{i,j}\right]\mathbb{E}\{Q_{i}(t)\}+nT+\frac{2\varepsilon s T}{\eta}  \nonumber\\& \ \ \ \ + \frac{\theta T}{\eta}+\eta ns(BV+1)^2  T + \frac{s}{\eta}\log\left(\frac{s}{\varepsilon}\right)+ \frac{n(BV+1)^2}{2}. \nonumber
\end{align}
\begin{proof}
Adding $-V\mathbb{E}\{\phi(\vec{\gamma}(t+1))\}$ to the result of Lemma~\ref{lemma:delta_lemma_1}, we have
\begin{align}\label{eqn:bef_fin_3}
    &\Delta(t)-V\mathbb{E}\{\phi(\vec{\gamma}(t+1))\}  \nonumber\\&\leq n-V\mathbb{E}\{\phi(\vec{\gamma}(t+1))\}+\sum_{i=1}^n\mathbb{E}\{\gamma_{i}(t+1)Q_{i}(t)\}  -  \sum_{i=1}^n\sum_{j=1}^m q_{i,j}\mathbb{E}\{Q_{i}(t)P_{i,j}(t)\} \nonumber\\& \leq_{(a)}  n-V\mathbb{E}\{\phi(\vec{\gamma}(t+1))\}+\sum_{i=1}^n\mathbb{E}\{\gamma_{i}(t+1)Q_{i}(t)\} -\sum_{i = 1}^n\sum_{j=1}^s\mathbb{E}\{Q_{i}(t) \hat{S}_{i,j}(t)\tilde{P}_{i,j}(t+1)\}\nonumber\\&\ \ \ \  +  \frac{1}{\eta}\mathbb{E}\{D(\vec{\tilde{P}}(t+1)\lVert \vec{P}(t))\} +  \eta ns(BV+1)^2,
\end{align} 
where (a) follows by Lemma~\ref{lemma:t_p_1_to_t_1}. Now notice that from the optimality of $\vec{\tilde{P}}(t+1), \vec{\gamma}(t+1)$ in \eqref{eqn:decision_1} (see Algorithm~\ref{algo:61}) with Lemma~\ref{lemma:push_back}-1, we have for any $\vec{\gamma} \in [0,1]^n$ and $\vec{P} \in \mathcal{S}^{\text{doub}}$,
\begin{align}
    &-V\phi(\vec{\gamma}(t+1))+\sum_{i=1}^n\gamma_{i}(t+1)Q_{i}(t)+\frac{1}{\eta}D(\vec{\tilde{P}}(t+1)\lVert \vec{P}(t))   - \sum_{i=1}^n\sum_{j=1}^sQ_{i}(t) \hat{S}_{i,j}(t)\tilde{P}_{i,j}(t+1) \nonumber\\& \leq -V\phi(\vec{\gamma}) + \sum_{i=1}^nQ_{i}(t)\left[\gamma_i-  \sum_{j=1}^s\hat{S}_{i,j}(t)P_{i,j}\right] + \frac{1}{\eta}D(\vec{P} \lVert \vec{P}(t))- \frac{1}{\eta}D(\vec{P} \lVert \vec{\tilde{P}}(t+1))
    \nonumber\\& \leq_{(a)} -V\phi(\vec{\gamma}) + \sum_{i=1}^nQ_{i}(t)\left[\gamma_i-  \sum_{j=1}^m\hat{S}_{i,j}(t)P_{i,j}\right]  + \frac{1}{\eta}D(\vec{P} \lVert \vec{P}(t))- \frac{1}{\eta}D(\vec{P} \lVert \vec{P}(t+1)) +   \frac{2\varepsilon s}{\eta}  \nonumber\\&\ \ \ \ + \frac{\theta}{\eta}
\end{align}
where (a) follows from Lemma~\ref{eqn:lemma_bound_lemma}.
Taking expectations of the above, we have
\begin{align}
    &-V\mathbb{E}\{\phi(\vec{\gamma}(t+1))\}+\sum_{i=1}^n\mathbb{E}\{\gamma_{i}(t+1)Q_{i}(t)\}  +\frac{1}{\eta}\mathbb{E}\{D(\vec{\tilde{P}}(t+1)\lVert \vec{P}(t))\}  \nonumber\\&\ \ \ \  - \sum_{i=1}^n\sum_{j=1}^m\mathbb{E}\{Q_{i}(t) \hat{S}_{i,j}(t)\tilde{P}_{i,j}(t+1)\} \nonumber\\& \leq  -V\phi(\vec{\gamma}) + \sum_{i=1}^n\mathbb{E}\left\{Q_{i}(t)\left[\gamma_i-  \sum_{j=1}^s\hat{S}_{i,j}(t)P_{i,j}\right]\right\} + \frac{1}{\eta}\mathbb{E}\{D(\vec{P} \lVert \vec{P}(t))\}\nonumber\\&\ \ \ \  - \frac{1}{\eta}\mathbb{E}\{D(\vec{P} \lVert \vec{P}(t+1))\}+   2\varepsilon s  + \frac{\theta}{\eta}\nonumber\\& =  -V\phi(\vec{\gamma}) + \sum_{i=1}^n\mathbb{E}\{Q_i(t)\}\left[\gamma_i-  \sum_{j=1}^mq_{i,j}P_{i,j}\right]   +  \frac{1}{\eta}\mathbb{E}\{D(\vec{P} \lVert \vec{P}(t))\}- \frac{1}{\eta}\mathbb{E}\{D(\vec{P} \lVert \vec{P}(t+1))\} \nonumber\\&\ \ \ \  +   \frac{2\varepsilon s}{\eta}  + \frac{\theta}{\eta}\nonumber
\end{align}
Substituting the above in \eqref{eqn:bef_fin_3}, we have
\begin{align}
     &\Delta(t)-V\mathbb{E}\{\phi(\vec{\gamma}(t+1))\} \nonumber\\& \leq  n -V\phi(\vec{\gamma}) + \sum_{i=1}^n\mathbb{E}\{Q_i(t)\}\left[\gamma_i-  \sum_{j=1}^mq_{i,j}P_{i,j}\right]  \nonumber\\&\ \ + \frac{1}{\eta}\mathbb{E}\{D(\vec{P} \lVert \vec{P}(t))\}- \frac{1}{\eta}\mathbb{E}\{D(\vec{P} \lVert \vec{P}(t+1))\} +   \frac{2\varepsilon s}{\eta}  + \frac{\theta}{\eta}+\eta ns(BV+1)^2
\end{align}
Summing the above for $t \in \{T_0,T_0+1,\dots,T+T_0-1\}$, we have
\begin{align}\label{eqn:bef_fin_5}
    \frac{1}{2}&\mathbb{E}\{\lVert \vec{Q}(T+T_0)\rVert^2\} -\frac{1}{2}\mathbb{E}\{\lVert \vec{Q}(T_0)\rVert^2\} -V\sum_{t=T_0}^{T+T_0-1}\mathbb{E}\{\phi(\vec{\gamma}(t+1))\} \nonumber\\& \leq -VT\phi(\vec{\gamma})  + \sum_{t=T_0}^{T+T_0-1}\sum_{i=1}^n\left[\gamma_{i}-\sum_{j=1}^mq_{i,j}P_{i,j}\right]\mathbb{E}\{Q_{i}(t)\} +nT+\frac{2\varepsilon s T}{\eta}  + \frac{\theta T}{\eta}\nonumber\\&\ \ \ \ +\eta ns(BV+1)^2 T + \frac{1}{\eta} \mathbb{E}\left\{D(\vec{P} \lVert \vec{P}(T_0))\right\} -\frac{1}{\eta} \mathbb{E}\left\{D(\vec{P} \lVert \vec{P}(T+T_0))\right\} \nonumber\\&\leq -VT\phi(\vec{\gamma})  + \sum_{t=T_0}^{T+T_0-1}\sum_{i=1}^n\left[\gamma_{i,j}-\sum_{j=1}^mq_{i,j}P_{i,j}\right]\mathbb{E}\{Q_{i}(t)\} +nT+\frac{2\varepsilon s T}{\eta}  + \frac{\theta T}{\eta}\nonumber\\&\ \ \ +\eta ns(BV+1)^2 T  + \frac{s}{\eta}\log\left(\frac{s}{\varepsilon}\right),
\end{align}
where for the last inequality, we have used Lemma~\ref{lemma:KL_u_bound_1}, and the fact that $\vec{P}(T_0) \in \mathcal{S}^{\text{doub}}_{\varepsilon/s}$. Rearranging the above, and using $\lVert \vec{Q}(T_0) \rVert^2 \leq n(BV+1)^2$ from Corollary~\ref{corr:queue_bound}, and $\lVert \vec{Q}(T+T_0)\rVert^2 \geq 0$ we are done.
\end{proof}
\end{lemma}
Next, we have the following lemma that combines iterates $\vec{\gamma}(T_0+1),\vec{\gamma}(T_0+2)\dots,\vec{\gamma}(T+T_0)$, and $\vec{X}(T_0),\vec{X}(T_0+1)$ $\dots,\vec{X}(T+T_0-1)$.
\begin{lemma}\label{lemma:gamma_to_X_1}
    We have for any $T,T_0 \in \mathbb{N}$, 
    \begin{align}
         &\phi\left(\frac{1}{T}\left(\sum_{t=T_0+1}^{T+T_0} \vec{\gamma}(t)\right)\right) \leq \phi\left(\frac{1}{T}\left(\sum_{t=T_0}^{T+T_0-1}\vec{X}(t)\right)\right) +\frac{\sqrt{n}B\lVert\vec{Q}(T+T_0)\rVert}{T}. \nonumber
    \end{align}  
\begin{proof}
Notice that
\begin{align}
    &\phi\left(\frac{1}{T}\left(\sum_{t=T_0+1}^{T+T_0} \vec{\gamma}(t)\right)\right) = \phi\left(\sum_{t=T_0}^{T+T_0-1} \frac{\vec{X}(t)}{T}+\sum_{t=T_0}^{T+T_0-1} \frac{\vec{\gamma}(t+1)-\vec{X}(t)}{T}\right) \nonumber\\&\leq_{(a)} \phi\left(\sum_{t=T_0}^{T+T_0-1} \frac{\vec{X}(t)}{T}   +\sum_{t=T_0}^{T+T_0-1} \frac{[\vec{Q}(t+1)-\vec{Q}(t)]}{T}\right) \nonumber\\&\leq_{(b)} \phi\left(\frac{1}{T}\left(\sum_{t=T_0}^{T+T_0-1} \vec{X}(t)\right)+\frac{\vec{Q}(T+T_0)}{T}\right) \nonumber\\&\leq_{(c)} \phi\left(\frac{1}{T}\left(\sum_{t=T_0}^{T+T_0-1} \vec{X}(t)\right)\right)+\frac{\sqrt{n}B\lVert\vec{Q}(T+T_0)\rVert}{T}, \nonumber
\end{align}
where (a) follows from the entrywise nondecreasing property of $\phi$, and the queuing equation~\eqref{eqn:queue_eqn}, for (b) we use $\vec{Q}(T_0) \geq 0$ and the entrywise nondecreasing property of $\phi$, and (c) follows since $\phi$ is $\sqrt{n}B$-Lipschitz continuous from Assumption~\textbf{A1}.  \footnote{Assumption~\textbf{A1} ensures that $\phi$ can be extended to an entrywise nondecreasing function that is $\sqrt{nB}$-Lipschitz continuous over the domain $[0,\infty)^n$.}
\end{proof}
\end{lemma}

Now, we are ready to establish Theorem~\ref{theorem:final_bound_1_}.

\begin{proof}[Proof of Theorem~\ref{theorem:final_bound_1_}]
Substituting $\vec{P}^*,\vec{\gamma}^*$ in Lemma~\ref{lemma:vital_lemma_1}, where $(\vec{P}^*,\vec{\gamma}^*)$ is the optimal solution of (P2) defined in Lemma~\ref{lemma:equiv_prob_lemma} we have 
\begin{align}\label{eqn:before_last_1}
    &VT\phi(\vec{\gamma}^*) -V\sum_{t=T_0}^{T+T_0-1}\mathbb{E}\{\phi(\vec{\gamma}(t+1))\}\nonumber\\&\leq  \sum_{t=T_0}^{T+T_0-1}\sum_{i=1}^n\left[\gamma^*_{i}-\sum_{j=1}^mq_{i,j}P^*_{i,j}\right]\mathbb{E}\{Q_{i}(t)\}+nT+\frac{2\varepsilon s T}{\eta}  + \frac{\theta T}{\eta}+\eta ns(BV+1)^2  T  \nonumber\\&\ \ \ \ + \frac{s}{\eta}\log\left(\frac{s}{\varepsilon}\right)+ \frac{n(BV+1)^2}{2}\nonumber\\& \leq   nT+\frac{2\varepsilon s T}{\eta}  + \frac{\theta T}{\eta}+\eta ns(BV+1)^2  T  + \frac{s}{\eta}\log\left(\frac{s}{\varepsilon}\right)+ \frac{n(BV+1)^2}{2}
\end{align}
where the last inequality follows since $\sum_{j=1}^mq_{i,j}P_{i,j}^* \geq \gamma_{i}^*$ for all $i \in [n]$ from the feasibility of $(\vec{P}^*,\vec{\gamma}^*)$ for (P2).
Now, we divide both sides of the above inequality by $VT$ and use the Jensen's inequality to obtain
\begin{align}
  \phi(&\vec{\gamma}^*) -\mathbb{E}\left\{\phi\left(\frac{1}{T}\sum_{t=T_0}^{T+T_0-1}\vec{\gamma}(t+1)\right)\right\}\nonumber\\&\leq_{(a)}   \frac{n}{V}+\frac{2\varepsilon s}{\eta V}  + \frac{\theta }{ V\eta}+\frac{\eta ns(BV+1)^2 }{V}  + \frac{s}{\eta VT}\log\left(\frac{s}{\varepsilon}\right)+ \frac{n(BV+1)^2}{2 VT}
\end{align}
Combining with Lemma~\ref{lemma:gamma_to_X_1}, we have 
\begin{align}
\phi(&\vec{\gamma}^*) -\mathbb{E}\left\{\phi\left(\frac{1}{T}\sum_{t=T_0}^{T+T_0-1}\vec{X}(t)\right)\right\} \nonumber\\&\leq_{(a)} \frac{n}{V}+\frac{2\varepsilon s}{\eta V}  + \frac{\theta }{ V\eta}+\frac{\eta ns(BV+1)^2 }{V}  + \frac{s}{\eta VT}\log\left(\frac{s}{\varepsilon}\right)+ \frac{n(BV+1)^2}{2 VT}  \nonumber\\&\ \ \ \ +\frac{\sqrt{n}B\mathbb{E}\{\lVert \vec{Q}(T+T_0)\rVert\}}{T} \nonumber\\&\leq_{(b)} \frac{n}{V}+\frac{2\varepsilon s}{\eta V}  + \frac{\theta }{ V\eta}+\frac{\eta ns(BV+1)^2 }{V}  + \frac{s}{\eta VT}\log\left(\frac{s}{\varepsilon}\right) + \frac{n(BV+1)^2}{2 VT} +\frac{nB(BV+1)}{T},  \nonumber
\end{align}
where (a) follows due to Lemma~\ref{lemma:gamma_to_X_1}, and (b) follows due to Corollary~\ref{corr:queue_bound}. Now, due to Lemma~\ref{lemma:equiv_prob_lemma}, we have $\phi(\vec{\gamma}^*) = \phi^* \geq \phi^{\text{opt}}$.  Using Jensen's inequality, we are done.
\end{proof}
\section{Adaptive MAC.CF Algorithm}\label{sec:slow_adaptive}
Now, we develop our second algorithm. The main difficulty in implementing the Algorithm~\ref{algo:61_} is the complexity of solving the inner problem~\eqref{eqn:decision_1}. In particular, the computational complexity of solving the above problem depends heavily on the structure of the matrices $\vec{P}(t),\vec{Q}(t),\vec{\hat{S}}(t)$~\cite{FRANKLIN1989717}. The goal of this section is to develop an adaptive algorithm that avoids the above computational complexities. The resulting algorithm admits efficient inner iterations, at the cost of a slower convergence rate.

The algorithm has the structure explained in Section~\ref{sec:intution}. However, instead of finding a doubly stochastic matrix by solving an inner problem, we find $\vec{\tilde{P}}(t)$ as a column stochastic matrix and a row stochastic matrix alternatively in odd and even slots. Unlike the previous algorithm, the above step can be solved explicitly. Then we obtain $\vec{P}(t)$ by approximating $\vec{\tilde{P}}(t)$ by a doubly stochastic matrix, using the rounding trick introduced in Sections~\ref{sec:round_sec}, after which we sample $\vec{Y}(t)$ from $\vec{P}(t)$ using Birkhoff-von Neumann Decomposition~\cite{birkhoff1946three}. In Algorithm~\ref{algo:61_}, we provide the algorithm for the task. The algorithm uses parameters $\eta>0,V>0,\varepsilon  \in (0,1/s]$
\begin{algorithm}
Initialize $\vec{\tilde{P}}(1) = \frac{1}{s}\vec{1}\vec{1}^{\top} \in \mathcal{S}^{\text{doub}}$ and the virtual queues $\vec{Q}(1)= \vec{0} \in \mathbb{R}^n$.\\
\For{each time slot $t \in \mathbb{N}$}{
Set $\vec{P}(t) = \text{ROUND}(\vec{\tilde{P}}(t))$ (See Section~\ref{sec:round_sec}. This function yields $\vec{P}(t) \in \mathcal{S}^{\text{doub}}_{\varepsilon/s}$ ).\\

Sample $\vec{Y}(t)$ from $\vec{P}(t)$ (see Section~\ref{sec:def_prel}-c), and receive $\vec{S}(t) \odot \vec{Y}(t)$ as feedback.\\
Compute the estimator $\vec{\hat{S}}(t)$ for $\vec{S}(t)$ using $\hat{S}_{i,j}(t) = S_{i,j}(t)Y_{i,j}(t)/P_{i,j}(t)$ for all $i \in [n]$, and $j \in [m]$.\\ 
Define
\begin{align}
    &\vec{\gamma}(t+1)= \arg\min_{\vec{\gamma} \in [0,1]^n}\Big{[} -V\phi(\vec{\gamma}) + \sum_{i=1}^n Q_{i}(t)\gamma_{i}\Big{]}
\end{align}\\
Find $\vec{\tilde{P}}(t+1) \in \mathcal{Q}$ using
\begin{align}\label{eqn:decision_1_}
    \vec{\tilde{P}}(t+1) &= \arg\min_{\vec{P} \in \mathcal{Q}}\Big{[} -\sum_{i=1}^n\sum_{j=1}^m Q_{i}(t) \hat{S}_{i,j}(t)P_{i,j}  + \frac{1}{\eta}D(\vec{P} \lVert \vec{\tilde{P}}(t)) \Big{]}, 
\end{align}
where $\mathcal{Q} = \mathcal{S}^{\text{col}}_{\varepsilon}$ if $t$ is even, and $\mathcal{Q} = \mathcal{S}^{\text{row}}_{\varepsilon}$ if $t$ is odd.\\
Update the virtual queues
\begin{align}\label{eqn:queue_eqn_}
    \vec{Q}(t+1) = \left[\vec{Q}(t) + \vec{\gamma}(t+1) - \vec{X}(t)\right]_+,
\end{align}
where $X_i(t) = \sum_{j=1}^m Y_{i,j}(t)S_{i,j}(t)$ for $i \in [n]$.\\
}
\caption{Adaptive MAC.CF (Parameters $\eta>0,V>0,\varepsilon  \in (0,1/s]$)}\label{algo:61_}
\end{algorithm}

In this algorithm, we employ the utility-based importance sampling estimator $\hat{S}_{i,j}(t) = \frac{S_{i,j}(t) Y_{i,j}(t)}{P_{i,j}(t)}$
of $q_{i,j}$. Moreover, the algorithm does not require mixing with the uniform distribution, since the inner problem~\eqref{eqn:decision_1} can be solved explicitly while enforcing the constraint that each entry is at least $\varepsilon$. The solution is discussed in Appendix~\ref{sec:intem_2}. Finding $\vec{\gamma}(t+1)$ is the same procedure as in Algorithm~\ref{algo:61}.

The following Theorem establishes the perforance bound of the Algorithm.
\begin{theorem}\label{theorem:final_bound_1}
Consider $T, T_0 \in \mathbb{N}$ and suppose that the success probabilities 
$q_{i,j}$ remain fixed on the interval $[T_0,\, T_0 + T - 1]$. Then, independently of their behavior outside this interval, for any algorithm parameters  
$\varepsilon \in (0,1/s]$, $\eta>0$, and $V>0$, Algorithm~\ref{algo:61_} yields
\begin{align}
    &\phi^{\text{opt}} -\phi\left(\mathbb{E}\left\{\frac{1}{T}\sum_{t=T_0}^{T+T_0-1}\vec{X}(t)\right\}\right) \nonumber\\&\leq \frac{n}{V}+\frac{9\eta nms^2 }{2\varepsilon V}(BV+1)^2    +\frac{\varepsilon nms (BV+1)}{V} +\frac{s}{\eta VT }\log\left(\frac{1}{\varepsilon}\right)   +\frac{n(BV+1)^2}{2 VT} +\frac{nB(BV+1)}{T}.  \nonumber
\end{align}

In particular, choosing  $\eta = \Theta(\log^{2/3}(T)/T)$, $\varepsilon = \Theta((\log(T)/T)^{1/3}) < 1/s$, and $V = \Theta(T^{1/3})$, we have 
\begin{align}
    \phi^{\text{opt}} -\phi\left(\mathbb{E}\left\{\frac{1}{T}\sum_{t=T_0}^{T+T_0-1}\vec{X}(t)\right\}\right) =  O\left(\sqrt[3]{\frac{\log(T)}{T}}\right). \nonumber
\end{align}
\end{theorem}
The basic intuition behind the algorithm is similar to that of 
Algorithm~\ref{algo:61}. The key difference lies in the separate 
treatment of even and odd iterations. The divergence term in~\eqref{eqn:decision_1_} ensures that 
$\vec{\tilde{P}}(t+1)$ remains close to $\vec{\tilde{P}}(t)$. 
Importantly, one of these matrices is row-stochastic, whereas the 
other is column-stochastic. Therefore, although we do not enforce 
double stochasticity at each iteration, the divergence term guarantees 
that a selected row-stochastic matrix remains close to being 
column-stochastic, and vice versa.

We first describe the ROUND function used in line~3 of the algorithm.
\subsection{ROUND function}\label{sec:round_sec}
We introduce ROUND, a technique adapted from the one introduced in~\cite{Altschuler2017} to approximate a nonnegative matrix by a matrix in the transport polytope. The technique can be readily extended to the Birkhoff polytope.

\noindent
\textbf{ROUND}($\vec{P}$) function for $\vec{P}\in \mathbb{R}^{s\times s}_+$  : 

\begin{enumerate}
\item 
Define the matrix $\vec{P}^{'}$ (row normalization of $\vec{P}$) using
\begin{align}\label{eqn:row_normalize}
    P^{'}_{i,j} = \begin{cases}
        \frac{P_{i,j}}{\sum_{l=1}^s P_{i,l}} & \text{ if } \sum_{l=1}^s P_{i,l}>1\\
        P_{i,j} & \text{ otherwise.}
    \end{cases}
\end{align}

\item
Define the matrix $\vec{P}^{''}$ (column normalization of $\vec{P}^{'}$) using
\begin{align}\label{eqn:column_normalize}
    P^{''}_{i,j} = \begin{cases}
        \frac{P^{'}_{k,j}}{\sum_{k=1}^s P^{'}_{k,j}} & \text{ if } \sum_{k=1}^s P^{'}_{k,j}>1\\
        P^{'}_{i,j} & \text{ otherwise.}
    \end{cases}
\end{align}

\item
Define the output matrix $\vec{Q}$,
\begin{align}\label{eqn:q_def_round}
    \vec{Q} = \begin{cases}
         \vec{P}^{''} + \frac{(\vec{1} - \vec{P}^{''}\vec{1})(\vec{1} - (\vec{P}^{''})^{\top}\vec{1})^{\top}}{C} & \text{ if } C \neq 0\\
          \vec{P}^{''} & \text{ otherwise, }
    \end{cases} 
\end{align}
where $C = \lVert \vec{1} - \vec{P}^{''}\vec{1}\rVert_1$.
\end{enumerate}
We establish the relationships between the input and the output matrices of the ROUND operation. A version of the following lemma is proved for the transport polytope in~\cite{Altschuler2017}. We adapt the idea for the Birkhoff polytope.
\begin{lemma}\label{lemma:algo_8}
The input $\vec{P} \in \mathbb{R}^{s \times s}_+$ and the output matrix $\vec{Q}$ of the ROUND function satisfy
\begin{enumerate}
    \item $\vec{Q} \in \mathcal{S}^{\text{doub}}$.
    \item If the input $\vec{P} \in \mathcal{S}_{\varepsilon}$, we have $\vec{Q} \in \mathcal{S}^{\text{doub}}_{\varepsilon/s}$.
    \item $\lVert \vec{P} - \vec{Q} \rVert_1 \leq 2\left(\lVert \vec{P}\vec{1} - \vec{1} \lVert_1 + \lVert \vec{P}^{\top}\vec{1} - \vec{1} \rVert_1\right)$.
\end{enumerate}
\begin{proof}
   See Appendix~\ref{app:algo_8}
\end{proof}
\end{lemma}

\subsection{Analysis of the Algorithm}


The rest of the section focuses on proving Theorem~\ref{theorem:final_bound_1}. Similar to Section~\ref{sec:fast_adaptive}, we consider the interval $[T_0:T_0+T-1]$, where the success probabilities $q_{i,j}$ remain fixed. We first notice that  Lemma~\ref{lemma:det_queue_bnd_1}, Corollary~\ref{corr:queue_bound}, and Lemma~\ref{lemma:delta_lemma_1} hold in the same form for Algorithm~\ref{algo:61_}. These results are used in the proof of Theorem~\ref{theorem:final_bound_1}. The intermediate problem~\eqref{eqn:decision_1_} admits an explicit solution for $\vec{\tilde{P}}(t+1)$, as discussed in Appendix~\ref{sec:intem_2}.

First, we introduce the following lemma that bounds the difference $ \vec{P}(t) - \vec{\tilde{P}}(t)$.
\begin{lemma}\label{lemma:round_lemma}
We have $\vec{P}(t) \in \mathcal{S}^{\text{doub}}_{\varepsilon/s}$ and 
\begin{align}
    \lVert \vec{P}(t) - \vec{\tilde{P}}(t) \rVert_1 \leq 2\lVert \vec{\tilde{P}}(t+1) - \vec{\tilde{P}}(t) \rVert_1. \nonumber
\end{align}
\begin{proof}
The fact that $\vec{P}(t) \in \mathcal{S}^{\text{doub}}_{\varepsilon/s}$ follows directly from Lemma~\ref{lemma:algo_8}-2.

For the other part, we only consider the case when $t$ is even. The case when $t$ is odd follows similarly.

\noindent
\textbf{Claim 1: } When $t$ is even, we have
\begin{align}
     \lVert \vec{P}(t) - \vec{\tilde{P}}(t) \rVert_1 \leq 2 \lVert \vec{\tilde{P}}(t)^{\top}\vec{1} - \vec{1} \rVert_1. \nonumber
\end{align}
This again follows by direct application of Lemma~\ref{lemma:algo_8}-4. (Notice that $\lVert \vec{\tilde{P}}(t)\vec{1} - \vec{1} \rVert_1 = 0$, since $ \vec{\tilde{P}}(t) \in \mathcal{S}^{\text{row}}$ when $t$ is even)

\noindent
\textbf{Claim 2: } When $t$ is even, we have
\begin{align}
     \lVert \vec{\tilde{P}}(t)^{\top}\vec{1} - \vec{1} \rVert_1 \leq \lVert \vec{\tilde{P}}(t+1) - \vec{\tilde{P}}(t) \rVert_1.  \nonumber
\end{align}
 Notice that 
\begin{align}
    \lVert \vec{\tilde{P}}(t+1) - \vec{\tilde{P}}(t) \rVert_1 &= \sum_{i=1}^s\sum_{j=1}^s|\tilde{P}_{i,j}(t+1) - \tilde{P}_{i,j}(t)|  \geq_{(a)}\sum_{j=1}^s\left\lvert\sum_{i=1}^s\tilde{P}_{i,j}(t+1) - \sum_{i=1}^s\tilde{P}_{i,j}(t)\right\rvert \nonumber\\& =  \lVert \vec{\tilde{P}}(t+1)^{\top}\vec{1} -\vec{\tilde{P}}(t)^{\top} \vec{1}\rVert_1 = \lVert \vec{\tilde{P}}(t)^{\top}\vec{1} -\vec{1}\rVert_1, \nonumber
\end{align}
where (a) follows from the triangle inequality and the last equality follows since  $\vec{\tilde{P}}(t+1) \in \mathcal{S}^{\text{col}}$ when $t$ is even.

Combining the two claims, we are done.
\end{proof}
\end{lemma}
Now, we first introduce the following lemma that bounds the error introduced by replacing the probabilities $q_{i,j}$ with $\hat{S}_{i,j}(t)$.
\begin{lemma}\label{lemma:t_p_1_to_t_1_}
We have for any $t \in [T_0,\, T_0 + T - 1]$,
\begin{align}
    -&\sum_{i=1}^n\sum_{j=1}^mq_{i,j}\mathbb{E}\left\{Q_{i}(t) P_{i,j}(t)\right\} \nonumber\\&\leq \frac{9\eta nms^2}{2\varepsilon}(BV+1)^2+\frac{1}{\eta}\mathbb{E}\left\{D(\vec{\tilde{P}}(t+1)\lVert \vec{\tilde{P}}(t))\right\}   - \sum_{i=1}^n\sum_{j=1}^m\mathbb{E}\left\{Q_{i}(t) \hat{S}_{i,j}(t)\tilde{P}_{i,j}(t+1)\right\}, \nonumber
\end{align}
where $\hat{S}_{i,j}(t)$ is defined in line~6 of Algorithm~\ref{algo:61_}.
\begin{proof}
Notice that
\begin{align}\label{eqn:prel_1}
     \sum_{i=1}^n\sum_{j=1}^m&\mathbb{E}\left\{Q_{i}(t) \hat{S}_{i,j}(t)\tilde{P}_{i,j}(t+1)\right\}  \nonumber\\&=  \sum_{i=1}^n\sum_{j=1}^m\mathbb{E}\left\{Q_{i}(t) \hat{S}_{i,j}(t)[\tilde{P}_{i,j}(t+1)-\tilde{P}_{i,j}(t)]\right\}\nonumber\\&\ \ \ \ +\sum_{i=1}^n\sum_{j=1}^m\mathbb{E}\left\{Q_{i}(t) \hat{S}_{i,j}(t)[\tilde{P}_{i,j}(t)-P_{i,j}(t)]\right\} +\sum_{i=1}^n\sum_{j=1}^m\mathbb{E}\left\{Q_{i}(t) \hat{S}_{i,j}(t)P_{i,j}(t)\right\}. 
\end{align}  
Now, we handle each of the above four terms separately.
Notice that
\begin{align}
    \sum_{i=1}^n&\sum_{j=1}^m\mathbb{E}\left\{Q_{i}(t) \hat{S}_{i,j}(t)[\tilde{P}_{i,j}(t+1)-\tilde{P}_{i,j}(t)]\right\} \nonumber\\&\leq_{(a)} \frac{3\eta s}{2}\mathbb{E}\left\{\sum_{i=1}^n\sum_{j=1}^m Q^2_i(t)\hat{S}^2_{i,j}(t)\right\}  +\frac{1}{6s\eta}\mathbb{E}\left\{\lVert \vec{\tilde{P}}(t+1)- \vec{\tilde{P}}(t)\rVert^2\right\}\nonumber\\&\leq \frac{3\eta s}{2}\mathbb{E}\left\{\sum_{i=1}^n\sum_{j=1}^m Q^2_i(t)\hat{S}^2_{i,j}(t)\right\} +\frac{1}{3\eta}\mathbb{E}\left\{D(\vec{\tilde{P}}(t+1)\lVert \vec{\tilde{P}}(t))\right\}, \nonumber
\end{align}
where for (a) we use $\frac{3\eta s}{2}\lVert \vec{a}\rVert^2 + \frac{1}{6\eta s}\lVert \vec{b}\rVert^2 \geq  \sum_{i=1}^k a_ib_i$ for $k$-dimensional vectors $\vec{a,b}$, and the last inequality follows from Lemma~\ref{lemma:KL_with_L1_1}. Next, Notice that
\begin{align}
    &\sum_{i=1}^n\sum_{j=1}^m\mathbb{E}\left\{Q_{i}(t) \hat{S}_{i,j}(t)[\tilde{P}_{i,j}(t)-P_{i,j}(t)]\right\} \nonumber\\&\leq 3\eta s\mathbb{E}\left\{\sum_{i=1}^n\sum_{j=1}^m Q^2_i(t)\hat{S}^2_{i,j}(t)\right\}+\frac{\mathbb{E}\left\{\lVert \vec{\tilde{P}}(t)- \vec{P}(t)\rVert^2_1\right\}}{12s\eta}\nonumber\\&\leq 3\eta s\mathbb{E}\left\{\sum_{i=1}^n\sum_{j=1}^m Q^2_i(t)\hat{S}^2_{i,j}(t)\right\} +\frac{\mathbb{E}\left\{\lVert \vec{\tilde{P}}(t+1)- \vec{\tilde{P}}(t)\rVert^2_1\right\}}{3s\eta} \nonumber\\&\leq 3\eta s\mathbb{E}\left\{\sum_{i=1}^n\sum_{j=1}^m Q^2_i(t)\hat{S}^2_{i,j}(t)\right\}  +\frac{2\mathbb{E}\left\{D(\vec{\tilde{P}}(t+1) \lVert \vec{\tilde{P}}(t))\right\}}{3\eta}, \nonumber
\end{align}
where the second inequality follows from Lemma~\ref{lemma:round_lemma}, and the last inequality follows from Lemma~\ref{lemma:KL_with_L1_1}. Next, notice that
\begin{align}
   &\sum_{i=1}^n\sum_{j=1}^m\mathbb{E}\left\{Q_{i}(t) \hat{S}_{i,j}(t)P_{i,j}(t)\right\} \nonumber\\&= \sum_{i=1}^n\sum_{j=1}^m \mathbb{E}\left\{Q_{i}(t) \frac{S_{i,j}(t)}{P_{i,j}(t)}P_{i,j}(t)Y_{i,j}(t)\right\} =   \sum_{i=1}^n\sum_{j=1}^mq_{i,j}\mathbb{E}\left\{Q_{i}(t) P_{i,j}(t)\right\}, \nonumber
\end{align}
Now, using the above in \eqref{eqn:prel_1} we have
\begin{align}\label{eqn:pres_rel_2}
    &\sum_{i=1}^n\sum_{j=1}^m\mathbb{E}\left\{Q_{i}(t) \hat{S}_{i,j}(t)\tilde{P}_{i,j}(t+1)\right\} \nonumber\\&\leq  \frac{9\eta s}{2}\mathbb{E}\left\{\sum_{i=1}^n\sum_{j=1}^m Q^2_i(t)\hat{S}^2_{i,j}(t)\right\} +\frac{1}{\eta}\mathbb{E}\left\{D(\vec{\tilde{P}}(t+1) \lVert \vec{\tilde{P}}(t))\right\} \nonumber\\&\ \ \ \ \ +\sum_{i=1}^n\sum_{j=1}^mq_{i,j}\mathbb{E}\left\{Q_{i}(t) P_{i,j}(t)\right\}.
\end{align}
But notice that
\begin{align}
   \mathbb{E}&\left\{\sum_{i=1}^n\sum_{j=1}^m Q^2_i(t)\hat{S}^2_{i,j}(t)\right\}=  \sum_{i=1}^n\sum_{j=1}^m \mathbb{E}\left\{Q^2_{i}(t)\frac{S_{i,j}(t)}{P^2_{i,j}(t)} Y_{i,j}(t)\right\} \nonumber\\&\leq_{(a)}  \frac{s}{\varepsilon}\sum_{i=1}^n\sum_{j=1}^m \mathbb{E}\left\{Q^2_{i}(t)\frac{S_{i,j}(t)}{P_{i,j}(t)} Y_{i,j}(t)\right\}= \frac{s}{\varepsilon}\sum_{i=1}^n\sum_{j=1}^mq_{i,j}\mathbb{E}\left\{Q^2_{i}(t)\right\}\leq_{(b)} \frac{s}{\varepsilon}\sum_{i=1}^n\sum_{j=1}^m\mathbb{E}\left\{Q^2_{i}(t)\right\} \nonumber\\& \leq \frac{nms}{\varepsilon}(BV+1)^2, \nonumber
\end{align}        
where (a) follows since $P_{i,j}(t) \geq \varepsilon/s$ from Lemma~\ref{lemma:round_lemma}, (b) follows since $q_{i,j} \leq 1$, and the last inequality follows due to Corollary~\ref{corr:queue_bound}.  Combining with \eqref{eqn:pres_rel_2}, we are done.
\end{proof}
\end{lemma}
Now, we introduce the following lemma that will be useful in deriving the final bounds.
\begin{lemma}\label{lemma:vital_lemma_1_}
We have for any $T,T_0 \in \mathbb{N}$, $\vec{\gamma} \in [0,1]^n$, and $\vec{P} \in \mathcal{S}^{\text{doub}}_{\varepsilon}$,
\begin{align}
    &VT\phi(\vec{\gamma}) -V\sum_{t=T_0}^{T+T_0-1}\mathbb{E}\{\phi(\vec{\gamma}(t+1))\}\nonumber\\& \leq  nT+\frac{9\eta nms^2 T}{2\varepsilon}(BV+1)^2 + \sum_{t=T_0}^{T+T_0-1}\sum_{i=1}^n\left[\gamma_{i}-\sum_{j=1}^mq_{i,j}P_{i,j}\right]\mathbb{E}\{Q_{i}(t)\}+ \frac{s}{\eta}\log\left(\frac{1}{\varepsilon}\right)\nonumber\\&\ \ \ \ + \frac{n(BV+1)^2}{2}. \nonumber
\end{align}
\begin{proof}

Adding $-V\mathbb{E}\{\phi(\vec{\gamma}(t+1))\}$ to the result of Lemma~\ref{lemma:delta_lemma_1}, we have
\begin{align}\label{eqn:bef_fin_3_}
    &\Delta(t)-V\mathbb{E}\{\phi(\vec{\gamma}(t+1))\}  \nonumber\\&\leq n-V\mathbb{E}\{\phi(\vec{\gamma}(t+1))\}+\sum_{i=1}^n\mathbb{E}\{\gamma_{i}(t+1)Q_{i}(t)\}  -  \sum_{i=1}^n\sum_{j=1}^m q_{i,j}\mathbb{E}\{Q_{i}(t)P_{i,j}(t)\} \nonumber\\& \leq_{(a)}  n-V\mathbb{E}\{\phi(\vec{\gamma}(t+1))\}+\sum_{i=1}^n\mathbb{E}\{\gamma_{i}(t+1)Q_{i}(t)\}  +\frac{9\eta nms^2}{2\varepsilon}(BV+1)^2  \nonumber\\&\ \ \ \ +\frac{1}{\eta}\mathbb{E}\left\{D(\vec{\tilde{P}}(t+1)\lVert \vec{\tilde{P}}(t))\right\}   - \sum_{i=1}^n\sum_{j=1}^m\mathbb{E}\left\{Q_{i}(t) \hat{S}_{i,j}(t)\tilde{P}_{i,j}(t+1)\right\},
\end{align} 
where (a) follows by Lemma~\ref{lemma:t_p_1_to_t_1_}. Now notice that from the optimality of $\vec{\tilde{P}}(t+1), \vec{\gamma}(t+1)$ in \eqref{eqn:decision_1_} (see Algorithm~\ref{algo:61_}) with Lemma~\ref{lemma:push_back}-1, we have for any $\vec{\gamma} \in [0,1]^n$ and $\vec{P} \in \mathcal{S}^{\text{doub}}_{\varepsilon}$,
\begin{align}
    &-V\phi(\vec{\gamma}(t+1))+\sum_{i=1}^n\gamma_{i}(t+1)Q_{i}(t)+\frac{1}{\eta}D(\vec{\tilde{P}}(t+1)\lVert \vec{\tilde{P}}(t))  - \sum_{i=1}^n\sum_{j=1}^mQ_{i}(t) \hat{S}_{i,j}(t)\tilde{P}_{i,j}(t+1) \nonumber\\& \leq -V\phi(\vec{\gamma}) + \sum_{i=1}^nQ_{i}(t)\left[\gamma_i-  \sum_{j=1}^m\hat{S}_{i,j}(t)P_{i,j}\right] + \frac{1}{\eta}D(\vec{P} \lVert \vec{\tilde{P}}(t))- \frac{1}{\eta}D(\vec{P} \lVert \vec{\tilde{P}}(t+1)). \nonumber
\end{align}
Taking expectations of the above, we have
\begin{align}
    &-V\mathbb{E}\{\phi(\vec{\gamma}(t+1))\}+\sum_{i=1}^n\mathbb{E}\{\gamma_{i}(t+1)Q_{i}(t)\}  +\frac{1}{\eta}\mathbb{E}\{D(\vec{\tilde{P}}(t+1)\lVert \vec{\tilde{P}}(t))\}  \nonumber\\&\ \ \ \  - \sum_{i=1}^n\sum_{j=1}^m\mathbb{E}\{Q_{i}(t) \hat{S}_{i,j}(t)\tilde{P}_{i,j}(t+1)\} \nonumber\\&\leq  -V\phi(\vec{\gamma}) + \sum_{i=1}^n\mathbb{E}\{Q_{i}(t)\gamma_i\}-  \sum_{i=1}^n\sum_{j=1}^m\mathbb{E}\{Q_{i}(t)\hat{S}_{i,j}(t)P_{i,j}\} + \frac{1}{\eta}\mathbb{E}\{D(\vec{P} \lVert \vec{\tilde{P}}(t))\}   \nonumber\\& \ \ \ \  - \frac{1}{\eta}\mathbb{E}\{D(\vec{P} \lVert \vec{\tilde{P}}(t+1))\}\nonumber\\& =  -V\phi(\vec{\gamma}) + \sum_{i=1}^n \gamma_{i}\mathbb{E}\{Q_{i}(t)\}-   + \frac{1}{\eta}\mathbb{E}\{D(\vec{P} \lVert \vec{\tilde{P}}(t))\} - \frac{1}{\eta}\mathbb{E}\{D(\vec{P} \lVert \vec{\tilde{P}}(t+1))\} \nonumber
\end{align}
Substituting the above in \eqref{eqn:bef_fin_3_}, we have
\begin{align}
     &\Delta(t)-V\mathbb{E}\{\phi(\vec{\gamma}(t+1))\} \nonumber\\& \leq  n +\frac{9\eta nms^2}{2\varepsilon}(BV+1)^2-V\phi(\vec{\gamma}) + \sum_{i=1}^n \left[\gamma_{i}-\sum_{j=1}^mq_{i,j}P_{i,j}\right]\mathbb{E}\{Q_{i}(t)\} + \frac{1}{\eta}\mathbb{E}\{D(\vec{P} \lVert \vec{\tilde{P}}(t))\}  \nonumber\\&\ \ \ \ - \frac{1}{\eta}\mathbb{E}\{D(\vec{P} \lVert \vec{\tilde{P}}(t+1))\}. \nonumber
\end{align}
Summing the above for $t \in \{T_0,T_0+1,\dots,T+T_0-1\}$, we have
\begin{align}\label{eqn:bef_fin_5_}
    \frac{1}{2}&\mathbb{E}\{\lVert \vec{Q}(T+T_0)\rVert^2\} -\frac{1}{2}\mathbb{E}\{\lVert \vec{Q}(T_0)\rVert^2\}  -V\sum_{t=T_0}^{T+T_0-1}\mathbb{E}\{\phi(\vec{\gamma}(t+1))\} \nonumber\\& \leq nT+\frac{9\eta nms^2 T}{2\varepsilon}(BV+1)^2 -VT\phi(\vec{\gamma})   + \sum_{t=T_0}^{T+T_0-1}\sum_{i=1}^n\left[\gamma_{i}-\sum_{j=1}^mq_{i,j}P_{i,j}\right]\mathbb{E}\{Q_{i}(t)\}\nonumber\\&\ \ \ + \frac{1}{\eta}\mathbb{E}\left\{D(\vec{P} \lVert \vec{\tilde{P}}(T_0))\right\} -\frac{1}{\eta}\mathbb{E}\left\{D(\vec{P} \lVert \vec{\tilde{P}}(T+T_0))\right\} \nonumber\\&\leq nT+\frac{9\eta nms^2 T}{2\varepsilon}(BV+1)^2 -VT\phi(\vec{\gamma})  + \sum_{t=T_0}^{T+T_0-1}\sum_{i=1}^n\left[\gamma_{i,j}-\sum_{j=1}^mq_{i,j}P_{i,j}\right]\mathbb{E}\{Q_{i}(t)\} \nonumber\\&\ \ \ \  + \frac{s}{\eta}\log\left(\frac{1}{\varepsilon}\right),
\end{align}
where for the last inequality, we have used Lemma~\ref{lemma:KL_u_bound_1}. Rearranging the above, and using $\lVert \vec{Q}(T_0) \rVert^2 \leq n(BV+1)^2$ from Corollary~\ref{corr:queue_bound}, and $\lVert \vec{Q}(T+T_0)\rVert^2 \geq 0$ we are done.
\end{proof}
\end{lemma}

Now, we are ready to establish Theorem~\ref{theorem:final_bound_1}

\begin{proof}[Proof of Theorem~\ref{theorem:final_bound_1}]
Substituting $\vec{P}^*(1-\varepsilon s)+\varepsilon \vec{1},\vec{\gamma}^*$ in Lemma~\ref{lemma:vital_lemma_1}, where $(\vec{P}^*,\vec{\gamma}^*)$ is the optimal solution of (P2) defined in Lemma~\ref{lemma:equiv_prob_lemma} and $\vec{1}$ is the all 1 matrix, we have 
\begin{align}
    &VT\phi(\vec{\gamma}^*) -V\sum_{t=T_0}^{T+T_0-1}\mathbb{E}\{\phi(\vec{\gamma}(t+1))\}\nonumber\\&\leq nT+\frac{9\eta nms^2 T}{2\varepsilon}(BV+1)^2 + \sum_{t=T_0}^{T+T_0-1}\sum_{i=1}^n\left(\gamma^*_{i}-\sum_{j=1}^mq_{i,j}P^*_{i,j}\right)\mathbb{E}\{Q_{i}(t)\}\nonumber\\&\ \ \ \ +\varepsilon\sum_{t=T_0}^{T+T_0-1}\sum_{i=1}^n\sum_{j=1}^m\left(q_{i,j}sP^*_{i,j}-q_{i,j}\right)\mathbb{E}\{Q_{i}(t)\}+\frac{s}{\eta}\log\left(\frac{1}{\varepsilon}\right)  +\frac{n(BV+1)^2}{2}
    \nonumber\\&\leq_{(a)} nT+\frac{9\eta nms^2 T}{2\varepsilon}(BV+1)^2    +\varepsilon nms T(BV+1) +\frac{s}{\eta}\log\left(\frac{1}{\varepsilon}\right)  +\frac{n(BV+1)^2}{2},
\end{align}
where (a) follows since $\sum_{j=1}^mq_{i,j}P_{i,j}^* \geq \gamma_{i}^*$ for all $i \in [n]$ from the feasibility of $(\vec{P}^*,\vec{\gamma}^*)$ for (P2), and Corollary~\ref{corr:queue_bound}. 
Now, we divide both sides of the above inequality by $VT$ and use the Jensen's inequality to obtain
\begin{align}
  \phi(&\vec{\gamma}^*) -\mathbb{E}\left\{\phi\left(\frac{1}{T}\sum_{t=T_0}^{T+T_0-1}\vec{\gamma}(t+1)\right)\right\}\nonumber\\&\leq_{(a)}  \frac{n}{V}+\frac{9\eta nms^2 }{2\varepsilon V}(BV+1)^2    +\frac{\varepsilon nms (BV+1)}{V} +\frac{s}{\eta VT }\log\left(\frac{1}{\varepsilon}\right)  +\frac{n(BV+1)^2}{2 VT}. \nonumber
\end{align}
Combining with Lemma~\ref{lemma:gamma_to_X_1}, we have 
\begin{align}
\phi(&\vec{\gamma}^*) -\mathbb{E}\left\{\phi\left(\frac{1}{T}\sum_{t=T_0}^{T+T_0-1}\vec{X}(t)\right)\right\} \nonumber\\&\leq_{(a)}\frac{n}{V}+\frac{9\eta nms^2 }{2\varepsilon V}(BV+1)^2    +\frac{\varepsilon nms (BV+1)}{V}+\frac{s}{\eta VT }\log\left(\frac{1}{\varepsilon}\right)  +\frac{n(BV+1)^2}{2 VT}\nonumber\\&\ \ \ \ +\frac{\sqrt{n}B\mathbb{E}\{\lVert \vec{Q}(T+T_0)\rVert\}}{T} \nonumber\\&\leq_{(b)} \frac{n}{V}+\frac{9\eta nms^2 }{2\varepsilon V}(BV+1)^2    +\frac{\varepsilon nms (BV+1)}{V}  +\frac{s}{\eta VT }\log\left(\frac{1}{\varepsilon}\right)  +\frac{n(BV+1)^2}{2 VT}\nonumber\\&\ \ \ \ +\frac{nB(BV+1)}{T},  \nonumber
\end{align}
where (a) follows due to Lemma~\ref{lemma:gamma_to_X_1}, and (b) follows due to Corollary~\ref{corr:queue_bound}. Now, due to Lemma~\ref{lemma:equiv_prob_lemma}, we have $\phi(\vec{\gamma}^*) = \phi^* \geq \phi^{\text{opt}}$. Combining with Jensen's inequality, we are done.
    
\end{proof}
\section{Special Cases}
In this section, we introduce several interesting cases of the general setting. The first is an adaptive algorithm for the single-channel case ($m=1$). This algorithm has both closed form solutions for iterations and fast convergence. Then we introduce a UCB-based non-adaptive algorithm with efficient implementation. Both algorithms will be used for comparisons. Finally, we consider the single channel ($m = 1$) special case with $\phi(\vec{x}) = \min\{x_1,x_2,\dots,x_n\}$. In this setting, a simple mechanism achieves convergence to the optimal value while adapting to channel variations, without requiring explicit estimation.
\subsection{Single Channel Algorithm} 
 The single channel case simplifies due to the absence of matching constraints. In particular, in each iteration, we assign a user to the channel, and the assigned user in the $t$-th time slot is sampled from $\vec{p}(t) \in \Delta^{n}$. For this section, we use the notation $S_i(t) = S_{i,1}(t)$, $Y_i(t) = Y_{i,1}(t)$. Algorithm~\ref{algo:62} summarizes the steps.
\begin{algorithm}
Initialize $\vec{p}(1) = \frac{1}{n}\vec{1}$ and the virtual queues $\vec{Q}(1) = \vec{0} \in \mathbb{R}^n$.\\
\For{each time slot $t \in \mathbb{N}$}{
Sample  $l_{t} \in [n]$ from $\vec{p}(t)$, set $\vec{Y}(t)$ as the $l_t$-th unit vector, and receive feedback.\\
Compute the estimator $\vec{\hat{S}}(t)$ for $\vec{S}(t)$ using $\hat{S}_{i}(t) = \frac{S_{i}(t)Y_{i}(t)}{p_{i}(t)}$ for all $i\in [n]$\\ 
Find
\begin{align}\label{eqn:decision_1__}
    &\vec{\gamma}(t+1)= \arg\min_{\vec{\gamma} \in [0,1]^n}\Big{[} -V\phi(\vec{\gamma}) + \sum_{i=1}^n Q_{i}(t)\gamma_{i}\Big{]}.\nonumber
    \nonumber\\&\vec{p}(t+1) = \arg\min_{\vec{p} \in \Delta^n_{\varepsilon}}\Big{[} -\sum_{i=1}^nQ_{i}(t) \hat{S}_{i}(t)p_{i}   + \frac{1}{\eta}D_{\text{KL}}(\vec{p} \lVert \vec{p}(t)) \Big{]},
\end{align}\\
Update $\vec{Q}(t+1) = \left[\vec{Q}(t) + \vec{\gamma}(t+1)- \vec{X}(t)\right]_+$.
}
\caption{Single-Channel Adaptive MAC (Parameters $V>0$, $\varepsilon \in (0,1/n])$, and $\eta>0$)}\label{algo:62}
\end{algorithm}

We first state the performance bound of the algorithm.

\begin{theorem}\label{theorem:final_bound}
Consider $T, T_0 \in \mathbb{N}$ and suppose that the success probabilities 
$q_{i,j}$ remain fixed on the interval $[T_0,\, T_0 + T - 1]$. For any parameters $\varepsilon \in (0,1/n)$, $\eta, V>0$, we have that,
\begin{align}
    \phi^{\text{opt}}& -\phi\left(\frac{1}{T}\sum_{t=T_0}^{T+T_0-1}\mathbb{E}\{\vec{X}(t)\}\right) \nonumber\\&\leq \frac{n}{V} +  \frac{n\varepsilon^2 e^{\frac{\eta (BV+1)}{\varepsilon}}}{\eta V} +\frac{\varepsilon n^2 (BV+1)}{V} +\frac{1}{\eta VT}\log\left(\frac{1}{\varepsilon} \right) + \frac{n(BV+1)^2}{2 VT}+\frac{nB(BV+1)}{T} \nonumber
\end{align}
In particular, using  $\eta = \Theta(\sqrt{\log(T)}/T)$, $\varepsilon = \Theta(\sqrt{\log(T)}/\sqrt{T}) < 1$, and $V = \Theta(T^{1/2})$, we have 
\begin{align}
    \phi^{\text{opt}} -\phi\left(\mathbb{E}\left\{\frac{1}{T}\sum_{t=T_0}^{T+T_0-1}\vec{X}(t)\right\}\right) =  O\left(\sqrt{\frac{\log(T)}{T}}\right). \nonumber
\end{align}
\end{theorem}

\subsubsection{Solving Problem~\eqref{eqn:decision_1__}}
Notice that finding $\vec{\gamma}(t+1)$ is the same problem as in the multi-channel case. In addition, the problem to solve for $\vec{p}(t+1)$ reduces to the problem (P3) discussed in Appendix~\ref{sec:intem_2}.
\subsubsection{Deterministic Problem}
 Consider the problem
\begin{align}
    &\underset{\vec{p},\vec{\gamma}}{\text{max }}\phi\left(\vec{\gamma}\right)\label{eqn:obj_new_pb}\\& \text{s.t. }
   \vec{p}\in \Delta_n\label{cons:simp_1}\\& \ \ \ \ \
   \vec{\gamma}\in [0,1]^n\label{cons:R_1}\\& \ \ \ \ \
    q_{i}p_i  \geq \gamma_{i} \ \forall i \in \{1,\dots,n\} \label{cons:pq_geq_gam_1},
\end{align}
Using an argument similar to Lemma~\ref{lemma:equiv_prob_lemma}, it can be shown that the optimal objective value of the above problem is at least $\phi^{\text{opt}}$. For this section, let $\phi^*$ denote the optimal objective value of the above problem. 

Let $(\vec{p}^*,\vec{\gamma}^*)$ denote an optimal solution of the above problem. If $q_1,q_2,\dots,q_n$ were known, the policy of sampling a user independently in each time slot using the distribution $\vec{p}^*$ achieves an objective value of $\phi^{*}$ for (P1). The subsequent analysis shows that even when $q_1,q_2,\dots,q_n$ are unknown, we can achieve an objective value arbitrarily close to $\phi^*$ for (P1), thereby showing $\phi^*= \phi^{\text{opt}}$.
\subsubsection{Analysis of the Algorithm}
Most of the results required for the error analysis are borrowed from the multi-channel case. In particular, notice that the deterministic queue bound Corollary~\ref{corr:queue_bound} holds in this case.

Define the drift $\Delta(t)$ as
\begin{align}
    \Delta(t)= \frac{1}{2}\mathbb{E}\{\lVert \vec{Q}(t+1) \rVert^2\} - \frac{1}{2}\mathbb{E}\{\lVert \vec{Q}(t) \rVert^2\}.
\end{align}
We have the following lemma.
\begin{lemma}\label{lemma:delta_lemma}
    We have that for all $t \in \{1,2,\dots\}$,
    \begin{align}
        \Delta(t)  \leq  n + \sum_{i=1}^n\mathbb{E}\{\gamma_i(t+1)Q_i(t)\} -  \sum_{i=1}^n q_i\mathbb{E}\{Q_i(t)p_i(t)\} \nonumber
    \end{align}
\begin{proof}
Proof follows repeating the same arguments as Lemma~\ref{lemma:delta_lemma_1}.
\end{proof}
\end{lemma}

\begin{lemma}\label{lemma:t_p_1_to_t}
For any $t \in [T_0:T+T_0-1]$, we have
\begin{align}
 -&\sum_{i = 1}^n q_i\mathbb{E}\{p_{i}(t)  Q_i(t)\}\nonumber\\&\leq - \sum_{i=1}^n  \mathbb{E}\{Q_i(t) \hat{S}_i(t)p_i(t+1)\}+\frac{1}{\eta}\mathbb{E}\{D_{\text{KL}}(\vec{p}(t+1)\lVert \vec{p}(t))\} +  \frac{n\varepsilon^2 e^{\frac{\eta (BV+1)}{\varepsilon}}}{\eta}
\end{align}
where $\hat{S}_i(t)$ is defined in line 4 of Algorithm~\ref{algo:62}.
\begin{proof}
Using Lemma~\ref{lemma:old}, we have
\begin{align}
    &\sum_{i=1}^n Q_i(t) \hat{S}_i(t)p_i(t+1) \nonumber\\&\leq_{(a)}  \frac{1}{\eta}\Bigg{[}D_{\text{KL}}(\vec{p}(t+1)\lVert \vec{p}(t)) + \ln\left(\sum_{i = 1}^n {p}_{i}(t) e^{\eta Q_i(t) \hat{S}_{i}(t)}\right)\Bigg{]}\nonumber\\&\leq_{(b)}  \frac{1}{\eta}\Bigg{[}D_{\text{KL}}(\vec{p}(t+1)\lVert \vec{p}(t)) + \ln\left(\sum_{i = 1}^n {p}_{i}(t) \left[1+ \eta Q_i(t) \hat{S}_{i}(t)+ \frac{\varepsilon^2e^{\frac{\eta (BV+1)}{\varepsilon}}}{ (BV+1)^2}Q^2_i(t)\hat{S}^2_i(t)\right]\right)\Bigg{]} 
    \nonumber\\&= \frac{1}{\eta}\Bigg{[}D_{\text{KL}}(\vec{p}(t+1)\lVert \vec{p}(t)) + \ln\left(1+\sum_{i = 1}^n \eta{p}_{i}(t)  Q_i(t) \hat{S}_{i}(t)+ \sum_{i = 1}^n\frac{\varepsilon^2e^{\frac{\eta (BV+1)}{\varepsilon}}}{(BV+1)^2}p_i(t)Q^2_i(t)\hat{S}^2_i(t)\right)\Bigg{]} \nonumber\\&\leq_{(c)} \frac{1}{\eta}\Bigg{[}D_{\text{KL}}(\vec{p}(t+1)\lVert \vec{p}(t)) + \sum_{i = 1}^n \eta{p}_{i}(t)  Q_i(t) \hat{S}_{i}(t)+ \sum_{i = 1}^n\frac{\varepsilon^2e^{\frac{\eta (BV+1)}{\varepsilon}}}{(BV+1)^2}p_i(t)Q^2_i(t)\hat{S}^2_i(t)\Bigg{]}\nonumber\\&=_{(d)} \frac{1}{\eta}D_{\text{KL}}(\vec{p}(t+1)\lVert \vec{p}(t)) + \sum_{i = 1}^n {p}_{i}(t)  Q_i(t) \hat{S}_{i}(t)+ \sum_{i = 1}^n\frac{\varepsilon^2 e^{\frac{\eta (BV+1)}{\varepsilon}}}{\eta}S_i(t)Y_i(t)\hat{S}_i(t)\nonumber\\&=_{(e)} \frac{1}{\eta}D_{\text{KL}}(\vec{p}(t+1)\lVert \vec{p}(t)) + \sum_{i = 1}^n {p}_{i}(t)  Q_i(t) \hat{S}_{i}(t)+ \frac{\varepsilon^2 e^{\frac{\eta (BV+1)}{\varepsilon}}}{\eta}\sum_{i = 1}^n\hat{S}_i(t)
\end{align}
where (a) follows from Lemma~\ref{lemma:old}, (b) follows combining Corollary~\ref{corr:queue_bound} and $\hat{S}_i(t) \leq 1/\varepsilon$ with $e^{x} \leq 1+ x + (e^c - c - 1)x^2/c^2$ for $x \in [0,c]$ for any $c>0$, (c) follows from $\ln(1+x) \leq x$, (d) follows from the definition of $\hat{S}_i(t)$, and the last inequality follows since $S_i(t),Y_i(t) \in \{0,1\}$.
Taking  the expected value conditioned on $\mathcal{H}(t)$, followed by taking unconditional expectations, we have 
\begin{align}
  &\sum_{i=1}^n \mathbb{E}\{Q_i(t) \hat{S}_i(t)p_i(t+1)\} \nonumber\\&\leq \frac{1}{\eta}\mathbb{E}\{D_{\text{KL}}(\vec{p}(t+1)\lVert \vec{p}(t))\} + \sum_{i = 1}^n q_i\mathbb{E}\{p_{i}(t)  Q_i(t)\}+ \frac{\varepsilon^2 e^{\frac{\eta (BV+1)}{\varepsilon}}}{\eta}\sum_{i=1}^n q_i\nonumber\\&\leq \frac{1}{\eta}\mathbb{E}\{D_{\text{KL}}(\vec{p}(t+1)\lVert \vec{p}(t))\} + \sum_{i = 1}^n q_i\mathbb{E}\{p_{i}(t)  Q_i(t)\}+ \frac{n\varepsilon^2 e^{\frac{\eta (BV+1)}{\varepsilon}}}{\eta}
\end{align}
where the last inequality follows since $q_i \in [0,1]$.
Rearranging, we are done.
\end{proof}
\end{lemma}
Now, we introduce the following lemma.
\begin{lemma}\label{lemma:vital_lemma}
We have that for any $T_0 \in \mathbb{N}$, $\vec{\gamma} \in [0,1]^n$, and $\vec{p} \in \Delta^n_{\varepsilon}$,
\begin{align}
    &VT\phi(\vec{\gamma})-V\sum_{t=T_0}^{T+T_0-1}\mathbb{E}\{\phi(\vec{\gamma}(t+1))\}\nonumber\\&\leq nT +  \frac{n\varepsilon^2 e^{\frac{\eta (BV+1)}{\varepsilon}}T}{\eta} + \sum_{t=T_0}^{T_0+T-1}\sum_{i=1}^n (\gamma_i-p_i q_i)\mathbb{E}\{Q_i(t)\}+\frac{1}{\eta}\log\left(\frac{1}{\varepsilon} \right) + \frac{n(BV+1)^2}{2}
\end{align}
\begin{proof}
Adding $-V\mathbb{E}\{\phi(\vec{\gamma}(t+1))\}$ to the result of Lemma~\ref{lemma:delta_lemma}, we have that,
\begin{align}\label{eqn:bef_fin}
    \Delta(t)&-V\mathbb{E}\{\phi(\vec{\gamma}(t+1))\}  \nonumber\\&\leq n-V\mathbb{E}\{\phi(\vec{\gamma}(t+1))\}+\sum_{i=1}^n\mathbb{E}\{\gamma_i(t+1)Q_i(t)\} -  \sum_{i=1}^n q_i\mathbb{E}\{Q_i(t)p_i(t)\} 
\end{align} 

Using Lemma~\ref{lemma:t_p_1_to_t} on the above, we have
\begin{align}\label{eqn:intem_1-1-}
    \Delta(t)&-V\mathbb{E}\{\phi(\vec{\gamma}(t+1))\}  \nonumber\\&\leq n-V\mathbb{E}\{\phi(\vec{\gamma}(t+1))\}+\sum_{i=1}^n\mathbb{E}\{\gamma_i(t+1)Q_i(t)\}   - \sum_{i=1}^n  \mathbb{E}\{Q_i(t)\hat{S}_i(t)p_i(t+1)\}\nonumber\\&\ \ \ \ +\frac{1}{\eta}\mathbb{E}\{D_{\text{KL}}(\vec{p}(t+1)\lVert \vec{p}(t))\} +  \frac{n\varepsilon^2 e^{\frac{\eta (BV+1)}{\varepsilon}}}{\eta}
\end{align} 
Notice that from the optimality of $(\vec{\gamma}(t+1),\vec{p}(t+1))$ in \eqref{eqn:decision_1__}, we have that for any $\vec{p} \in \Delta^n_{\varepsilon}$ and $\vec{\gamma} \in [0,1]^n$
\begin{align}
    &-V\phi(\vec{\gamma}(t+1))+\sum_{i=1}^n\mathbb{E}\{\gamma_i(t+1)Q_i(t)\}-\sum_{i=1}^n\mathbb{E}\{ Q_i(t) \hat{S}_i(t)p_i(t+1)\}\nonumber\\&\ \ \ \ +\frac{1}{\eta}\mathbb{E}\{D_{\text{KL}}(\vec{p}(t+1)\lVert \vec{p}(t))\} \nonumber\\& \leq -V\phi(\vec{\gamma}) + \sum_{i=1}^n \gamma_i\mathbb{E}\{Q_i(t)\}- \sum_{i=1}^np_i\mathbb{E}\{Q_i(t) \hat{S}_i(t)\}+\frac{1}{\eta}\mathbb{E}\{D_{\text{KL}}(\vec{p}\lVert \vec{p}(t))\}\nonumber\\&\ \ \ \  -\frac{1}{\eta}\mathbb{E}\{D_{\text{KL}}(\vec{p}\lVert \vec{p}(t+1))\}\nonumber\\&= -V\phi(\vec{\gamma}) + \sum_{i=1}^n (\gamma_i-p_i q_i)\mathbb{E}\{Q_i(t)\}+\frac{1}{\eta}\mathbb{E}\{D_{\text{KL}}(\vec{p}\lVert \vec{p}(t))\} -\frac{1}{\eta}\mathbb{E}\{D_{\text{KL}}(\vec{p}\lVert \vec{p}(t+1))\}
\end{align}
where the inequality follows from Lemma~\ref{lemma:push_back}-2. Using the above in \eqref{eqn:intem_1-1-}, we have for any $\vec{p} \in \Delta^n_{\varepsilon}$  and $\vec{\gamma} \in [0,1]^n$
\begin{align}
    \Delta(t)&-V\mathbb{E}\{\phi(\vec{\gamma}(t+1))\}  \nonumber\\&\leq n +  \frac{n\varepsilon^2 e^{\frac{\eta (BV+1)}{\varepsilon}}}{\eta}-V\phi(\vec{\gamma}) + \sum_{i=1}^n (\gamma_i-p_i q_i)\mathbb{E}\{Q_i(t)\}+\frac{1}{\eta}\mathbb{E}\{D_{\text{KL}}(\vec{p}\lVert \vec{p}(t))\} \nonumber\\&\ \ \ \ -\frac{1}{\eta}\mathbb{E}\{D_{\text{KL}}(\vec{p}\lVert \vec{p}(t+1))\}
\end{align} 
Now, we add the above for $t \in [T_0:T_0+T-1]$ to get
\begin{align}
    &\frac{1}{2}\mathbb{E}\{\lVert \vec{Q}(T+T_0)\rVert^2\} -\frac{1}{2}\mathbb{E}\{\lVert \vec{Q}(T_0)\rVert^2\}-V\sum_{t=T_0}^{T+T_0-1}\mathbb{E}\{\phi(\vec{\gamma}(t+1))\} \nonumber\\&\leq nT +  \frac{n\varepsilon^2 e^{\frac{\eta (BV+1)}{\varepsilon}}T}{\eta}-VT\phi(\vec{\gamma}) + \sum_{t=T_0}^{T_0+T-1}\sum_{i=1}^n (\gamma_i-p_i q_i)\mathbb{E}\{Q_i(t)\}+\frac{1}{\eta}\mathbb{E}\{D_{\text{KL}}(\vec{p}\lVert \vec{p}(T_0))\} \nonumber\\&\ \ \ \ -\frac{1}{\eta}\mathbb{E}\{D_{\text{KL}}(\vec{p}\lVert \vec{p}(T + T_0))\}
    \nonumber\\&\leq nT +  \frac{n\varepsilon^2 e^{\frac{\eta (BV+1)}{\varepsilon}}T}{\eta}-VT\phi(\vec{\gamma}) + \sum_{t=T_0}^{T_0+T-1}\sum_{i=1}^n (\gamma_i-p_i q_i)\mathbb{E}\{Q_i(t)\}+\frac{1}{\eta}\mathbb{E}\{D_{\text{KL}}(\vec{p}\lVert \vec{p}(T_0))\} \nonumber\\&\ \ \ \ -\frac{1}{\eta}\mathbb{E}\{D_{\text{KL}}(\vec{p}\lVert \vec{p}(T + T_0))\}\nonumber\\&\leq_{(a)} nT +  \frac{n\varepsilon^2 e^{\frac{\eta (BV+1)}{\varepsilon}}T}{\eta}-VT\phi(\vec{\gamma}) + \sum_{t=T_0}^{T_0+T-1}\sum_{i=1}^n (\gamma_i-p_i q_i)\mathbb{E}\{Q_i(t)\}+\frac{1}{\eta}\log\left(\frac{1}{\varepsilon} \right) 
\end{align}
where (a) follows by $D_{\text{KL}}(\vec{p}\lVert \vec{s}) \leq \log(1/\varepsilon)$ for $\vec{p} \in \Delta^n$ and $\vec{s} \in \Delta^{n}_{\varepsilon}$ (Proof follows the same steps as Lemma~\ref{lemma:KL_u_bound_1}). Rearranging and using $\lVert Q(T_0)\rVert ^2 \leq n (BV+1)^2$ (follows from Corollary~\eqref{corr:queue_bound}), we are done.
\end{proof}
\end{lemma}
Now, we are ready to establish Theorem~\ref{theorem:final_bound}.

\begin{proof}[Proof of Theorem~\ref{theorem:final_bound}]
Substituting $(1-\varepsilon n)\vec{p}^* +\varepsilon,\vec{\gamma}^*$ in Lemma~\ref{lemma:vital_lemma}, where $\vec{p}^*,\vec{\gamma}^*$ is the optimal solution of Problem~\eqref{eqn:obj_new_pb}-\eqref{cons:pq_geq_gam_1}, we have that,
\begin{align}\label{eqn:before_last}
    &VT\phi^{*} -V\sum_{t=T_0}^{T+T_0-1}\mathbb{E}\{\phi(\vec{\gamma}(t+1))\}\nonumber\\&\leq nT +  \frac{n\varepsilon^2 e^{\frac{\eta (BV+1)}{\varepsilon}}T}{\eta} + \sum_{t=T_0}^{T_0+T-1}\sum_{i=1}^n (\gamma^*_i-p^*_i q_i)\mathbb{E}\{Q_i(t)\}+\varepsilon n\sum_{t=T_0}^{T_0+T-1}\sum_{i=1}^n\mathbb{E}\{Q_i(t)\}\nonumber\\&\ \ \ \ +\frac{1}{\eta}\log\left(\frac{1}{\varepsilon} \right) + \frac{n(BV+1)^2}{2}\nonumber\\&\leq_{(a)} nT +  \frac{n\varepsilon^2 e^{\frac{\eta (BV+1)}{\varepsilon}}T}{\eta} + \varepsilon n^2 (BV+1)T +\frac{1}{\eta}\log\left(\frac{1}{\varepsilon} \right) + \frac{n(BV+1)^2}{2}
\end{align}
where (a) follows since $q_ip_i^* \geq \gamma_i^*$ (since $(\vec{p}^*,\vec{\gamma}^*)$ is feasible for problem~\eqref{eqn:obj_new_pb}-\eqref{cons:pq_geq_gam_1}), and Corollary~\eqref{corr:queue_bound}. Dividing both sides by $VT$ and using Jensen's inequality, we have
\begin{align}
    &\phi^{*} -\mathbb{E}\left\{\phi\left( \frac{1}{T}\sum_{t=T_0}^{T+T_0-1}\vec{\gamma}(t+1)\right)\right\}\nonumber\\&\leq \frac{n}{V} +  \frac{n\varepsilon^2 e^{\frac{\eta (BV+1)}{\varepsilon}}}{\eta V} +\frac{\varepsilon n^2 (BV+1)}{V} +\frac{1}{\eta VT}\log\left(\frac{1}{\varepsilon} \right) + \frac{n(BV+1)^2}{2 VT}
\end{align}
Combining above with Lemma~\ref{lemma:gamma_to_X_1}, we have that,
\begin{align}
&\phi^{\text{opt}} -\mathbb{E}\left\{\phi\left(\frac{1}{T}\sum_{t=T_0}^{T+T_0-1}\vec{X}(t)\right)\right\} \nonumber\\&\leq_{(a)}\frac{n}{V} +  \frac{n\varepsilon^2 e^{\frac{\eta (BV+1)}{\varepsilon}}}{\eta V} +\frac{\varepsilon n^2 (BV+1)}{V} +\frac{1}{\eta VT}\log\left(\frac{1}{\varepsilon} \right) + \frac{n(BV+1)^2}{2 VT} +\frac{\sqrt{n}B\mathbb{E}\{\lVert\vec{Q}(T+T_0)\rVert\}}{T} \nonumber\\&\leq_{(b)} \frac{n}{V} +  \frac{n\varepsilon^2 e^{\frac{\eta (BV+1)}{\varepsilon}}}{\eta V} +\frac{\varepsilon n^2 (BV+1)}{V} +\frac{1}{\eta VT}\log\left(\frac{1}{\varepsilon} \right) + \frac{n(BV+1)^2}{2 VT}+\frac{nB(BV+1)}{T}, \nonumber
\end{align}
where (a) follows due to Lemma~\ref{lemma:gamma_to_X_1}, and (b) follows due to Corollary~\eqref{corr:queue_bound}. Now, using Jensen's inequality, we are done.
\end{proof}
\subsection{Non-Adaptive MAC}
In this section, we present a UCB-based algorithm for the link selection problem in MAC (Algorithm~\ref{algo:8}), after which we move on to the analysis. For $t \in [s]$, we set $\vec{Y}(t)$ arbitrarily such that each user-channel pair is explored exactly once during the interval $[s]$. For each $t >s$, we compute permutation matrix $\vec{Y}(t)$, an auxiliary vector $\vec{\gamma}(t)$, and a virtual queue $\vec{Q}(t) = [Q_1(t),\dots,Q_n(t)]$ similar to Algorithm~\ref{algo:61}. The algorithm uses parameters $\delta_{s}, \delta_{s+1},\dots$ that satisfy $\delta_t > 0$ for all $t \in \{s,s+1,\dots\}$. For each $i \in [n]$, $j \in [m]$, we define,
\begin{align}\label{eqn:ucb_term}
    &n_{i,j}(t) = \sum_{\tau = 1}^t Y_{i,j}(t),\ \forall t \in \mathbb{N}\nonumber\\&
    \hat{S}_{i,j}(t) = \begin{cases}
        \frac{\sum_{\tau = 1}^t Y_{i,j}(t)S_{i,j}(t)}{n_{i,j}(t)} &  \text{ if } n_{i,j}(t)>0\\
        0 & \text{ otherwise}
    \end{cases}, \ \forall t \in \mathbb{N}\nonumber\\
    & f_{i,j}(t) = \sqrt{\frac{\log\left(\frac{n_{i,j}(t)[n_{i,j}(t)+1]}{\delta_t}\right)}{2n_{i,j}(t)}},\ \forall t \in \{s,s+1,\dots\} \nonumber\\
    & \text{UCB}_{i,j}(t) = \hat{S}_{i,j}(t)+ f_{i,j}(t),\ \forall t \in \{s,s+1,\dots\}
\end{align}
\begin{algorithm}
For $t \in [s]$, choose $\vec{Y}(t) \in \{0,1\}^{n \times m}$ arbitrarily such that each user-channel link is explored exactly once.\\

Initialize $ \vec{Q}(s+1) = 0$.\\
\For{each time slot $t \in \{s+1,s+2,\dots,\}$}{
  Find $\vec{\gamma}(t)$ and $\vec{Y}(t)$ by solving
  \begin{align}\label{eqn:intem_prob}
      &\vec{\gamma}(t) = \arg\min_{\vec{\gamma} \in [0,1]^n}-V\phi(\vec{\gamma})+\sum_{i=1}^N Q_i(t)\gamma_i, \\
      &\vec{Y}(t) = \arg\max_{\vec{Y} \in \mathcal{S}^{\text{doub}}}\sum_{i=1}^n\sum_{j=1}^m Q_i(t)\text{UCB}_{i,j}(t-1)Y_{i,j}, \nonumber
  \end{align}
  where $\text{UCB}_{i,j}(t-1)$ is defined in \eqref{eqn:ucb_term}.\\
  Receive feedback $\vec{Y}(t) \odot \vec{S}(t)$.\\
  Update the queues
  \begin{align}\label{eqn:queue_eqn_2}
      Q_i(t+1) = [Q_i(t)+\gamma_i(t) - X_{i}(t)]_+,
  \end{align}
  $\forall \ i \in [n]$, where $X_i(t) = \sum_{j=1}^m Y_{i,j}(t)S_{i,j}(t)$.\\

}
\caption{UCB MAC}\label{algo:8}
\end{algorithm}

We first state the performance bound of Algorithm~\ref{algo:8}.
\begin{theorem}\label{thm:ucb_bound}
 Given a time horizon $T \in \{s+1,s+2,\dots\}$, $V> 0$, and $\delta_{s},\delta_{s+1},\dots,\delta_{T-1} > 0$, we have 
 \begin{align}\label{eqn:second}
      \phi^{\text{opt}}& -  \phi\left(\mathbb{E}\left\{\frac{1}{T}\sum_{t=1}^T \vec{X}(t)\right\}\right) \nonumber\\&\leq  \frac{n}{V}  + \frac{2(BV+1)}{V(T-s)}\sum_{t=s+1}^T \sum_{i=1}^n\sum_{j=1}^m\mathbb{E}\left\{Y_{i,j}(t)f_{i,j}(t-1)\right\}   + \frac{2(2V\phi^{\text{max}}+nBV+n)nm}{V(T-s)}\sum_{t=s}^{T-1}\delta_t \nonumber\\&\ \ \ \ +\frac{nB(BV+1)}{T-s}+ \frac{(\sqrt{n}B+1)s}{T}, 
 \end{align}
 where $f_{i,j}(t)$ is defined in \eqref{eqn:ucb_term} and $\phi^{\text{max}} = \max_{\vec{\gamma} \in [0,1]^n} \phi(\vec{\gamma})$. In particular using $\delta_t = 1/t$ for all $t \geq s$, we have 
 \begin{enumerate}
     \item For all $V> 0$,
     \begin{align}
          &\phi^{\text{opt}} -  \phi\left(\mathbb{E}\left\{\frac{1}{T}\sum_{t=1}^T \vec{X}(t)\right\}\right) \nonumber\\&\leq \frac{n}{V}  + \frac{2\sqrt{6}nm(BV+1)\sqrt{T \log(T)}}{V(T-s)} + \frac{2(2V\phi^{\text{max}}+nBV+n)nm(\log(T)+1)}{V(T-s)}\nonumber\\&\ \ \ \ +\frac{nB(BV+1)}{T-s}+ \frac{(\sqrt{n}B+1)s}{T}. \nonumber
     \end{align}
     \item Assume $T \geq 2s$. Using $V = \Theta(\sqrt{T})$, we have 
     \begin{align}
    \phi^{\text{opt}}  -\phi\left(\mathbb{E}\left\{\frac{1}{T}\left(\sum_{t=1}^{T} \vec{X}(t)\right)\right\}\right) = \mathcal{O}\left(\sqrt{\frac{\log(T)}{T}}\right),\nonumber
\end{align}
and $\mathcal{O}$ hides the dependence on all parameters but $T$.
 \end{enumerate}
 \end{theorem}
\subsubsection{Solving Problem~\eqref{eqn:intem_prob}}
We first focus on solving the inner problem~\eqref{eqn:intem_prob}. Notice that the problem to solve to obtain $\vec{\gamma}(t)$ is the same as for Algorithm~\ref{algo:61}. The problem to solve to obtain $\vec{Y}(t)$ has a classic max-weight structure (hence, $\vec{Y}(t)$ is a permutation matrix). Hence, we can use the Hungarian algorithm~\cite{Kuhn1955Hungarian} to obtain the solution.

\subsubsection{Analysis of the Algorithm}
Now, we focus on proving Theorem~\ref{thm:ucb_bound}.
 Since the queue updates and the problem to solve to obtain $\vec{\gamma}(t)$ are the same as in Algorithm~\ref{algo:61}, we have the same deterministic queue bound. We formally state this in the following lemma.
\begin{lemma}\label{lemma:det_queue_bnd}
We have for all $t \in \{s+1,s+2,\dots\}$ and $i \in [n]$, $Q_i(t)  \leq BV + 1$
\end{lemma}

Define the collection of \emph{good events} $\Omega_{s+1},\Omega_{s+2},\dots$ as
 \begin{align}\label{eqn:good_event}
     \Omega_t = \Bigg{\{}&\text{UCB}_{i,j}(t-1)-2f_{i,j}(t-1) \leq q_{i,j} \leq \text{UCB}_{i,j}(t-1),\forall i \in [n], j\in [m] \Bigg{\}},
 \end{align}
 where $\text{UCB}_{i,j}(t-1)$ and $f_{i,j}(t-1)$ are defined in \eqref{eqn:ucb_term}.
We have the following lemma.
\begin{lemma}\label{lemma:good_events}
    For each $t \in \{s+1,s+2,\dots,\}$, we have $\mathbb{P}\{\Omega_t\} \geq 1  - 2nm \delta_{t-1}$.
\begin{proof}
    The proof follows directly by applying the Hoeffding inequality with union bound. See~\cite{lattimore_szepesvári_2020} for more details.
\end{proof}
\end{lemma}
Define the drift $\Delta_t$
\begin{align}
    \Delta_t = \frac{1}{2}\mathbb{E}\{\lVert \vec{Q}(t+1)\rVert^2 - \lVert \vec{Q}(t)\rVert^2\}\nonumber
\end{align}
and history $\mathcal{H}(t)$ as
\begin{align}\label{eqn:history}
    \mathcal{H}(t) = \{&\vec{Y}(1),\dots, \vec{Y}(t-1),\vec{Y}(1)\odot\vec{S}(1),\dots,\vec{Y}(t-1)\odot\vec{S}(t-1)\},
\end{align}
for each $t \in \{s+1,s+2,\dots\}$. Notice that $\vec{Y}(t),\vec{\gamma}(t),\vec{Q}(t)$ are $\mathcal{H}(t)$-measurable.
\begin{lemma}\label{lemma:drift_lemma}
We have for each $t \in \{s+1,s+2,\dots\}$
\begin{align}
    \Delta(t) \leq n + \mathbb{E}\left\{\sum_{i=1}^nQ_i(t)\left[\gamma_i(t) -\sum_{j=1}^mq_{i,j}Y_{i,j}(t)\right] \right\}. \nonumber
\end{align}
\begin{proof}
Notice that from the queuing equation~\eqref{eqn:queue_eqn_2}, we have 
\begin{align}
    Q^2_i(t+1) &\leq (Q_i(t) + \gamma_i(t) -X_i(t))^2  \leq Q^2_i(t) + \gamma_i^2(t) +X^2_i(t)+ 2Q_i(t)[\gamma_i(t) -X_i(t)] \nonumber\\& \leq Q^2_i(t) + 2 + 2Q_i(t)[\gamma_i(t) -X_i(t)]\nonumber
\end{align} 
for all $i \in [n]$. Summing the above for $i \in [n]$, we have 
\begin{align}
    \lVert \vec{Q}(t+1)\rVert^2 &\leq \lVert \vec{Q}(t)\rVert^2 + 2n  + 2 \sum_{i=1}^n Q_i(t)[\gamma_i(t) -X_i(t)]. \nonumber
\end{align}
Taking the expectations conditioned on the history $\mathcal{H}(t)$ defined in \eqref{eqn:history}, we have 
\begin{align}
    \mathbb{E}\{\lVert \vec{Q}(t+1)\rVert^2 |\mathcal{H}(t) \} &\leq \lVert \vec{Q}(t)\rVert^2+ 2n + 2\sum_{i=1}^nQ_i(t)[\mathbb{E}\{\gamma_i(t) |\mathcal{H}(t)\} -\mathbb{E}\{X_i(t)|\mathcal{H}(t)\}]  \nonumber\\&= \lVert\vec{Q}(t)\rVert^2 + 2n + 2\sum_{i=1}^nQ_i(t)\left[\gamma_i(t) -\sum_{j=1}^mq_{i,j}Y_{i,j}(t)\right],\nonumber
\end{align}
where the last equality follows since $\vec{Y}(t),\vec{\gamma}(t),\vec{Q}(t)$ are $\mathcal{H}(t)$-measurable. Taking expectations, we have the result.
 \end{proof}
 \end{lemma}
Fix a time horizon $T \in \{s+1,s+2,\dots\}$. Then we have the following result. 
\begin{lemma}\label{eqn:most_imp_lemma}
Given a time horizon $T \in \{s+1,s+2,\dots\}$, $V> 0$, and $\delta_{s},\delta_{s+1},\dots,\delta_{T-1} > 0$, we have
\begin{align}
    \phi^{\text{opt}} - \mathbb{E}\left\{ \phi\left(\frac{1}{T-s}\sum_{t=s+1}^T\vec{\gamma}(t)\right)\right\} &\leq \frac{n}{V} + \frac{2(BV+1)}{V(T-s)}\sum_{t=s+1}^T \sum_{i=1}^n\sum_{j=1}^m\mathbb{E}\left\{Y_{i,j}(t)f_{i,j}(t-1)\right\} \nonumber\\&\ \ \ \  + \frac{2(2V\phi^{\text{max}}+nBV+n)nm}{V(T-s)}\sum_{t=s}^{T-1}\delta_t, \nonumber
\end{align}
 where $f_{i,j}(t)$ is defined in \eqref{eqn:ucb_term}.
\begin{proof}
We begin with the following two claims. 

\noindent
\textbf{Claim 1: } We have that,
\begin{align}
    &-V\phi(\vec{\gamma}(t))+\sum_{i=1}^nQ_i(t)\left[\gamma_i(t) -\sum_{j=1}^mq_{i,j}Y_{i,j}(t)\right] \leq V\phi^{\text{max}}+nBV+n \nonumber
\end{align}
\begin{proof}
    The proof immediately follows from Lemma~\ref{lemma:det_queue_bnd} and the definition of $\phi^{\text{max}}$ in Assumption~\textbf{A1}.
\end{proof}

\noindent
\textbf{Claim 2: } We have that,
\begin{align}
&\mathbb{E}\left\{-V\phi(\vec{\gamma}(t))+\sum_{i=1}^nQ_i(t)\left[\gamma_i(t) -\sum_{j=1}^mq_{i,j}Y_{i,j}(t)\right] \Bigg{|} \Omega_t\right\} 
\nonumber\\& \leq -V\phi^{\text{opt}} +2 (BV+1)\mathbb{E}\left\{\sum_{i=1}^n\sum_{j=1}^m f_{i,j}(t-1)Y_{i,j}(t)\Bigg{|} \Omega_t\right\}\nonumber
\end{align}
\begin{proof}
From the definition of the \emph{good event} $\Omega_t$ in  \eqref{eqn:good_event},  we have
\begin{align}\label{eqn:111}
    &\mathbb{E}\left\{\sum_{i=1}^nQ_i(t)\sum_{j=1}^mq_{i,j}Y_{i,j}(t)\Bigg{|} \Omega_t\right\} \nonumber \\& \geq \mathbb{E}\left\{\sum_{i=1}^nQ_i(t)\sum_{j=1}^m \text{UCB}_{i,j}(t-1)Y_{i,j}(t)\Bigg{|} \Omega_t\right\}  -2\mathbb{E}\left\{\sum_{i=1}^nQ_i(t)\sum_{j=1}^m f_{i,j}(t-1)Y_{i,j}(t)\Bigg{|} \Omega_t\right\}\nonumber \\& \geq \mathbb{E}\left\{\sum_{i=1}^nQ_i(t)\sum_{j=1}^m \text{UCB}_{i,j}(t-1)M_{i,j}(t)\Bigg{|} \Omega_t\right\}  \nonumber\\&\ \ \ \ -2(BV+1)\mathbb{E}\left\{\sum_{i=1}^n\sum_{j=1}^m f_{i,j}(t-1)Y_{i,j}(t)\Bigg{|} \Omega_t\right\}
\end{align}
where the last inequality follows from Lemma~\ref{lemma:det_queue_bnd}. Also, notice that,
\begin{align}\label{eqn:112}
   &\mathbb{E}\left\{-V\phi(\vec{\gamma}(t))+\sum_{i=1}^nQ_i(t)\gamma_i(t)\Bigg{|} \Omega_t\right\} -\mathbb{E}\left\{\sum_{i=1}^nQ_i(t)\sum_{j=1}^m\text{UCB}_{i,j}(t-1)M_{i,j}(t) \Bigg{|} \Omega_t\right\}\nonumber\\& \leq_{(a)} \mathbb{E}\left\{-V\phi(\vec{\gamma}^*)+\sum_{i=1}^nQ_i(t)\gamma_i^*\Bigg{|} \Omega_t\right\}  -\mathbb{E}\left\{\sum_{i=1}^nQ_i(t)\sum_{j=1}^m\text{UCB}_{i,j}(t-1)P_{i,j}^* \Bigg{|} \Omega_t\right\}\nonumber\\& \leq_{(b)} \mathbb{E}\left\{-V\phi(\vec{\gamma}^*)+\sum_{i=1}^nQ_i(t)\gamma_i^*\Bigg{|} \Omega_t\right\} -\mathbb{E}\left\{\sum_{i=1}^nQ_i(t)\sum_{j=1}^mq_{i,j}P_{i,j}^* \Bigg{|} \Omega_t\right\} \nonumber\\&\leq_{(c)} -V \phi^{*} \leq -V\phi^{\text{opt}}
\end{align}
where (a) follows from the optimality of $\vec{\gamma}(t),\vec{Y}(t)$ for the intermediate problem \eqref{eqn:intem_prob} in Algorithm~\ref{algo:8} (recall $(\vec{P}^*,\vec{\gamma}^*)$ is the optimal solution of P2); (b) follows from the definition of the \emph{good event} $\Omega_t$ in  \eqref{eqn:good_event}; (c) follows from  the constraint \eqref{cons:pq_geq_gam}; the last inequality follows from Lemma~\ref{lemma:equiv_prob_lemma}. Adding the inequalities \eqref{eqn:111} and \eqref{eqn:112} and rearranging, we are done.
\end{proof}

Combining claim 1 and claim 2, and the law of total probability, we have that,
\begin{align}\label{eqn:one_last}
    &\mathbb{E}\left\{-V\phi(\vec{\gamma}(t))+\sum_{i=1}^nQ_i(t)\left[\gamma_i(t) -\sum_{j=1}^mq_{i,j}Y_{i,j}(t)\right] \right\}  \nonumber\\& \leq ( V\phi^{\text{max}}+nBV+n)\mathbb{P}\{\Omega_t^c\}-V\phi^{\text{opt}}\mathbb{P}\{\Omega_t\} \nonumber\\&\ \ \ \ +2 (BV+1)\mathbb{E}\left\{\sum_{i=1}^n\sum_{j=1}^m f_{i,j}(t-1)Y_{i,j}(t)\Bigg{|} \Omega_t\right\}\mathbb{P}\{\Omega_t\} \nonumber\\& \leq_{(a)} ( V\phi^{\text{max}}+V\phi^{\text{opt}}+nBV+n)\mathbb{P}\{\Omega_t^c\}-V\phi^{\text{opt}} +2 (BV+1)\mathbb{E}\left\{\sum_{i=1}^n\sum_{j=1}^m f_{i,j}(t-1)Y_{i,j}(t)\right\} \nonumber\\& \leq_{(b)} 2( 2V\phi^{\text{max}}+nBV+n)nm\delta_{t-1}-V\phi^{\text{opt}} +2 (BV+1)\mathbb{E}\left\{\sum_{i=1}^n\sum_{j=1}^m f_{i,j}(t-1)Y_{i,j}(t)\right\} 
\end{align}
where (a) follows from the fact that for a nonnegative random variable $\vec{X}$ and an event $\mathcal{H}$ we have $\mathbb{E}\{X|\mathcal{H}\}\mathbb{P}\{\mathcal{H}\} \leq \mathbb{E}\{X\}$ and (b) follows since $\phi^{\text{opt}} \leq \phi^{\text{max}}$ and Lemma~\ref{lemma:good_events}. Adding \eqref{eqn:one_last} to the result of Lemma~\ref{lemma:drift_lemma} and rearranging, we have
\begin{align}
    &\Delta(t)-V\mathbb{E}\{\phi(\vec{\gamma}(t))\} \nonumber\\&\leq n - V\phi^{\text{opt}} +2 (BV+1)\mathbb{E}\left\{\sum_{i=1}^n\sum_{j=1}^m f_{i,j}(t-1)Y_{i,j}(t)\right\} +2( 2V\phi^{\text{max}}+nBV+n)nm\delta_{t-1} \nonumber
\end{align} 
Now, we sum the above for $t \in [s+1,T]$ to obtain
 \begin{align}
    &\frac{1}{2}\mathbb{E}\{\lVert \vec{Q}(T+1)\rVert^2-\lVert \vec{Q}(s+1)\rVert^2\}-V\sum_{t=s+1}^T \mathbb{E}\{\phi(\vec{\gamma}(t))\}\nonumber\\&\leq n(T-s) - V(T-s)\phi^{\text{opt}}+2(BV+1)\sum_{t=s+1}^T\mathbb{E}\left\{\sum_{i=1}^n\sum_{j=1}^mf_{i,j}(t-1)Y_{i,j}(t)\right\}\nonumber\\& \ \ \ \ +2(2V\phi^{\text{max}}+nBV+n)nm\sum_{t=s}^{T-1}\delta_t. \nonumber
 \end{align}
  Using $\lVert \vec{Q}(T+1)\rVert^2 \geq 0$ and $\lVert \vec{Q}(s+1)\rVert^2 = 0$, we have
 \begin{align}
     &V(T-s)\phi^{\text{opt}} - V\sum_{t=s+1}^T \mathbb{E}\{\phi(\vec{\gamma}(t))\}\nonumber\\&\leq n(T-s) +2(BV+1)\sum_{t=s+1}^T\mathbb{E}\left\{\sum_{i=1}^n\sum_{j=1}^mf_{i,j}(t-1)Y_{i,j}(t)\right\}\nonumber\\& \ \ \ \ +2(2V\phi^{\text{max}}+nBV+n)nm\sum_{t=s}^{T-1}\delta_t. \nonumber
 \end{align}
Dividing both sides by $V(T-s)$, and using Jensen's inequality, we are done.
\end{proof}
\end{lemma}
Now we have the following lemma that combines $\vec{\gamma}(s+1),\vec{\gamma}(s+2),\dots,\vec{\gamma}(T)$ with $\vec{X}(s+1),\vec{X}(s+2),\dots,\vec{X}(T)$.
\begin{lemma}\label{lemma:gamma_to_X_2}
We have 
\begin{align}
 &\phi\left(\frac{1}{T-s}\left(\sum_{t=s+1}^{T} \vec{\gamma}(t)\right)\right) \leq  \phi\left(\frac{1}{T-s}\left(\sum_{t=s+1}^{T} \vec{X}(t)\right)\right)+\frac{nB(BV+1)}{T-s}. \nonumber
\end{align}
\begin{proof}
The proof uses the same argument as Lemma~\ref{lemma:gamma_to_X_1}.
\end{proof}
\end{lemma}
Now, we are ready to prove Theorem~\ref{thm:ucb_bound}.
\begin{proof}[Proof of Theorem~\ref{thm:ucb_bound}]
  First, notice that
\begin{align}\label{eqn:22}
    &\left\lvert\phi\left(\frac{1}{T-s}\sum_{t=s+1}^T \vec{X}(t)\right) - \phi\left(\frac{1}{T}\sum_{t=1}^T \vec{X}(t)\right) \right\rvert\nonumber\\&\leq \sqrt{n}B \left\lVert \frac{1}{T-s}\sum_{t=s+1}^T \vec{X}(t) - \frac{1}{T}\sum_{t=1}^T \vec{X}(t)\right\rVert \sqrt{n}B\left\lVert \frac{1}{T}\sum_{t=1}^s \vec{X}(t)+\frac{s}{T(T-s)}\sum_{t=s+1}^T \vec{X}(t)\right\rVert\nonumber\\&\leq \frac{\sqrt{n}B}{T}\sum_{t=1}^s \left\lVert  \vec{X}(t)\right\rVert+\frac{s}{T(T-s)}\sum_{t=s+1}^T \left\lVert\vec{X}(t)\right\rVert \leq \frac{\sqrt{n}sB}{T}+\frac{s}{T} = \frac{(\sqrt{n}B+1)s}{T},
\end{align}
where the first inequality follows since $\phi$ is $\sqrt{n}B$-Lipschitz continuous from Assumption~\textbf{A1}.

Combining Lemma~\ref{lemma:gamma_to_X_2} and Lemma~\ref{eqn:most_imp_lemma}, we have 
 \begin{align}\label{eqn:11}
        &\phi^{\text{opt}} - \mathbb{E}\left\{ \phi\left(\frac{1}{T-s}\sum_{t=s+1}^T \vec{X}(t)\right)\right\} \nonumber\\&\leq \frac{n}{V}  + \frac{2(BV+1)}{V(T-s)}\sum_{t=s+1}^T \sum_{i=1}^n\sum_{j=1}^m\mathbb{E}\left\{Y_{i,j}(t)f_{i,j}(t-1)\right\}   + \frac{2(2V\phi^{\text{max}}+nBV+n)nm}{V(T-s)}\sum_{t=s}^{T-1}\delta_t \nonumber\\&\ \ \ \ +\frac{nB(BV+1)}{T-s}.
    \end{align}
Combining \eqref{eqn:22} and \eqref{eqn:11} and using the Jensen's inequality, we have \eqref{eqn:second}. 

To prove part 1, first fix $i \in [n]$ and $j \in [m]$. When $\delta_{t-1} = 1/(t-1)$ for all $t \in [s+1:T]$, we have 
\begin{align}\label{eqn:fij_bnd}
    f_{i,j}(t-1) &= \sqrt{\frac{\log\left(\frac{n_{i,j}(t-1)[n_{i,j}(t-1)+1]}{\delta_{t-1}}\right)}{2n_{i,j}(t-1)}}  \leq \sqrt{\frac{3\log\left(T\right)}{2n_{i,j}(t-1)}}.  
\end{align}
Also, $n_{i,j}(t)-n_{i,j}(t-1) = 1$ if and only if $Y_{i,j}(t) = 1$. Hence, we have 
\begin{align}\label{eqn:one_bef_last}
    &\mathbb{E}\left\{\sum_{t=s+1}^T \frac{Y_{i,j}(t)}{\sqrt{n_{i,j}(t-1)}}\right\} = \mathbb{E}\left\{\sum_{k=1}^{n_{i,j}(T-1)-1} \sqrt{\frac{1}{k}}\right\} \leq \mathbb{E}\left\{\sum_{k=1}^{T} \sqrt{\frac{1}{k}}\right\} \leq 2\sqrt{T}
\end{align}
where for the last inequality we have used $\sum_{\tau=1}^t \frac{1}{\sqrt{\tau}} \leq 2\sqrt{t}$ for all $t \geq 1$.  Combining \eqref{eqn:fij_bnd} and \eqref{eqn:one_bef_last}, we have
\begin{align}
    \sum_{t=s+1}^T \mathbb{E}\{Y_{i,j}(t)f_{i,j}(t-1)\} \leq \sqrt{6T\log(T)} \nonumber
\end{align}
Now substituting the above in \eqref{eqn:second}, we have 
\begin{align}
&\phi^{\text{opt}} -  \phi\left(\mathbb{E}\left\{\frac{1}{T}\sum_{t=1}^T \vec{X}(t)\right\}\right) \nonumber\\&\leq  \frac{n}{V}  + \frac{2\sqrt{6}nm(BV+1)\sqrt{T \log(T)}}{V(T-s)}  + \frac{2(2V\phi^{\text{max}}+nBV+n)nm}{V(T-s)}\sum_{t=s}^{T-1}\frac{1}{t} \nonumber\\&\ \ \ \ +\frac{nB(BV+1)}{T-s}+ \frac{(\sqrt{n}B+1)s}{T}.
\end{align}
Part 1 follows from above after using the fact that $\sum_{\tau=1}^t \frac{1}{\tau} \leq \ln(t)+1$ for all $t \geq 1$.

For part 2, notice that since $T \geq 2s$, from part 1 we have 
 \begin{align}
          &\phi^{\text{opt}} -  \phi\left(\mathbb{E}\left\{\frac{1}{T}\sum_{t=1}^T \vec{X}(t)\right\}\right) \nonumber\\&\leq   \frac{n}{V}  + \frac{4\sqrt{6}nm(BV+1) \sqrt{\log(T)}}{V\sqrt{T}}  + \frac{4(2V\phi^{\text{max}}+nBV+n)nm(\log(T)+1)}{VT}\nonumber\\&\ \ \ \ +\frac{2nB(BV+1)}{T}+ \frac{(\sqrt{n}B+1)s}{T}, \nonumber
     \end{align}
     which is clearly $\mathcal{O}\left(\sqrt{\frac{\log(T)}{T}}\right)$ when $V = \Theta(\sqrt{T})$.  
\end{proof}

\subsection{Single-Channel Case with Minimum Utility}\label{sec:sing_with_min}
Consider the single-channel case $m = 1$ with utility function $\phi(\vec{x}) = \min\{x_1, \dots, x_n\}$. Observe that if $q_i = 0$ for some $i \in [n]$, then $\phi\left(\mathbb{E}\{\vec{\bar{X}}(T)\}\right) = 0$ under any assignment policy. Consequently, $\phi^{\text{opt}} = 0$. Therefore, without loss of generality, we assume that $q_i > 0$ for all $i \in [n]$.

If the values $q_1,\dots,q_n$ are known, an optimal policy for (P1) is to sample user $i$ with probability $p_i^* = \frac{1/q_i}{\sum_{k=1}^n (1/q_k)}$. Under this policy, $\mathbb{E}[X_i(t)] = q_i p_i^* 
= \frac{1}{\sum_{k=1}^n (1/q_k)}$ for all $i \in [n]$. Hence, we have
$\phi^{\mathrm{opt}} 
= \frac{1}{\sum_{k=1}^n (1/q_k)}$.

Now consider the following procedure:
\begin{itemize}
    \item Sample a user uniformly at random.
    \item Continue to serve the user until a successful transmission.
    \item Resample a user uniformly at random and repeat.
\end{itemize}
This mechanism does not require knowledge of the $q_i$. Fix a user $i \in [n]$. Let $(T_j, R_j)_{j \ge 1}$ denote the renewal process where, $T_j$ is the number of slots required until the $j$-th sampled user achieves a success, $R_j = 1$ if the $j$-th sampled user is $i$, and $R_j = 0$ otherwise. Note that the sequence ${(T_j, R_j)}_{j \ge 1}$ is independent and identically distributed. Since users are sampled uniformly at random, and conditional on sampling user $k$, the transmission duration is geometrically distributed with mean $1/q_k$, we have $\mathbb{E}\{R_1\} = \frac{1}{n}$ and $\mathbb{E}[T_1] = \frac{1}{n} \sum_{k=1}^n \frac{1}{q_k}$. By the renewal reward theorem, with probability~1,
\begin{align}
\lim_{T \to \infty} 
\frac{1}{T} \sum_{t=1}^T X_i(t)
=
\frac{\mathbb{E}[R_1]}{\mathbb{E}[T_1]}
=
\frac{1}{\sum_{k=1}^n (1/q_k)} = \phi^{\text{opt}}.
\end{align}
Thus every user achieves throughput $\phi^{\text{opt}}$, and therefore
\begin{align}
\lim_{T \to \infty} \min\{\bar{X}_1(T),\dots,\bar{X}_n(T)\}
=
\phi^{\mathrm{opt}}
\end{align}
with probability 1. Hence, the mechanism converges to the optimal value without requiring knowledge of $q_1,\dots,q_n$.

Suppose now that the values $q_i$ change at some time. Because geometric service times are memoryless, the residual service time at the change point has the same distribution as a fresh geometric random variable with the new parameter. Therefore, from the change point onward, the system evolves as a delayed renewal process under the new parameters, and the renewal argument still applies. Hence, the process is adaptive. 
\section{Simulations}
We consider eight simulation scenarios. All experiments were conducted on a virtual machine equipped with an Intel Xeon 2.20 GHz CPU (2 virtual CPUs). The algorithms were implemented in Python. In the first four scenarios, we use the smooth utility function $\phi(\mathbf{x}) = \sum_{i=1}^n \log(1 + b_i x_i)$,
where $b_i > 0$ are fixed constants. In scenarios 5 and 6, we consider the nonsmooth utility function $\phi(\mathbf{x}) = a\sum_{i=1}^nx_i + b\min(x_1, x_2, \dots, x_n)$, where $a,b>0$, are fixed constants. In the last two scenarios, we use the utility function $b\min(x_1, x_2, \dots, x_n)$, where $a,b>0$, is a fixed constant.

We run each algorithm for $T = 10^6$ time slots.  
The channel success probabilities are changed once at time $T_0 = 5 \times 10^5$, in order to evaluate the adaptiveness of the algorithms.  
The algorithms are neither restarted nor informed of this change. To examine performance before and after the change point, we plot the evolution of the objective applied to empirical averages of the observed vectors $\vec{X}(t)$.

\begin{figure*}[t]
\centering
\subfloat[Scenario 1]{
    \includegraphics[width=0.31\linewidth]{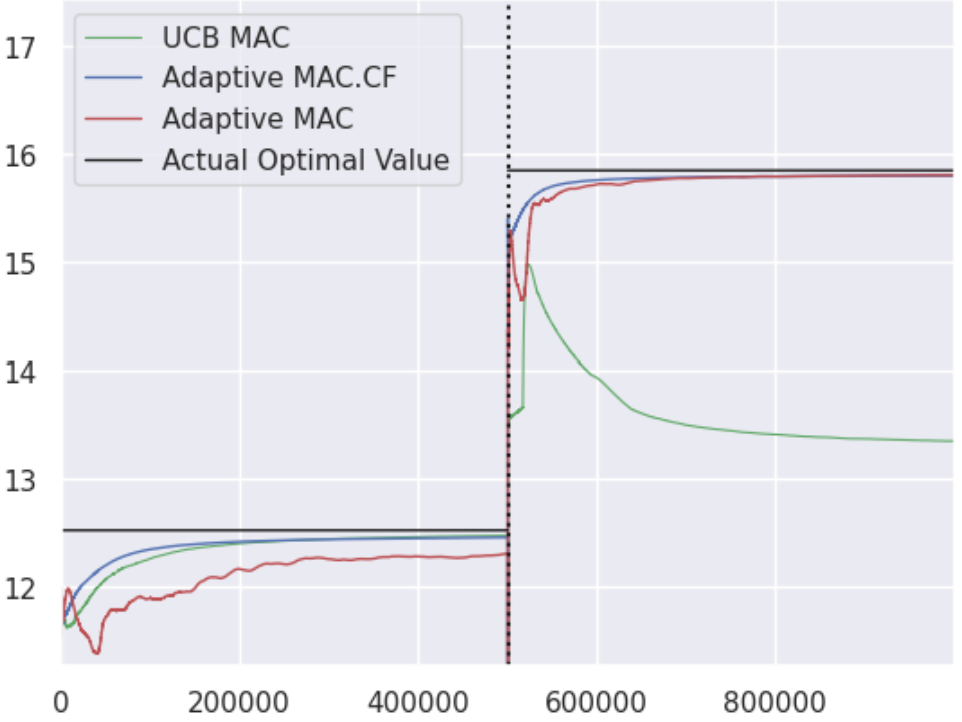}
    \label{fig:sim1}
}\hfill
\subfloat[Scenario 2]{
    \includegraphics[width=0.31\linewidth]{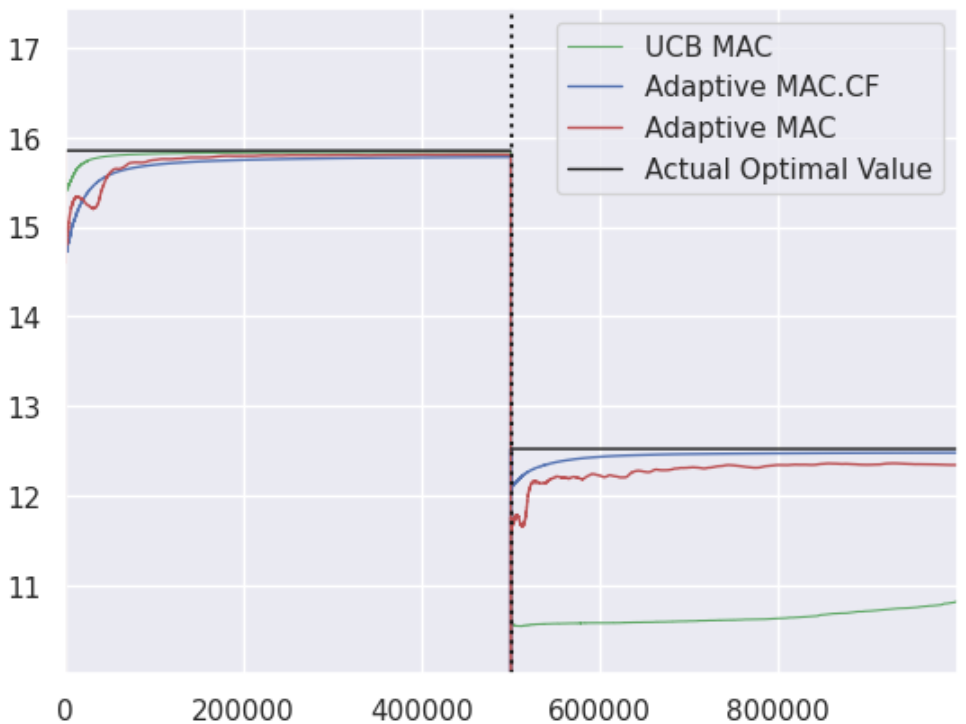}
    \label{fig:sim3}
}\hfill
\subfloat[Scenario 3]{
    \includegraphics[width=0.31\linewidth]{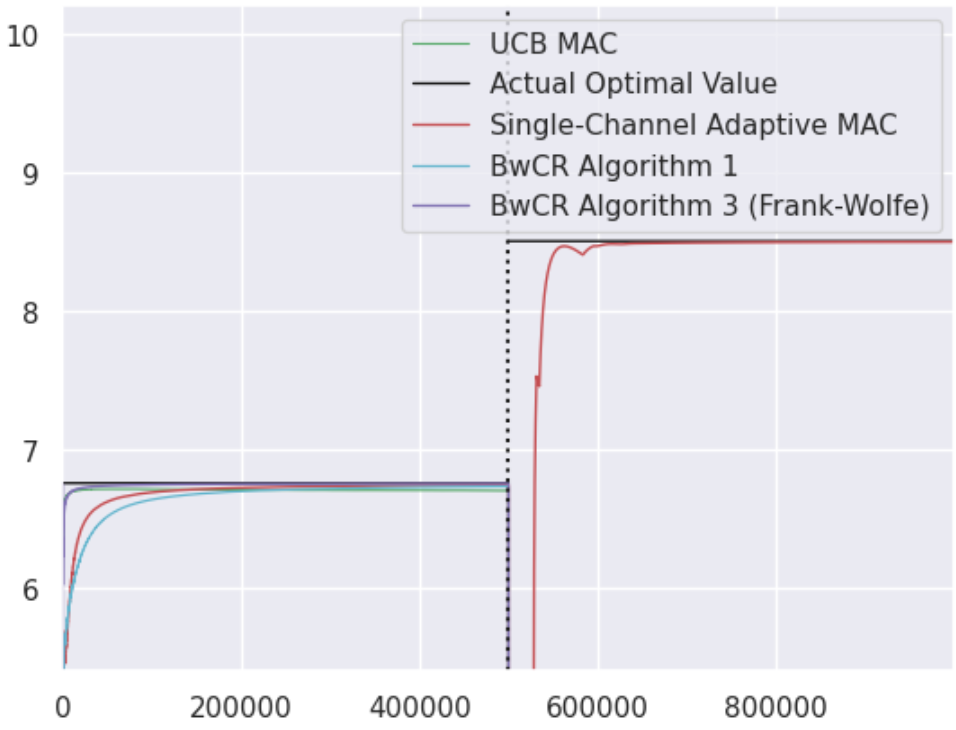}
    \label{fig:sim5}
}\hfill
\subfloat[Scenario 4]{
    \includegraphics[width=0.31\linewidth]{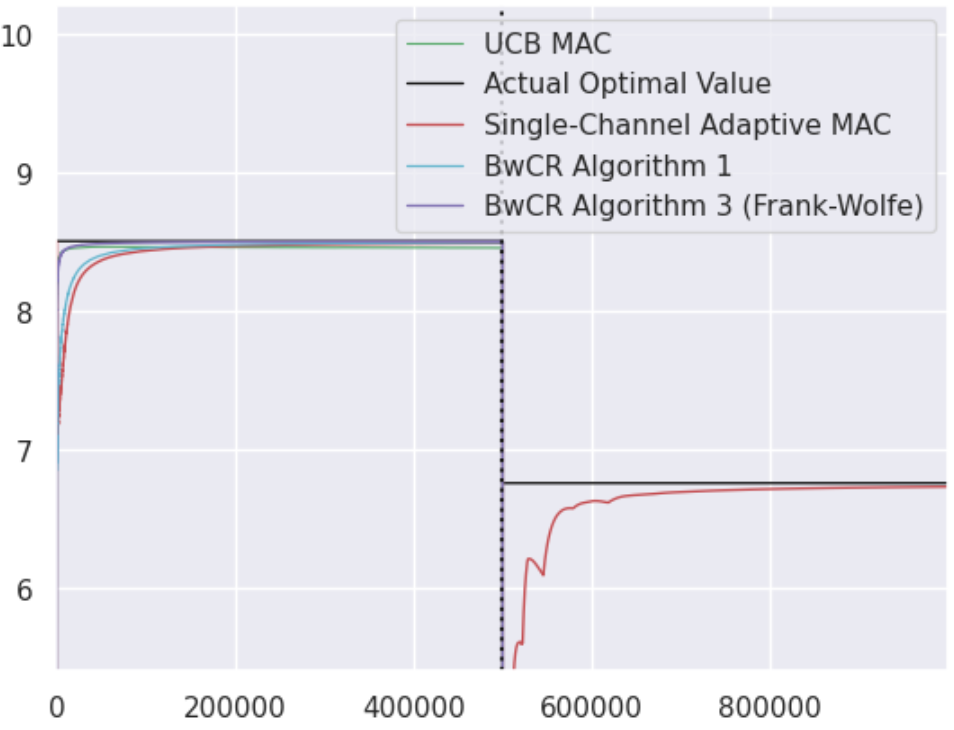}
    \label{fig:sim7}
}
\subfloat[Scenario 5]{
    \includegraphics[width=0.31\linewidth]{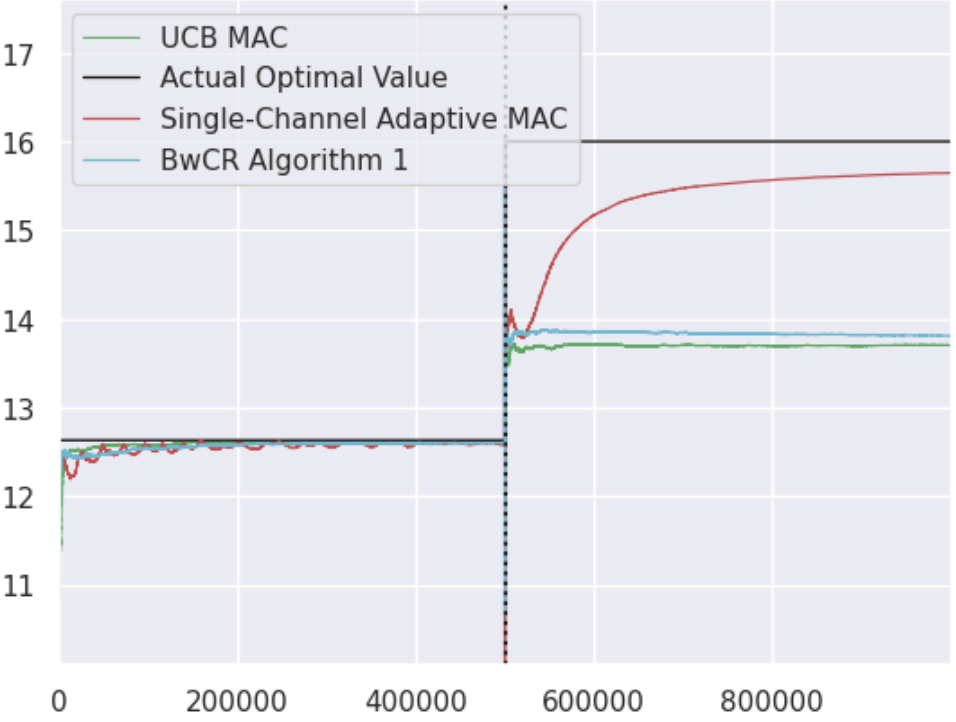}
    \label{fig:sim2}
}
\hfill
\subfloat[Scenario 6]{
    \includegraphics[width=0.31\linewidth]{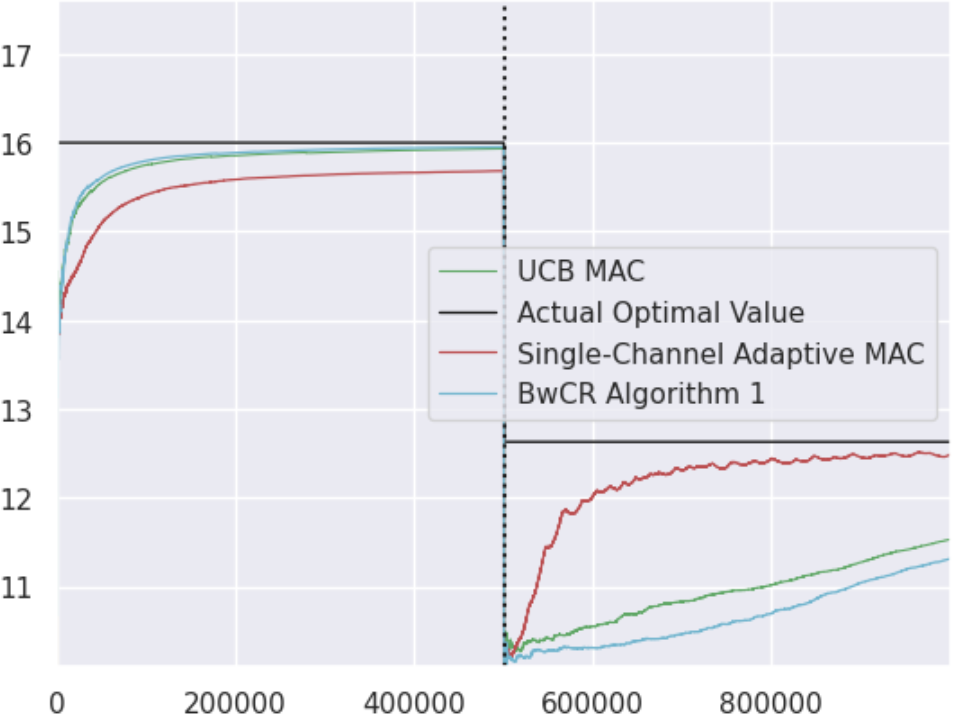}
    \label{fig:sim4}
}
\hfill
\subfloat[Scenario 7]{
    \includegraphics[width=0.31\linewidth]{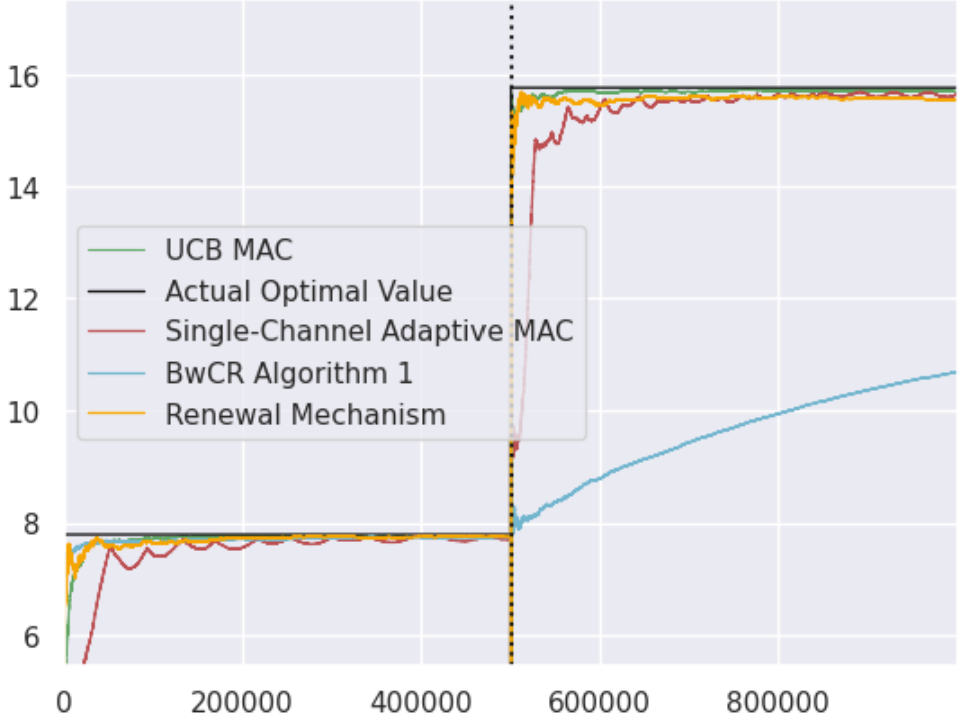}
    \label{fig:sim6}
}
\subfloat[Scenario 8]{
    \includegraphics[width=0.31\linewidth]{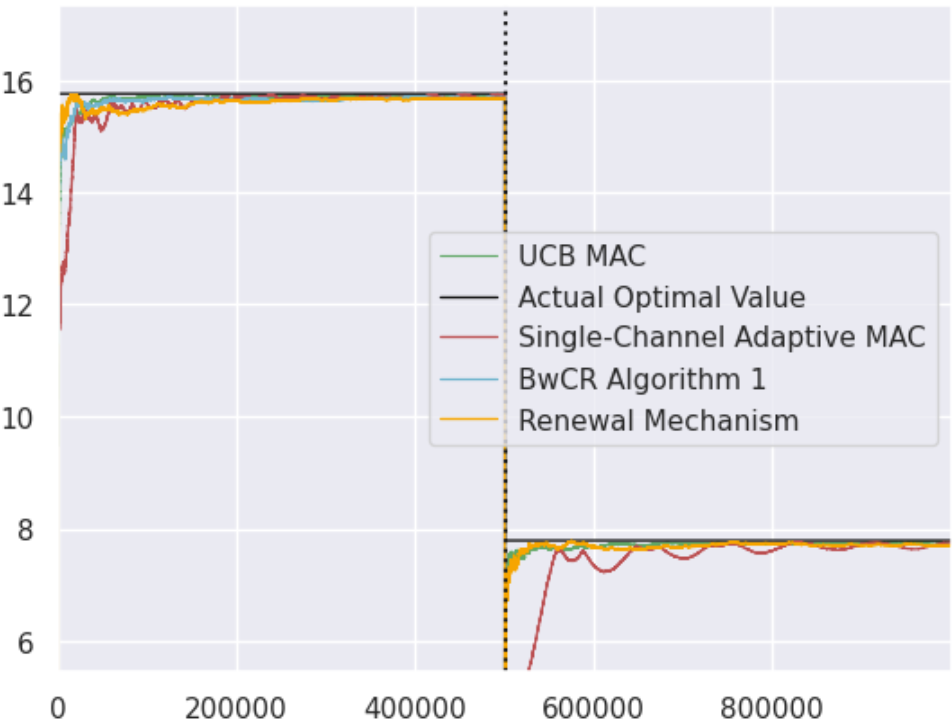}
    \label{fig:sim8}
}
\caption{Evolution of the objective 
$\Phi(t)$ defined in \eqref{eqn:Phi_1}.
A vertical dashed line marks $T_0 = 5 \times 10^5$, the point at which the success probabilities change.}
\label{fig:adaptation}
\end{figure*}

We compute
\begin{align}\label{eqn:Phi_1}
\Phi(t)
= \begin{cases}
    \phi\!\left(
\frac{1}{t}
\sum_{\tau=1}^{t}
\mathbf{X}(\tau)
\right),
& \text{ if } t < T_0\\
\phi\!\left(
\frac{1}{t - T_0 + 1}
\sum_{\tau=T_0}^{t}
\mathbf{X}(\tau)
\right) & \text{ if } t \geq T_0
\end{cases}
\end{align}
We then plot $\Phi(t)$ as a function of $t$. This representation allows us to separately examine the convergence behavior before the change and the transient adaptation after the change. In each plot, we also include the optimal utility value obtained by solving Problem~(P1) using the corresponding channel success probabilities. Figure~\ref{fig:adaptation} denotes the plots. In each plot, the y-axis is zoomed in to clearly highlight the performance differences among the algorithms. If the curve corresponding to a particular algorithm is not visible in the second half of the plot, it indicates that the algorithm has not adapted sufficiently to reach the new optimal value.
\captionsetup[subfloat]{labelformat=empty}

In Scenarios 1 and 2, we consider a multi-channel setting with 
$n=4$ and $m=3$. In Scenario 1, the success probabilities $q_{i,j}$ in the two 
halves of the simulation are chosen randomly.
In Scenario 2, we interchange these $q_{i,j}$ values. Here, we compare the Adaptive MAC (Algorithm~\ref{algo:61}), Adaptive MAC.CF (Algorithm~\ref{algo:8}), and UCB-MAC (Algorithm~\ref{algo:8}) algorithms. Notice that the adaptive algorithms have decent performance in the first half of the simulation. Although UCB-MAC has faster convergence in the first half, it does not adapt to changes in success probabilities in the second half. An interesting observation is that, despite the stronger worst-case guarantee of the Adaptive MAC algorithm, the Adaptive MAC.CF algorithm appears to outperform it in simulation.

In Scenarios 3 and 4, we consider a single-channel setting with 
$n=4$. We assume that some links may become unavailable without 
the controller’s knowledge; in that case, the corresponding 
$q_{i,j}$ is set to zero. When a link is available, its 
success probability is $q_{i,j}=0.4$. Within each half of the simulation horizon, the availability 
status of each link remains fixed. Here, for the adaptive algorithm, we use the Single-Channel Adaptive MAC (Algorithm~\ref{algo:62}). We also compare with the Algorithms introduced in~\cite{Agrawal2014} (Algorithm 1 of~\cite{Agrawal2014}, and Frank-Wolfe-based Algorithm 3 of~\cite{Agrawal2014}). In this case, the adaptation capability of the adaptive MAC algorithm can be clearly seen. In fact, in Scenario 3, the UCB-MAC algorithm performs extremely 
poorly in the second half of the simulation, resulting in an 
objective value that drops close to zero.

For Scenario 5 and Scenario 6, we again consider a single channel scenario.  Recall that the utility function in this case is $\phi(\vec{x}) = a\sum_{i=1}^n x_i +b\min(x_1,x_2,\dots,x_n)$. We compare Single-Channel Adaptive MAC, UCB-MAC, and Algorithm 1 of~\cite{Agrawal2014}. Notice that even for the nonsmooth utility, the adaptive algorithm adapts fast in the second half of the simulation.

In Scenarios 7 and 8, we consider the single-channel setting with utility $\phi(\vec{x}) = b \min(x_1, x_2, \dots, x_n)$. We also plot the simple renewal mechanism described in Section~\ref{sec:sing_with_min}.
In these scenarios, UCB-MAC also adapts in the second half of the simulation. Interestingly, we observe that UCB-MAC does not rely on explicit UCB-based estimation in this case. Instead, it effectively implements a mechanism that closely resembles the renewal mechanism.

In Table~\ref{tab:scenario_comparison}, we report the average runtime per time slot of the algorithms under different scenarios. We observe that Adaptive MAC.CF has faster iterations than Adaptive MAC. However, the UCB-based non-adaptive algorithm outperforms both adaptive algorithms in terms of runtime. In the single-channel scenario, the adaptive algorithm achieves runtimes comparable to those of the UCB MAC algorithm. Finally, the UCB-based non-adaptive algorithm outperforms the algorithms proposed in~\cite{Agrawal2014} in terms of empirical runtime complexity. 
\begin{table*}[t]
\centering
\caption{Average runtime per time slot(in milliseconds) of each algorithm under each scenario. For the Adaptive MAC algorithm, we also report (in parenthesis) the average number of Sinkhorn iterations per main iteration. }
\label{tab:scenario_comparison}
\begin{tabular}{|l|c|c|c|c|c|c|c|c|}
\hline
\textbf{Algorithm} 
& \textbf{Scenario 1} 
& \textbf{Scenario 2} 
& \textbf{Scenario 3} 
& \textbf{Scenario 4} 
& \textbf{Scenario 5} 
& \textbf{Scenario 6} 
& \textbf{Scenario 7} 
& \textbf{Scenario 8} \\
\hline
Adaptive&3.878   &3.849 & - & - & - & - & - &- \\
MAC &(12.982)  &(12.054)  & - & - & - & - & - &- \\
\hline
Adaptive& 2.655& 2.433  & - & - & - & -  & - &- \\
MAC.CF & &   &  &  &  &   &  & \\
\hline
UCB MAC &	0.207&0.211&	0.157&	0.171&	0.0682 &	0.07&0.0651& 0.0655   \\
\hline
Single &  &  &  &  & &   &  & \\
Channel& - & - & 0.242 &	0.27&	0.0729&	0.0744 & 0.0706 & 0.0715\\
MAC & &  & & &  &	 &  & \\
\hline
BwCR& & & & & & & & \\
Algorithm 1&-&- &9.224&	9.556&	0.0691&	0.0773 &0.0672&0.0655\\
\cite{Agrawal2014}& & & & & & & & \\
\hline
BwCR & & & & & & & &\\
Algorithm 3&-  &-  & 0.17 &0.171  & - & - & - &-\\
\cite{Agrawal2014} & & & & & & & &\\
\hline
\end{tabular}
\end{table*}
\section{Conclusions}
This paper focuses on the problem of designing algorithms for automatic link selection in multi-channel multiple access systems with transmission failures. In particular, we solve a network utility maximization problem with matching constraints and bandit feedback on transmission outcomes. We propose two adaptive algorithms: the first achieves faster convergence, while the second admits a more efficient implementation. We also study several special cases with additional structural properties. Extending the framework to adversarial settings and incorporating more general combinatorial constraints are directions for future work.

{\appendices
\section{SINKHORN function}\label{app:sinkhorn}
We introduce SINKHORN proceure on positive $s \times s$ matrices. 

\noindent
\textbf{SINKHORN$(\vec{R})$}\\
\noindent
Initialize $\vec{R}^1 = \vec{R}$.\\
For each iteration $\tau \in \{1,2\dots, \}$
\begin{enumerate}
    \item{Row normalization:} If $\tau$ is even set $\vec{R}^{\tau +1}$ as 
    \begin{align}
        R_{i,j}^{\tau+1} = \frac{R_{i,j}^{\tau}}{\sum_{k=1}^s R_{i,k}^{\tau}}
    \end{align}
    \item{Column normalization:} Else set $\vec{R}(\tau +1)$ as 
    \begin{align}
        R_{i,j}^{\tau+1} = \frac{R_{i,j}^{\tau}}{\sum_{k=1}^s R_{k,j}^{\tau}}
    \end{align}
\end{enumerate}
 Given $\vec{X} \in \mathbb{R}^{s \times s}$, and $\vec{Q} \in (0,\infty)^{s\times s}$, SINKHORN studied in~\cite{FRANKLIN1989717} is an iterative process to solve problems of the type
\begin{align}\label{eqn:1010110101}
    \min_{\vec{P} \in \mathcal{S}^{\text{doub}}} \sum_{i=1}^s \sum_{j=1}^s X_{i,j}P_{i,j}+ D(\vec{P}\lVert \vec{Q})
\end{align}
We provide the following lemma adapted from \cite{FRANKLIN1989717}
\begin{lemma}\label{lemma:sink_lemma}
Given $\epsilon>0$, running SINKHORN($\vec{R}$), where $\vec{R}$ is defined by $R_{i,j} = Q_{i,j}\exp(-X_{i,j})$ for sufficiently many iterations (say $U$ iterations) generates a matrix $\vec{R}^U$ such that
\begin{align}\label{eqn:019999}
        \exp(-\epsilon) \leq \frac{R^U_{i,j}}{P^*_{i,j}} \leq \exp(\epsilon)
\end{align}
where $\vec{P}^*$ is the solution of Problem~\eqref{eqn:1010110101}.
\end{lemma}
\section{Finding $\vec{\hat{P}}(t+1)$ in Problem~\eqref{eqn:decision_1}}\label{app:intem_al_2}
Finding $\vec{\hat{P}}(t+1)$ can be summarized under the following steps.
\begin{enumerate}
    \item Define matrix $\vec{R} \in (0,\infty)^{s\times s}$ by $R_{i,j} = P_{i,j}(t)\exp(\eta Q_i(t)\hat{S}_{i,j}(t))$
    \item Run SINKHORN(\vec{R}) for $U$ iteration to get $\vec{R}^U$ such that
    \begin{align}\label{eqn:101010101}
   e^{-\theta/(3s)} \leq  \frac{R^U_{i,j}}{\tilde{P}_{i,j}(t+1)} \leq e^{\theta/(3s)}
\end{align}
This is possible due to  Lemma~\ref{lemma:sink_lemma}. See Appendix~\ref{app:sinkhorn} for details.
\item Set $\vec{\hat{P}}(t+1) = \text{ROUND}(\vec{R}^U)$. See Section~\ref{sec:round_sec} for details.
\end{enumerate}
We have the following lemma that establishes the required property of $\vec{\hat{P}}(t+1)$.
\begin{lemma}
    We have that $\vec{\hat{P}}(t+1)$ found above is a $(\theta/s)$-approximation of $\vec{\tilde{P}}(t+1)$
\begin{proof}
First, notice that for each $i \in [s]$
\begin{align}
    \sum_{k=1}^s R_{i,k}^U \leq \sum_{k=1}^s \tilde{P}_{i,j}(t+1)e^{\theta/(3s)} = e^{\theta/(3s)},
\end{align}
where the inequality follows from~\eqref{eqn:101010101}, and the equality follows since $\vec{\tilde{P}}(t+1)$ is doubly stochastic by definition. Similarly, for each $j \in [s]$,
we have $\sum_{k=1} ^ s R_{k,j}^U \leq e^{\theta/(3s)}$.

First, from Lemma~\ref{lemma:algo_8}, we have that $\vec{\hat{P}}(t+1) \in \mathcal{S}^{\text{doub}}$. Also, from the description of the ROUND function in Section~\ref{sec:round_sec}, notice that 
\begin{align}\label{eqn:hahaha}
    \hat{P}_{i,j}(t+1) &\geq \frac{R_{i,j}^U}{(\sum_{k=1}^s R_{i,k}^U)(\sum_{k=1}^s R_{k,j}^U)} \geq_{(a)} \frac{R_{i,j}^U}{ e^{2\theta/(3s)}}  \geq \frac{\tilde{P}_{i,j}(t+1)}{ e^{\theta/s}}
\end{align}
where (a) follows from the previous two inequalities, and (b) follows from \eqref{eqn:101010101}. Hence, we are done.
\end{proof}  
\end{lemma}

\section{Proof of Lemma~\ref{lemma:algo_8}}\label{app:algo_8}
Before proving this, notice that due to the normalization steps 1 and 2 of the ROUND function, we have $\vec{1} \geq \vec{P}^{''} \vec{1}$, and $\vec{1} \geq (\vec{P}^{''})^{\top}\vec{1}$, where the inequalities are taken entrywise. Hence, we have $\vec{Q}\geq \vec{P}^{''} \geq 0$. Now, we prove each part separately.

\noindent
1) First, suppose $\lVert \vec{1} - \vec{P}^{''}\vec{1} \rVert_1 = 0$. Then $\vec{P}^{''} \in \mathcal{S}^{\text{row}}$ and $\vec{Q} = \vec{P}^{''}$ (by the definition of $\vec{Q}$ in \eqref{eqn:q_def_round}). Also, due to the normalization of the columns in step 2, each column of $ \vec{P}^{''}$ has a sum of at most 1. However, notice that since $\vec{P}^{''}  \in \mathcal{S}^{\text{row}}$, the sum of its entries is $n$. Hence, the sum of each column must be exactly 1. Hence, $\vec{Q} =  \vec{P}^{''} \in \mathcal{S}^{\text{doub}}$ as desired. Next, suppose $\lVert \vec{1} - \vec{P}^{''}\vec{1} \rVert_1 > 0$. Then we have
\begin{align}
     \vec{Q}\vec{1} &=  \vec{P}^{''}\vec{1}+ \frac{(\vec{1} - \vec{P}^{''}\vec{1})(\vec{1} - (\vec{P}^{''})^{\top}\vec{1})^{\top} \vec{1}}{\lVert \vec{1} - \vec{P}^{''}\vec{1}\rVert_1} \nonumber\\& =_{(a)} \vec{P}^{''}\vec{1}+\frac{(\vec{1}\vec{1}^{\top} - \vec{P}^{''}\vec{1}\vec{1}^{\top}- \vec{1}\vec{1}^{\top}\vec{P}^{''}+\vec{P}^{''}\vec{1}\vec{1}^{\top}\vec{P}^{''})\vec{1}}{\vec{1}^{\top}(\vec{1} - \vec{P}^{''}\vec{1})} \nonumber\\&  = \vec{P}^{''}\vec{1}+\frac{\vec{1}\vec{1}^{\top}\vec{1} - \vec{P}^{''}\vec{1}\vec{1}^{\top}\vec{1}- \vec{1}\vec{1}^{\top}\vec{P}^{''}\vec{1}+\vec{P}^{''}\vec{1}\vec{1}^{\top}\vec{P}^{''}\vec{1}}{\vec{1}^{\top}\vec{1} - \vec{1}^{\top}\vec{P}^{''}\vec{1}} \nonumber\\& =  \vec{P}^{''}\vec{1}+\frac{s\vec{1} - s\vec{P}^{''}\vec{1}- \vec{1}\vec{1}^{\top}\vec{P}^{''}\vec{1}+\vec{P}^{''}\vec{1}\vec{1}^{\top}\vec{P}^{''}\vec{1}}{\vec{1}^{\top}\vec{1} - \vec{1}^{\top}\vec{P}^{''}\vec{1}}\nonumber\\& =  \vec{P}^{''}\vec{1}+\frac{(\vec{1} - \vec{P}^{''}\vec{1})(s- \vec{1}^{\top}\vec{P}^{''}\vec{1})}{s - \vec{1}^{\top}\vec{P}^{''}\vec{1}} = \vec{1}, \nonumber
\end{align}
where in (a) we have used $\vec{1} \geq \vec{P}^{''}\vec{1}$ when simplifying the denominator. The claim $\vec{1}^{\top}\vec{Q} = \vec{1}^{\top}$ follows repeating the same argument.

\noindent
2) We will prove the case $\vec{P} \in \mathcal{S}^{\text{row}}_{\varepsilon}$. The other case follows similarly. Since $\vec{P} \in \mathcal{S}^{\text{row}}_{\varepsilon}$, from \eqref{eqn:row_normalize}, we have $P^{'}_{i,j} = P_{i,j}$. Also, since $\vec{P} \in \mathcal{S}^{\text{row}}_{\varepsilon}$ we have $P_{i,j} \leq 1$ for all $i,j \in \{1,\dots,s\}$. Hence $P^{'}_{i,j} \leq 1$ for all $i,j \in \{1,\dots,s\}$. Hence, from \eqref{eqn:column_normalize}, we have $P^{''}_{i,j} \geq  P{'}_{i,j}/s = P_{i,j}/s \geq \varepsilon/s$. Since we already know $\vec{Q} \geq \vec{P}^{''}$, we have $Q_{i,j} \geq \varepsilon/s$ for all $i,j$. Part 1 shows $Q \in \mathcal{S}^{\text{doub}}$, so $Q \in \mathcal{S}^{\text{doub}}_{\varepsilon/s}$. So we are done.

\noindent
3) We first prove the following claim

\noindent
\textbf{Claim: } We have that $\lvert \vec{Q} - \vec{P}^{''}\rVert_1 \leq s - \lVert \vec{P}^{''}\rVert_1$

\begin{proof}
  First, suppose $\lVert \vec{1} - \vec{P}^{''}\vec{1} \rVert_1 = 0$. Then we have $\lVert \vec{Q} - \vec{P}^{''}\rVert_1 = 0$. Also, due to the scaling, we have $s - \lVert\vec{P}^{''}\rVert_1 \geq 0$. Hence, we have $\lVert \vec{Q} - \vec{P}^{''}\rVert_1 = 0 \leq s - \lVert\vec{P}^{''}\rVert_1$ and we are done. Next, suppose $\lVert \vec{1} - \vec{P}^{''}\vec{1} \rVert_1> 0$. Notice that
\begin{align}
     \lVert \vec{Q} - \vec{P}^{''}\rVert_1 &= \frac{\lVert (\vec{1} - \vec{P}^{''}\vec{1})(\vec{1} - (\vec{P}^{''})^{\top}\vec{1})^{\top} \lVert_1}{\lVert \vec{1} - \vec{P}^{''}\vec{1}\rVert_1}  = \frac{\lVert \vec{1}\vec{1}^{\top} - \vec{P}^{''}\vec{1}\vec{1}^{\top}- \vec{1}\vec{1}^{\top}\vec{P}^{''}+\vec{P}^{''}\vec{1}\vec{1}^{\top}\vec{P}^{''} \lVert_1}{\lVert \vec{1} - \vec{P}^{''}\vec{1}\rVert_1} \nonumber\\& =_{(a)} \frac{\vec{1}^{\top}\left( \vec{1}\vec{1}^{\top} - \vec{P}^{''}\vec{1}\vec{1}^{\top}- \vec{1}\vec{1}^{\top}\vec{P}^{''}+\vec{P}^{''}\vec{1}\vec{1}^{\top}\vec{P}^{''} \right)\vec{1}}{\vec{1}^{\top}\left(\vec{1} - \vec{P}^{''}\vec{1}\right)} \nonumber\\& = \frac{\vec{1}^{\top}\vec{1}\vec{1}^{\top}\vec{1} - \vec{1}^{\top}\vec{P}^{''}\vec{1}\vec{1}^{\top}\vec{1}- \vec{1}^{\top}\vec{1}\vec{1}^{\top}\vec{P}^{''}\vec{1}+\vec{1}^{\top}\vec{P}^{''}\vec{1}\vec{1}^{\top}\vec{P}^{''}\vec{1}}{\vec{1}^{\top}\vec{1} - \vec{1}^{\top}\vec{P}^{''}\vec{1}} \nonumber\\& = \frac{s^2 - 2s\lVert\vec{P}^{''}\rVert_1+\lVert\vec{P}^{''}\rVert_1^2}{s - \lVert\vec{P}^{''}\rVert_1}  = s - \lVert\vec{P}^{''}\rVert_1, \nonumber
\end{align}
where (a) follows since $\vec{1} - \vec{P}^{''}\vec{1} \geq 0$, and $\vec{1} - (\vec{P}^{''})^{\top}\vec{1} \geq 0$, as a result of which $(\vec{1} - \vec{P}^{''}\vec{1})(\vec{1} - (\vec{P}^{''})^{\top}\vec{1})^{\top} \geq 0$.   
\end{proof}

Now we prove part 3.
Denote by $\mathcal{U} = \left\{ i \in [s]: \sum_{j=1}^s P_{i,j} > 1\right\} $ and $\mathcal{U}^c  = [s] \setminus \mathcal{U}$. 
Notice that
\begin{align}\label{eqn:first}
    &\lVert \vec{P}\vec{1} - \vec{1}\rVert_1 + \lVert \vec{P}\rVert_1 - s = \sum_{i=1}^s \left\lvert \sum_{j=1}^s P_{i,j}-1\right\rvert+\sum_{i=1}^s \sum_{j=1}^s P_{i,j}-s \nonumber\\&= \sum_{i \in \mathcal{U}}\left( \sum_{j=1}^s P_{i,j}-1\right) +\sum_{i \in \mathcal{U}^c}\left( 1-\sum_{j=1}^s P_{i,j}\right) +\sum_{i=1}^s \sum_{j=1}^s P_{i,j}-s \nonumber\\&= 2\sum_{i \in \mathcal{U}} \sum_{j=1}^s P_{i,j} - |\mathcal{U}| +|\mathcal{U}^c|-s \nonumber\\&= 2\left(\sum_{i \in \mathcal{U}} \sum_{j=1}^s P_{i,j} - |\mathcal{U}|\right)= 2\sum_{i \in \mathcal{U}} \left(\sum_{j=1}^s P_{i,j} - 1\right) \nonumber\\& =_{(a)}  2(\rVert \vec{P} \rVert_1 - \lVert \vec{P}^{'} \rVert_1) =  2(\rVert \vec{P} \rVert_1 - \lVert \vec{P}^{''} \rVert_1)-2(\lVert \vec{P}^{'} \rVert_1 - \lVert \vec{P}^{''} \rVert_1 )\nonumber\\& =_{(b)}  2(\rVert \vec{P} \rVert_1 - \lVert \vec{P}^{''} \rVert_1)-2\left(\sum_{j=1}^s \left[\sum_{i=1}^s P^{'}_{i,j}-1\right]_+ \right)\nonumber\\& \geq_{(c)}   2(\rVert \vec{P} \rVert_1 - \lVert \vec{P}^{''} \rVert_1)-2\left(\sum_{j=1}^s \left[\sum_{i=1}^s P_{i,j}-1\right]_+ \right)\nonumber\\& \geq   2(\rVert \vec{P} \rVert_1 - \lVert \vec{P}^{''} \rVert_1)-2\left(\sum_{j=1}^s \left\lvert\sum_{i=1}^s P_{i,j}-1\right\rvert \right)\nonumber\\& \geq   2(\rVert \vec{P} \rVert_1 - \lVert \vec{P}^{''} \rVert_1)-2\lVert \vec{P}^{\top}\vec{1} - \vec{1}\rVert_1
\end{align}
where (a) and (b) follow from the definitions of $\vec{P}^{'}, \vec{P}^{''}$ in \eqref{eqn:row_normalize} and \eqref{eqn:column_normalize}, respectively; (c) follows since $P_{i,j} \geq P^{'}_{i,j}$ due to scaling. Rearranging the above inequality, we have \begin{align}\label{eqn:p_minus_q_bnd}
   &\lVert \vec{P} \rVert_1 - 2\lVert\vec{P}^{''} \rVert_1+s \leq \lVert \vec{P}\vec{1} - \vec{1} \lVert_1 + 2\lVert \vec{P}^{\top}\vec{1} - \vec{1} \rVert_1. 
\end{align}
Now, to complete the proof notice that
\begin{align}\label{eqn:p_min_q}
    \lVert \vec{P} - \vec{Q} \rVert_1 &\leq \lVert \vec{P} - \vec{P}^{''} \rVert_1+\lVert \vec{Q} - \vec{P}^{''} \rVert_1  =_{(a)} \lVert \vec{P}\rVert_1 - \lVert \vec{P}^{''} \rVert_1+\lVert \vec{Q} - \vec{P}^{''}\rVert_1 \nonumber\\& \leq_{(b)} \lVert \vec{P}\rVert_1 - 2\lVert \vec{P}^{''}\rVert_1 +s \leq_{(c)} \lVert \vec{P}\vec{1} - \vec{1} \lVert_1 + 2\lVert \vec{P}^{\top}\vec{1} - \vec{1} \rVert_1 \nonumber\\& \leq  2\left(\lVert \vec{P}\vec{1} - \vec{1} \lVert_1 + \lVert \vec{P}^{\top}\vec{1} - \vec{1} \rVert_1\right),
\end{align}
where (a) follows since $\vec{P}^{''} \leq \vec{P}$ due to the scaling, (b) follows due to the claim, and (c) follows from \eqref{eqn:p_minus_q_bnd}.

\section{Finding $\vec{\tilde{P}}(t+1)$ in Problem~\eqref{eqn:decision_1_}}\label{sec:intem_2}
We only consider the case where $t$ is even. The case where $t$ is odd can be solved similarly. Notice that we can separately solve for each column of $\vec{\tilde{P}}(t+1)$. To solve for the $j$-th column of $\vec{\tilde{P}}(t+1)$, we define $\vec{x},\vec{y} \in \mathbb{R}^s$, where 
\begin{align}
    x_i = \begin{cases}
       \eta Q_i(t)\hat{S}_{i,j}(t) & \text{ if } i \in [n], j\in [m]\\
        0 & \text{ otherwise}
    \end{cases}\nonumber
\end{align}
and $\vec{y}$ is the $j$-th column of  $\vec{\tilde{P}}(t)$. The problem to be solved is 
\begin{align}
  \text{(P3:) } &\underset{\vec{p}}{\text{min}}-\sum_{i=1}^s x_i p_i+ D_{\text{KL}}(\vec{p}\lVert \vec{y}) \nonumber\\& \text{s.t.  }\vec{p} \in \Delta^s_{\varepsilon},\nonumber
\end{align}
It should be noted that (P3) has a classic structure that is solved in~\cite{Huang2024}. In particular, we define $\vec{z}$, where $z_i = y_i\exp(x_i)$. First, assume that $\vec{z}$ is sorted in the increasing order. Then it can be shown that there exists $i \in [0:s-1]$ such that the vector $\vec{u}^i \in \mathbb{R}^s$ given by,
\begin{align}
   u^i_j = \begin{cases}
       \varepsilon & \text{ if } j \leq i\\
       \frac{z_j}{\sum_{l = i+1}^s z_l}(1-\varepsilon i) & \text{ if } j > i\\
   \end{cases} \nonumber
\end{align}
satisfies, $u^i_j \geq \varepsilon $ for all $j \in [i+1:s]$. Then it can be shown that $\vec{u}^i$ is the solution to (P3). Hence, solving (P3) amounts to calculating $\vec{u}^i$ for each $i \in [0:s-1]$ and checking the above condition. 
\bibliographystyle{IEEEtran}
\bibliography{main}
\end{document}